\documentclass[hyper,10pt,paper]{article}
\usepackage{jheppub}
\usepackage[usenames,dvipsnames,table]{xcolor}
\usepackage{graphicx,amsmath,amssymb,multirow,array,bm,mathrsfs,bbold,bbm,float}
\usepackage{epsf,amsfonts,slashed}
\usepackage[numbers,sort&compress]{natbib}
\usepackage[bbgreekl]{mathbbol}

\title{
Holographic Turbulence in a Large Number of Dimensions
}
\author[*]{Moshe Rozali}
\author[\dagger]{, Evyatar Sabag}
\author[\dagger]{, and Amos Yarom}
\affiliation[*]{Department of Physics and Astronomy, University of British Columbia,\\
Vancouver, BC V6T 1Z1, Canada\\}

\affiliation[\dagger]{Department of Physics, Technion, Haifa 32000, Israel}

\abstract{
We consider relativistic hydrodynamics  in the limit where the number of spatial dimensions is very large. We show that under certain restrictions, the resulting equations of motion simplify significantly. Holographic theories in a large number of dimensions satisfy the aforementioned restrictions and their dynamics are captured by hydrodynamics with a naturally truncated derivative expansion. Using analytic and numerical techniques we analyze two and three-dimensional turbulent flow of such fluids in various regimes and its relation to geometric data. 
}
\begin{document}
\maketitle

\section{Introduction and Summary}

Turbulent fluid flow is prevalent in every-day phenomena. Yet, while common, it is difficult to characterize it in a quantitative manner. Recently, the gauge-gravity duality has provided a novel geometric means to study relativistic fluid flow of strongly coupled, large N gauge theories, \cite{Bhattacharyya:2008jc}. Indeed,  in a suitable parameter range, and given appropriate initial conditions, such flows have been shown to exhibit (decaying) turbulent behavior \cite{Adams:2013vsa,green2014holographic}.  This development hints at a fascinating set of connections between turbulence and the dynamics of black hole horizons in general relativity.

A connection between turbulence and event horizons of asymptotically AdS black holes opens the possibility of using geometric tools to study turbulence. Indeed, various pioneering works have suggested a connection between the dynamics of the event horizon and a Kolmogorov-type cascade \cite{Adams:2013vsa,green2014holographic,Eling:2015mxa}. Unfortunately, the structure of time-dependent black hole horizons is not fully understood, which is one  of the reasons this research program has not yet come to fruition. In what follows we continue exploring these connections using newly developed tools for studying gravity in a large number of dimensions. More specifically, we consider the turbulent behavior of event horizons of asymptotically AdS spaces, in the limit where the number of dimensions, $d$, is very large. 

There is an extensive, ongoing, program to understand general relativity in the limit where the number of dimensions is large \cite{Emparan:2013moa,Emparan:2014jca,Emparan:2014aba,Emparan:2015rva,Emparan:2015hwa,Bhattacharyya:2015dva,Emparan:2015gva,Bhattacharyya:2015fdk,Andrade:2015hpa,Emparan:2016sjk,Herzog:2016hob,Bhattacharyya:2016nhn,Dandekar:2016fvw,Dandekar:2016jrp,Bhattacharyya:2017hpj}. The works of \cite{Emparan:2015rva,Andrade:2015hpa,Herzog:2016hob,Bhattacharyya:2017hpj,Miyamoto:2017ozn} in particular address the large $d$ limit of asymptotically AdS spaces. To obtain a large $d$ limit, one considers a metric ansatz in $d=n+p+1$ spacetime dimensions, where the dynamics depend on $p+1$ dimensions. The remaining number of dimensions, $n$, is taken to be infinite. This allows one to relate the large $d$ limit of gravity to gravity in $p+1$ dimensions.  Since our goal is to relate gravity at large $d$  to hydrodynamic flow, we obtain the $d\to\infty$ limit through a novel but circuitous path which will eventually allow a direct relation between horizon dynamics and fluid dynamics.

We start our analysis by considering the equations of motion of relativistic (conformal) hydrodynamics in the limit where the number of dimensions is large. %In this work we take a somewhat different angle on the large $d$ limit of gravity. Instead of considering a gravitational system in the large $d$ limit, we study the equations of motion of (conformal) relativistic hydrodynamics when the number of dimensions becomes large.
Generically, relativistic hydrodynamics does not simplify in such a limit. 
Nevertheless, as we show in section \ref{S:largedhydro}, if we scale the time and space coordinates appropriately, tune the transport coefficients of the theory to lie in an appropriate subset of parameter space, and work in a suitable fluid frame, the hydrodynamic equations significantly simplify. In appropriate variables they are given by \eqref{E:fulleom}.

Given the simplicity of the hydrodynamic equations in an appropriately scaled coordinate system, we turn our attention to asymptotically AdS gravity in the same limit. We find that our scaling laws for the coordinates are precisely those introduced in \cite{Emparan:2015gva} to study the Einstein equations in the large $d$ limit. Additionally, the holographic transport coefficients satisfy precisely the relations required for the aforementioned simplification of the large $d$ hydrodynamic equations. Indeed, as we show in section \ref{S:holography} the large $d$ Einstein equations of \cite{Emparan:2015gva} are precisely the large $d$ hydrodynamic equations of motion, albeit in an unusual, but well motivated, fluid frame.

Since we were compelled to scale our coordinate system, the large $d$ limit of the relativistic hydrodynamic equations of motion bear a striking
%While our starting point involved relativisitc fluid flow, to obtain simple equations of motion in the limit of large $d$ we were compelled to scale the space and time coordinates in  suitable manner. As a result, the large $d$ limit of the hydrodynamic equations bear a striking 
resemblance to the (non-relativistic) Navier-Stokes equations. More precisely, they are a variant of the compressible version of the Navier-Stokes equations. As such, they may be studied using traditional tools of non relativistic fluid dynamics. As we show in section \ref{S:fluidflow} when the flow is subsonic the Kolmogorov or Kraichnan scaling laws are expected to emerge and an inverse cascade should be exhibited when the system mimics two dimensional (sustained) turbulent fluid flow. In fact, even when the Mach number for the flow is not small we find  evidence for an inverse cascade-like behavior. 

Following the discussion in section \ref{S:fluidflow} we turn to a numerical analysis of the equations of motion in section \ref{S:numerical}. By using an appropriate initial condition we generate flow which exhibits decaying turbulence. We then compare various stages of the flow to expectations made in section \ref{S:fluidflow} for two and three spatial dimensions and to an analysis of holographic (decaying) turbulence in AdS${}_4$ \cite{Adams:2013vsa,green2014holographic}.

Given our improved understanding and analytic control over the relation of the fluid flow to the dynamics of the event horizon (for subsonic flows in particular), we study, in section \ref{S:geometry}, a proposed relation between the horizon curvature power spectrum and the hydrodynamic energy power spectrum \cite{Adams:2013vsa}. While these quantities are linearly related in the regime of small Mach number, the relation between these two quantities becomes obscure as the Mach number increases.

Finally, we present a summary of our findings in section \ref{S:Outlook} where we also discuss possible extensions and puzzles associated with out results.

%We end this exposition with a summary and an outlook in section \ref{S:Outlook}.

\vspace{10pt}
\section{The Large $d$ Limit of Hydrodynamics}
\label{S:largedhydro}

Hydrodynamics is a universal low energy effective description of many-body systems. Following a Wilsonian type of coarse graining, the equations of motion of relativistic hydrodynamics may be characterized by a handful of fields: a unit normalized velocity field $u^{\mu}(x)$, a temperature field $T(x)$ and, in the presence of charged matter, a chemical potential $\mu(x)$. When working in the limit where gradients of the hydrodynamic fields are small compared to the inverse mean free path, all observables may be expressed as local functions of the hydrodynamic fields. These expressions are referred to as constitutive relations.

In a $d$ spacetime dimensional conformal theory in flat space (and in the absence of charge, anomalies or parity breaking terms) the constitutive relations for the stress tensor take the form
\begin{equation}\label{eq:stressHydro}
        T^{\mu \nu}=T^{\mu \nu}_{(0)}+T^{\mu \nu}_{(1)}+T^{\mu \nu}_{(2)}+\mathcal{O}(\partial^3)\,,
\end{equation}
where
\begin{align}
\begin{split}
\label{E:constitutive}
        T^{\mu \nu}_{(0)} =&  \epsilon(T) u^{\mu}u^{\nu} + \frac{\epsilon(T)}{d-1} P^{\mu \nu}\,,
        \qquad
        T^{\mu \nu}_{(1)}  =  -2\eta(T) \sigma^{\mu \nu}\,,\\
        T_{(2)}^{\mu\nu} =& \lambda_{1}(T) u^{\lambda} \mathcal{D}_{\lambda} \sigma^{\mu\nu}+\lambda_{2}(T) \left(\sigma^{\lambda\mu}\sigma_{\lambda}^{\;\,\nu}-\frac{\sigma^{\alpha\beta}\sigma_{\alpha\beta}}{d-1}P^{\mu\nu}\right)   \\
                & +\lambda_{3}(T) \left(\omega^{\mu\lambda}\sigma_{\lambda}^{\;\,\nu}+\omega^{\nu\lambda}\sigma_{\lambda}^{\;\,\mu}\right)+\lambda_{4}(T)\left(\omega^{\mu\lambda}\omega^{\nu}{}_{\lambda}  -\frac{\omega^{\alpha\beta}\omega_{\alpha\beta}}{d-1}P^{\mu\nu}\right) \,,
\end{split}
\end{align}
and
\begin{align}
\begin{split}
\label{E:tensorstructures}
        P^{\mu\nu} & =  g^{\mu\nu}+{u}^{\mu}{u}^{\nu},\\
        \sigma^{\mu\nu} & =  \frac{1}{2}P^{\mu\alpha}P^{\nu\beta}\left(\partial_{\alpha}{u}_{\beta}+\partial_{\beta}{u}_{\alpha}\right)-\frac{1}{d-1}P^{\mu\nu}\partial_{\alpha}{u}^{\alpha},\\
        \omega^{\mu\nu} & =  \frac{1}{2}P^{\mu\alpha}P^{\nu\beta}\left(\partial_{\alpha}{u}_{\beta}-\partial_{\beta}{u}_{\alpha}\right), \\
        \mathcal{D}_{\lambda}\sigma^{\mu\nu}  & =  P^{\mu\alpha}P^{\nu\beta}\partial_{\lambda}\sigma_{\alpha\beta}+\frac{\partial_{\alpha}{u}^{\alpha}}{d-1}\sigma^{\mu\nu}\,,
\end{split}
\end{align}
and we have set the speed of light to unity.
%where $\tilde{u}^\mu = u^\mu/c$, with $c$ the speed of light. 
See \cite{Haack:2008cp,Bhattacharyya:2008mz} for details.

Following the standard reasoning behind the construction of effective field theories, the constitutive relations \eqref{E:constitutive} are the most general ones allowed, compatible with the conformal symmetry of the underlying theory, with a local version of the second law, and taking into account useful features of the derivative expansion. Indeed, in writing down \eqref{E:constitutive} we have used the equations of motion at order $n-1$ in derivatives to simplify the constitutive relations at order $n$. In addition, we have used a particular definition of the fluid velocity and energy density, usually referred to as the Landau frame, in order to further simplify the relations \eqref{E:constitutive}. Since these features of the derivative expansion will be essential in our study of the large $d$ limit of hydrodynamics, we describe them in some detail in what follows.

In the absence of charge, the equations of motion for the hydrodynamic fields are nothing but the conservation equations for the stress tensor 
\begin{equation}
\label{E:conservation}
        \partial_{\mu}T^{\mu\nu}=0\,, 
\end{equation}  
which need to be solved perturbatively in a derivative expansion. For instance, at leading order in derivatives one has
\begin{align}
\begin{split}
\label{E:zerothorder}
        {u}^{\mu} \partial_{\mu} \epsilon + d\, \epsilon  \partial_{\mu} {u}^{\mu} & = 0 
        \\
        \frac{d}{d-1}  \epsilon  {u}^{\mu}\partial_{\mu} {u}^{\nu}  + P^{\mu\nu}\partial_{\mu} \epsilon  & = 0\,.
\end{split}
\end{align}
Therefore, when constructing $T_{(1)}^{\mu\nu}$ it is possible to replace ${u}^{\mu} \partial_{\mu} \epsilon$ and ${u}^{\mu}\partial_{\mu} {u}^{\nu}$ with $\partial_{\mu} {u}^{\mu}$ and $P^{\mu\nu}\partial_{\mu} \epsilon$. The same reasoning applies to the construction of $T_{(2)}^{\mu\nu}$; at $n$'th order in the derivative expansion, the available tensor structures from which the stress tensor may be constructed is reduced due to the equations of motion at lower orders.

It is possible to modify the relations \eqref{E:constitutive} for $T_{(1)}^{\mu\nu}$ and $T_{(2)}^{\mu\nu}$  by an appropriate field redefinition of $T$ and ${u}^{\mu}$. The constitutive relations \eqref{E:constitutive} were written in the Landau frame, defined via
\begin{equation}
        {u}_{\mu}T^{\mu\nu} = {u}_{\mu}T_{(0)}^{\mu\nu}  = - \epsilon(T) {u}^{\nu}\,.
\end{equation}
From a physical standpoint the Landau frame is defined such that $u^{\mu}$ points in the direction of the energy flux, and the relation between $\epsilon(T)$ and $T$ does not get modified by derivative corrections.
%where $\epsilon(T)$ was given in \eqref{E:scaling}. 
When the system is out of equilibrium, the Landau frame is only one of many choices for possible definitions for $T$ and $u^{\mu}$. Indeed, working perturbatively in the derivative expansion one may always redefine the temperature and velocity field via, e.g.,
\begin{align}
\begin{split}
\label{E:framesimple}
         \epsilon_{new}(T_{new}) &= \epsilon(T) - k_1 {u}^{\mu}\partial_{\mu}\epsilon(T)   + \mathcal{O}(\partial^2)  \\
         \epsilon(T)\, {u}^{\mu}_{new} & = \epsilon(T) {u}^{\mu} - c_1 \partial^{\mu}\epsilon(T) +  \mathcal{O}(\partial^2) \,,
\end{split}
\end{align}
where the $c$'s and $k$'s are real numbers.%\footnote{In writing \eqref{E:framesimple} we have omitted the Levi-Civitta tensor which may be used in this context in three spacetime dimensions.} Such a field redefinition will modify the constitutive relations but leave physical observables unchanged.

The equations of motion for the velocity field and temperature, which follow from inserting the constitutive relations \eqref{E:constitutive} into the conservation equation \eqref{E:conservation}, are lengthy and cumbersome. While the zeroth order equations of motion \eqref{E:zerothorder} are relatively short and simple, writing the first order equations in terms of $u^{\mu}$ is a formidable task, let alone the second order equations. In the remainder of this section we will study the equations of motion in the limit where the number of dimensions $d$ becomes very large. We will see that with some restrictions, to be elaborated on below, the large $d$ limit of these hydrodynamic equations become easier to handle.

Before working out the hydrodynamic equations, let us first consider thermodynamic equilibrium in the large $d$ limit. In equilibrium, the temperature and velocity field are constant so that the stress tensor is given exactly by $T^{\mu\nu}_{(0)}$. Since the velocity is constant, we may conveniently choose a coordinate system where ${u}^{\mu}\partial_{\mu} = \partial_t$. If we assume, without loss of generality, that the energy density $\epsilon$ is $\mathcal{O}(d^0)$ and insist that the pressure term, proportional to $P^{\mu\nu}$, not be suprressed relative to the energy density, then we must work in a coordinate systems scaled so that the spatial coordinates carry an extra factor of $1/\sqrt{d}$ relative to the time coordinate: $ds^2 = -dt^2 + \sum_i (dx^i)^2/d$. %This leaves the speed of sound finite in the large $d$ limit. 
For the same reasoning, we parameterize the velocity field of a boosted thermally equilibrated system by $u^{\mu} = (1,\,{\beta}^i)/\sqrt{1-\beta^i\delta_{ij}\beta^j/d}$. Note that our definition of $\beta^i$ differs from the conventional definition by a factor of $\sqrt{d}$, $\beta^i = \sqrt{d}\beta_{conventional}^i$.

With this parameterization in mind, let us now go back to \eqref{E:conservation}. Conservation of energy and momentum provide $d$ equations for the $d$ hydrodynamical variables $T$ and $u^{\mu}$. In order for the large $d$ limit to provide a useful proxy to a finite $p+1$ dimensional system we must take the large $d$ limit in such a way that the number of dynamical equations remains $p+1$. To do so, we use an ansatz where the dynamical variables do not depend on $n$ out of the $d=1+p+n$ coordinates. Thus, we use
\begin{equation}
\label{E:largedmetric}
        {d}s^2 = - {d}t^2 + \frac{\delta_{ab}}{n} {d}\zeta^a {d}\zeta^b + \frac{ {d}\vec{\chi}_\perp^2}{n}
\end{equation}
where $\vec{\chi}_\perp$ denotes the $n\gg 1$ coordinates on which $T$ and $u^{\mu}$ do not depend, and Latin indices run over the remaining $p$ spatial components. The velocity field in this coordinate system is given by
\begin{equation}
\label{E:ulargen}
        u^\mu = \frac{\left(1,\beta^a(t,\zeta),\,\vec{0}\right)}{\sqrt{1-\frac{\beta^b(t,\zeta)\beta^c(t,\zeta)\delta_{bc}}{1+p+n}}}\,.
\end{equation}
In the remainder of this work we will raise and lower Latin indices with $\delta^{ab}$ and $\delta_{ab}$ respectively.

Keeping with this notation, the tensor structures \eqref{E:tensorstructures} evaluate to
\begin{align}
\begin{split}
\label{E:structures}
        \sigma^{\mu\nu} & =  \frac{n}{2}\delta^{\mu a}\delta^{\nu b}\left(\partial_{a}\beta_{b} + \partial_{b}\beta_{a}\right) + \mathcal{O}\left(n^{0}\right),\\
        \omega^{\mu\nu} & =  \frac{n}{2}\delta^{\mu a}\delta^{\nu b}\left(\partial_{a}\beta_{b} - \partial_{b}\beta_{a}\right) + \mathcal{O}\left(n^{0}\right), \\
        u^\lambda \mathcal{D}_\lambda\sigma^{\mu\nu} & =  \frac{n}{2}\delta^{\mu a}\delta^{\nu b}\left(\partial_{t}+\beta^{c}\partial_{c}\right)\left(\partial_{a}\beta_{b}+\partial_{b}\beta_{a}\right)+\mathcal{O}\left(n^{0}\right)\,.
\end{split}
\end{align}
Inserting \eqref{E:structures} into \eqref{E:constitutive}, we find that in order for the shear viscosity and second order transport coefficients to contribute to the dynamics they must both scale as $1/d$ relative to $\epsilon$. Recall that dimensional analysis and conformal invariance imply that for any finite $d$
\begin{equation}
\label{E:scaling}
        \epsilon \propto T^d\,,
        \qquad
        \eta \propto T^{d-1}\,,
        \qquad
        \lambda_i \propto T^{d-2}\,,
\end{equation}
where the proportionality constants are dimensionless numbers. Thus, we may write
\begin{equation}
\label{E:r0hydro}
	\eta  \propto \frac{1}{d} \epsilon^{1-1/d} \\
	        = \frac{ \epsilon \xi}{d} \left(\epsilon \xi^d\right)^{-1/d} \\
	        \xrightarrow[d\gg 1]{} \frac{\epsilon \xi}{d}
\end{equation}
where $\xi$ is an emergent length scale which is determined by requiring that $\epsilon^{1/d} \xi$ remain finite in the large $d$ limit.
Thus, in what follows we will use the parameterization
\begin{equation}
\label{E:transportscaling}
        \eta = \frac{h_0}{n} \xi \epsilon + \mathcal{O}(n^{-2})
        \qquad
        \lambda_i = \frac{2 \ell_i}{n} \xi^2 \epsilon + \mathcal{O}(n^{-2})\,,
\end{equation}
where $h_0$ and $\ell_i$ are dimension independent numbers. In the majority of this work we will set $\xi=1$. It will be reintroduced where necessary.

The large $d$ stress-energy tensor is now given by
\begin{equation}
\label{eq:stressHydroLn}
        T^{\mu \nu} = \epsilon\left(\begin{array}{cc}
        1 & \beta^{b}\\
        \beta^{a} & \beta^{a}\beta^{b}+\delta^{ab}+\tilde{T}^{ab}_{(1)}+\tilde{T}^{ab}_{(2)}+\mathcal{O}\left(\partial^{3}\right)
        \end{array}\right)+\mathcal{O}\left(n^{-1}\right)
\end{equation}
with,
\begin{align}
\begin{split}
\label{E:Tmnstructure}
        \tilde{T}^{ab}_{(1)} = & -h_0\left(\partial^{a}\beta^{b}+\partial^{b}\beta^{a}\right),\\
        \tilde{T}^{ab}_{(2)}  = & \ell_{1}\left(\partial_{t}+\beta^{c}\partial_{c}\right)\left(\partial^{a}\beta^{b}+\partial^{b}\beta^{a}\right)
                +\frac{\ell_{2}}{2}\left(\partial^{a}\beta^{c}+\partial^{c}\beta^{a}\right)\left(\partial^{b}\beta_{c}+\partial_{c}\beta^{b}\right)  \\
                & +\frac{\ell_{3}}{2}\left( \left(\partial^{ a}\beta^{c}-\partial^{c}\beta^{ a}\right)\left(\partial^{b}\beta_{c}+\partial_{c}\beta^{b}\right) + \left({a} \leftrightarrow {b}\right)\right)
                +\frac{\ell_{4}}{2}\left(\partial^{a}\beta^{c}-\partial^{c}\beta^{a}\right)\left(\partial^{b}\beta_{c}-\partial_{c}\beta^{b}\right)\,.
\end{split}
\end{align}
%where we have set $\ell=1$. For the most part of this work we will set $c=1$ and $\ell=1$. However, at places such as \eqref{E:dlparams} we will reintroduce factors of $c$ and $r_0$ to emphasize various features of the hydrodynamic equations. 

The equations of motion \eqref{E:conservation} which result from the constitutive relations \eqref{E:Tmnstructure} are still somewhat unwieldy. To simplify them further we will use our freedom to perform field redefinitions of the velocity and temperature fields in order to remove all third order derivative terms from the equations of motion. The most general frame transformation we can carry out %can be parameterized by \eqref{E:zerothorder} to first order in derivatives. To 
at second order in derivatives and in the large $d$ limit is given by
\begin{align}
\begin{split}
\label{E:generalframe}
        \epsilon \to & \, \epsilon + k_{1}\epsilon\partial_{b}\beta^{b} 
                +  k_{2}\partial_{b}\partial^{b}\epsilon + k_{3}\epsilon\left(\partial_{b}\beta^{b}\right)^{2} + k_{4}\epsilon\partial_{b}\beta^{c}\partial_{c}\beta^{b} + k_{5}\frac{\partial_{b}\epsilon\partial^{b}\epsilon}{\epsilon} + k_{6}\epsilon\partial_{b}\beta_{c}\partial^{b}\beta^{c}  + \mathcal{O}(\partial^3)\\
        \beta^a \to & \, \beta^{a} + c_{1}\frac{\partial^{a}\epsilon}{\epsilon} 
                 + c_{2}\partial_{b}\beta^{b}\frac{\partial^{a}\epsilon}{\epsilon} + c_{3}\partial^{a}\beta^{b}\frac{\partial_{b}\epsilon}{\epsilon} + c_{4}\partial^{a}\partial_{b}\beta^{b} + c_{5}\frac{\partial_{b}\epsilon}{\epsilon}\partial^{b}\beta^{a} + c_{6}\partial_{b}\partial^{b}\beta^{a} + \mathcal{O}(\partial^3) \,.
\end{split} 
\end{align}
Inserting \eqref{E:generalframe} into \eqref{eq:stressHydroLn} and utilizing the derivative expansion, c.f., \eqref{E:zerothorder}, we find that if
\begin{subequations}
\label{E:no2order}
\begin{equation}
\label{E:largedconstitutive}
        \ell_2 + 2 \ell_3 + \ell_4 = 0\,,
        \qquad
        \hbox{and}
        \qquad
        \ell_1 - \ell_2 - \ell_3 = 0\,,
\end{equation}
then the second order terms in the equations of motion vanish (i.e., no contributions from second order terms in the stress tensor), provided we choose a frame where
\begin{equation}
\label{E:largedframe}
        k_2 = \ell_1 + \ell_4
        \qquad
        -2 c_1 h_0 = c_1^2 + 2 \ell_1
        \qquad
        c_5 = 2 (\ell_1 + \ell_4) 
        \qquad
        c_6 = \ell_1 + \ell_4\,,
\end{equation}
\end{subequations}
and the remaining $c_i$'s and $k_i$'s vanish.

In what follows we will refer to \eqref{E:largedframe} as the large $d$ frame. To iterate our finding in the previous paragraph, when \eqref{E:no2order} are satisfied, the equations of motion take the form \begin{align}
\begin{split}
\label{E:fulleom}
        \partial_t \epsilon + c_1 \partial_a \partial^a \epsilon  & = - \partial_a j^a \\
        \partial_t j^a - h_0 \, \partial_b \partial^b j^a & = -\partial^a \epsilon - \partial_b \left( \frac{j^a j^b}{\epsilon} \right) - 
                 \left(\frac{c_1}{2} - \frac{\ell_1}{c_1}\right) \partial_{b}\left(\frac{j^{a}}{\epsilon}\partial^{b}\epsilon - \epsilon\partial^{a}\left(\frac{j^{b}}{\epsilon}\right)\right)
\end{split}
\end{align}
where we have defined
\begin{equation}
        j^a = \beta^a\,\epsilon\,,
\end{equation}
and $c_1$ is given by the quadratic equation in \eqref{E:largedframe}.
The stress tensor whose divergence gives these equations of motion is given by
\begin{align}
\label{E:generalstressenergy}
\begin{split}
         T_{(0)}^{\mu\nu} &= \begin{pmatrix} \epsilon & j^a \\ j^b  & \delta^{ab} \epsilon + \frac{j^a j^b}{\epsilon} \end{pmatrix}
         \qquad
         T_{(1)}^{\mu\nu} = \begin{pmatrix} 0 & c_1 \partial^a \epsilon \\ c_1 \partial^b \epsilon & (c_1 + h_0) \left(  \frac{j^a}{\epsilon} \partial^b \epsilon + \frac{j^b}{\epsilon} \partial^a \epsilon \right)  -h_0 \left(  \partial^a j^b  + \partial^b j^a \right)  \end{pmatrix} \\
        T_{(2)}^{\mu\nu} & = \begin{pmatrix} 0 & 0 \\ 0 & c_1^2 \partial^a \partial^b \epsilon \end{pmatrix} + (\ell_1 + \ell_4) \partial_c \partial^c T_{(0)}^{\mu\nu}\,.
\end{split}
\end{align}
We emphasize that even though $\partial_{\mu} T_{(2)}^{\mu\nu} \neq 0$, the equations of motion are only second order in derivatives due to cancellations coming from the perturbative expansion.

\vspace{10pt}
\section{Holography in a Large Number of Dimensions}
\label{S:holography}

Our final result, equation \eqref{E:fulleom}, is, evidently, a simplified version of the full equations of motion. In order to obtain it we had to restrict our transport coefficients to satisfy \eqref{E:transportscaling} and \eqref{E:largedconstitutive}. Both equations impose restrictions on the transport coefficients when the number of dimensions is large. Equation \eqref{E:largedconstitutive} implies that the four transport coefficients of the theory are restricted to lie on a plane in parameter space while equation \eqref{E:transportscaling} implies a particular scaling of the transport coefficients with $d$ when $d$ is large. Since the large $d$ limit is used as a proxy for finite dimensional system one may, naively, enforce by hand a $1/d$ scaling of transport coefficients for large $d$ so that both \eqref{E:transportscaling} and \eqref{E:largedconstitutive} will be satisfied. Notwithstanding, it seems likely that such a behavior will affect the radius of convergence of the large $d$ expansion. It is then somewhat surprising that the relations \eqref{E:largedconstitutive} are valid in a holographic setup. There, we find that
\begin{equation}
        \lambda_1-\lambda_2 - \lambda_3 = 0
        \qquad
        \lambda_2 + 2\lambda_3 -\lambda_4 = \mathcal{O}\left(d^{-3}\right)
\end{equation}
see \cite{Haack:2008cp,Bhattacharyya:2008mz}. Note that the first of the two relations is satisfied in any number of dimensions \cite{Haack:2008cp} even in the presence of matter \cite{Haack:2008xx} and also in other instances \cite{Kanitscheider:2009as}. As we will now demonstrate the constraints \eqref{E:transportscaling} are also satisfied in a holographic context.

The constraints \eqref{E:transportscaling} and \eqref{E:largedconstitutive} guarantee that the equations of motion will take the simplified form \eqref{E:fulleom}. Therefore, if they are satsified, and we work perturbatively in the derivative expansion, equations \eqref{E:fulleom} must emerge from a holographic analysis \cite{Bhattacharyya:2008jc}. We find that not only do \eqref{E:fulleom} emerge, they are also valid to all orders in the derivative expansion. Put differently, the Einstein equations at large $d$ are equivalent to large $d$ hydrodynamics naturally truncated at second order in derivatives. In what follows we will show this property of the Einstein equations explicitly.

Consider the $D=d+1$ space-time dimensional Einstein-Hilbert action
\begin{equation}
        S = \int \sqrt{g} \left(R + (D-1)(D-2) \right) d^Dx\,,
\end{equation}
where we have set the radius of AdS space to unity.
Following \cite{Emparan:2015gva}, we wish to take the large $D$ limit of this action while retaining translation invariance in most of the spatial directions. To this end, we use the ansatz
\begin{equation}
        ds^{2}=2dt\left(-A\left(t,r,z^{i}\right)dt+dr-F_{a}\left(t,r,z^{i}\right)dz^{a}\right)+G_{ab}\left(t,r,z^{i}\right)dz^{a}dz^{b}+G_{\perp}\left(t,r,z^{i}\right)d\vec{x}_{\perp}^{2}
\end{equation}
where $a,b=1,\ldots,p$, $D=p+n+2$ and $\vec{x}_{\perp}$ is $n$ dimensional. To obtain a boundary metric as in \eqref{E:largedmetric} we make the replcaements
\begin{equation}
        z^{a}=\frac{\zeta^{a}}{\sqrt{n}}
        \qquad
        \vec{x}_{\perp}=\frac{\vec{\chi}_{\perp}}{\sqrt{n}}
\end{equation}
as well as
\begin{equation}
        \tilde{F}_{a} = \frac{F_{a}}{\sqrt{n}}
        \qquad
        \tilde{G}_{ab} = \frac{G_{ab}}{n}
        \qquad
        \tilde{G}_{\perp} = \frac{G_{\perp}}{n}\,,
\end{equation}
and choose boundary conditions such that 
\begin{equation}
        A=r^2\left(\frac{1}{2}+\mathcal{O}\left(r^{-n}\right)\right)
        \quad
        \tilde{F}_{a}=\frac{r^2}{n}\left(\mathcal{O}\left(r^{-n}\right)\right)
        \quad
        \tilde{G}_{ab}=\frac{r^2}{n}\left(\delta_{ab}+\mathcal{O}\left(r^{-n}\right)\right)
        \quad
        \tilde{G}_{\perp}=\frac{r^2}{n}\left(1+\mathcal{O}\left(r^{-n}\right)\right)\,.
\end{equation}
In addition, we rescale the radial coordinate such that
\begin{equation}
\label{E:bigR}
        R=\left(\frac{r}{r_0}\right)^{n}
\end{equation}
is finite in the large $d$ limit. Note that $r_0$ serves as a reference length scale. In what follows we will set $r_0=1$ for clarity. In the new coordinate system the line element takes the form
\begin{equation}
\label{E:finalLE}
        ds^{2}=2dt\left(-A\left(t,R,\zeta^{c}\right)dt+\frac{R^{\frac{1}{n}-1}}{n}dR-\tilde{F}_{a}\left(t,R,\zeta^{c}\right)d\zeta^{a}\right)+\tilde{G}_{ab}\left(t,R,\zeta^{c}\right)d\zeta^{a}d\zeta^{b}+\tilde{G}_{\perp}\left(t,R,\zeta^{c}\right)d\vec{\chi}_{\perp}^{2}\,.
\end{equation}

The scaling used to obtain \eqref{E:finalLE} is precisely that of \cite{Emparan:2015gva}. See also \cite{Herzog:2016hob}. Inserting \eqref{E:finalLE} into the Einstein equations and solving order by order in $n$ we find that
\begin{equation}
        A =\frac{1}{2}-\frac{a\left(\zeta^{\mu}\right)}{2R} + \mathcal{O}\left(n^{-1}\right)
        \qquad  
        \tilde{F}_a  =\frac{f_{a}\left(\zeta^{\mu}\right)}{nR}+ \mathcal{O}\left(n^{-2}\right)
\end{equation}       
\begin{equation}
        \tilde{G}_{ab}=\frac{\delta_{ab}}{n} + \mathcal{O}(n^{-2})
        \qquad
        \tilde{G}_{\perp}= \frac{1}{n} + \mathcal{O}(n^{-2})
\end{equation}
where the functions $a$ and $f_a$ must satisfy the constraint equations
\begin{eqnarray}
\label{E:constraint}
        \partial_{t}a-\partial_{b}\partial^{b}a & = & -\partial_{b}f^{b}\nonumber \\
        \partial_{t}f_{a}-\partial_{b}\partial^{b}f_{a} & = & -\partial_{a}a-\partial_{b}\left(\frac{f_{a}f^{b}}{a}\right)\,.
        \label{collEOM}
\end{eqnarray}

As expected, equations \eqref{E:constraint} and \eqref{E:fulleom} match as long as we set
\begin{equation}
\label{E:h0c0l1}
        h_0=1\,,
        \qquad
        c_1=-1\,,
        \qquad
        \ell_1= \frac{1}{2}
\end{equation}
and make the identifications
\begin{equation}
        a = \epsilon
        \qquad
        f_a = j_a\,.
\end{equation}
Thus, in the large $D$ limit the Einstein equations precisely reproduce the large $d$ limit of the equations of motion for relativistic hydrodynamics. The associated stress tensor is given in \eqref{E:generalstressenergy}.

Before proceeding let us pause to consider the relation between the scale $r_0$ associated with the horizon, \eqref{E:bigR}, and the scale $\xi$ generated by taking the large $d$ limit of hydrodynamics \eqref{E:r0hydro}. Recall that for any finite $d$ the (local) Hawking temperature associated with the event horizon is
\begin{equation}
	\frac{1}{a^{1/d} r_0} = \frac{d}{4\pi T}\,,
\end{equation}
and that in the absence of charge the entropy density is given by
\begin{equation}
\label{E:entropy1}
	s = \frac{\partial P}{\partial T} = \frac{a}{T}
\end{equation}
where $P$ is the pressure and $\epsilon=a$ the energy density. Using the holographic result for the shear viscosity to entropy density ratio \cite{Kovtun:2004de} we find
\begin{equation}
\label{E:holographiceta}
	\eta = \frac{a^{1-1/d}}{d r_0}\,.
\end{equation}
Comparing \eqref{E:holographiceta} to \eqref{E:r0hydro} and using \eqref{E:h0c0l1} we obtain
\begin{equation}
\label{E:xir0}
	\xi = \frac{1}{r_0}\,.
\end{equation}

Given \eqref{E:xir0} the expression for the entropy density \eqref{E:entropy1} takes the form
\begin{equation}
\label{E:entropy2}
	s = \frac{4 \pi \xi}{d} a
\end{equation}
in the large $d$ limit and in our current conventions.\footnote{Note that using the Bekenstein-Hawking formula $S=A/4 G_N$ we obtain $s = \frac{r_0^{p+n}}{4 G_N n^{n/2}} a$. Equation \eqref{E:entropy2} agrees with this result up to an overall normalization which was chosen when we set $\epsilon = a$ in \eqref{E:holographiceta}.}  
Since we have a good handle over the dynamics of the area element of the event horizon we can test various proposals regarding its behavior in the presence of turbulence, and analyze the geometric structure associated with turbulent flows.

\vspace{10pt}
\section{Analysis of Large $d$ Fluid Flows}
\label{S:fluidflow}

Thus far we have described hydrodynamics using the variables $\epsilon$ and $\beta^a$. Since we are working in the large $d$ frame \eqref{E:largedframe} these variables are somewhat different from the energy density and velocity field as described in the Landau frame. Indeed, $\epsilon$ and $\beta^a$ coincide with the energy density and fluid velocity of the Landau frame only in equilibrium. Otherwise, they are related to the Landau frame variables through \eqref{E:generalframe}. Nevertheless, if we rewrite the equations of motion \eqref{E:fulleom} in terms of $\epsilon$ and $\beta^a = j^a/\epsilon$, we find
\begin{align}
\begin{split}
\label{eq:EOM}
        \partial_{t}\epsilon + \left(\vec{\beta}\cdot \vec{\nabla}\right) \epsilon + \epsilon \vec{\nabla} \cdot \vec{\beta}  & = \nabla^2 \epsilon \\
        \partial_{t}\vec{\beta}+\left(\vec{\beta}\cdot\nabla\right)\vec{\beta} + \frac{\vec{\nabla} \epsilon}{\epsilon} & =  \nabla^2 \vec{\beta} +  2\left(\frac{\vec{\nabla} \epsilon}{\epsilon}\cdot\vec{\nabla}\right)\vec{\beta} \,.
\end{split}
\end{align}

In order to understand the dynamics associated with equations \eqref{eq:EOM} and their relation to the compressible Navier-Stokes equations, it is useful to switch to dimensionless variables,
\begin{equation} 
        \vec{u} = \frac{\vec{\beta}}{\sqrt{n} U}\,,
        \quad
        p = \frac{\epsilon}{E}\,,
        \quad
        \frac{\partial}{\partial t} \to \frac{L_0}{\sqrt{n} U}\frac{\partial}{\partial t}\,,
        \quad
        \vec{{\nabla}} \to L_0 {\vec{\nabla}},
\end{equation}
where $L_0$ is a characteristic length scale of the system, $\sqrt{n}U$ a characteristic velocity , and $E$ a characteristic energy density. The factor of $\sqrt{n}$ in the definition of the characteristic velocity arises from the scaling we used for $\beta^a$ in \eqref{E:ulargen}. 
In terms of these dimensionless variables, equations \eqref{eq:EOM} take the form
\begin{align}
\begin{split}
\label{eq:EOMdimless}
        \partial_{t}p +\left(\vec{u}\cdot\vec{\nabla}\right)p + p\vec{\nabla}\cdot\vec{u} &=  \frac{1}{Re}\nabla^{2} p  \\
        \partial_{t}\vec{u}+\left(\vec{u}\cdot\vec{\nabla}\right)\vec{u} + \frac{\vec{\nabla} p}{M^2 p} & = \frac{1}{Re}\nabla^{2} \vec{u} + \frac{2}{Re}\left(\frac{\vec{\nabla} p}{p}\cdot\vec{\nabla}\right)\vec{u} \,,
\end{split}
\end{align}
where, in analogy with the compressible Navier-Stokes equations, we have defined a Reynolds number $Re=\sqrt{n}L_0U$ and a Mach number $M=\sqrt{n}U$.
If we reinsert the dimensionful parameters $c$ and $\xi$ we find
\begin{equation}
\label{E:dlparams}
        Re=\frac{\sqrt{n}L_0U}{\xi c}=\frac{L_0U}{\xi c_s}
        \qquad
        M = \frac{\sqrt{n}U}{c}=\frac{U}{c_s}\,.
\end{equation}
where $c_s$ is the speed of sound.

When the Mach number is small we may expand $\vec{u}$ and $p$ in powers of the Mach number,
%When the Reynolds number is large the right hand side of both equations in \eqref{eq:EOMdimless} is negligible, and we obtain a variant of the dissapationless compressible Navier-Stokes equation. If, in addition, the Mach number is small, and we expand $\vec{u}$ and $\epsilon$ in powers of the Mach number,
\begin{equation}
        \vec{u} = \sum_{n=0} M^n \vec{u}_{(n)}
        \qquad
        p = \sum_{n=2} M^n p_{(n)}\,.
\end{equation}
On a manifold with no boundary $p_{(0)}$ and $p_{(1)}$ are constant and $u_{(0)}$ and $p_{(2)}$ satisfy the incompressible Navier-Stokes equation,
\begin{equation}
\label{E:lowMach}
        \vec{\nabla} \cdot \vec{u}_{(0)} = 0
        \qquad
        \partial_{t}\vec{u}_{(0)}+\left(\vec{u}_{(0)} \cdot\vec{\nabla}\right)\vec{u}_{(0)} + \frac{\vec{\nabla} p_{(2)}}{\epsilon_{(0)}}  = \frac{1}{Re} \nabla^2 \vec{u}_{(0)}\,.
\end{equation}
with $p_{(0)}$ the density and  $p_{(2)}$ the pressure.
Equations \eqref{E:lowMach} are precisely the Navier-Stokes equations for an incompressible fluid.
Thus,
in the limit of small Mach number we may use the exhaustive machinery developed for incompressible flow to study solutions of \eqref{E:lowMach}. While this is textbook material (see for instance \cite{davidson2015turbulence}) let us remind the reader of the salient features of such flows. 

We define the total energy and enstrophy via
\begin{equation}
\label{E:EnergyEnstrophyDef}
        E_I =\frac{1}{2} \int  |\vec{u}_{(0)}|^2 d^{p}x\,,
        \qquad
        \Omega_I = \frac{1}{2}  \int \omega_{(0)ij}\omega_{(0)}^{ij} d^px
\end{equation}
with $\omega_{(0)ij} = \partial_i u_{(0)\,j} - \partial_j u_{(0)\,i}$ the vorticity two-form. The evolution equations for $E_I$ and $\Omega_I$ are given by
\begin{subequations}
\label{E:icconservation}
\begin{align}
        \frac{\partial}{\partial t} E_I &= -\frac{1}{Re} \Omega_I \\
\label{E:icenstrophy}
        \frac{\partial}{\partial t}\Omega_I &= \int \omega_{(0)}^{ij}\omega_{(0)jk}\sigma_{(0)}^{ki} d^px - \frac{1}{Re} P_I
\end{align}
\end{subequations}
where we have introduced the non relativistic shear tensor, $\sigma_{(0)}^{ij}$ and the Palinstrophy, $P_I$,
\begin{equation}
        \sigma_{(0)ij} = \partial_i u_{(0)\,j} + \partial_j u_{(0)\,i}\,,
        \qquad
        P_I = \frac{1}{2}\int \partial_k \omega_{(0)ij} \partial^k \omega_{(0)}^{ij} d^dp\,.
\end{equation}
We once again emphasize that the energy density is given by $T^{00}$ via \eqref{eq:stressHydroLn} and that $E_I$ is the would be energy of the analog Navier-Stokes equation in the incompressible limit. Nevertheless, we shall, with some abuse of language, refer to $E_I$ as the energy and to $\Omega_I$ as the enstrophy. 

Note that $\Omega_I \geq 0$. Thus, the energy $E_I$ can only decrease. On the other hand, $\Omega_I$ itself can increase or decrease depending on the sign of  the first term on the right hand side of \eqref{E:icenstrophy}. The dynamics associated with this term is often referred to as vortex stretching. There is ample phenomenological evidence and various imperfect arguments that vortex stretching increases the enstrophy for turbulent incompressible flow. Indeed, one usually posits that, for sustained turbulent flow, $\lim_{Re\to\infty}\Omega_I/Re = e_0 > 0$. The appearance of an emergent scale $e_0$ at large Reynolds number implies an energy cascade. To see this consider the two point function
\begin{equation}
        Q(\vec{r}) = \frac{ \int \vec{u}_{(0)}(\vec{x}) \cdot u_{(0)}(\vec{x} + \vec{r}) d^p x}{\int d^p x}
\end{equation}
and its Fourier transform
\begin{equation}
        \widehat{Q}(\vec{k}) = \frac{1}{(2\pi)^p} \int Q(\vec{r}) e^{-i \vec{k} \cdot \vec{r}} d^pr\,.
\end{equation}
We define the energy density $E(k)$ via
\begin{equation}
        E(k) = \frac{1}{2} \int \widehat{Q}(\vec{k}) k^{p-1} d\theta_k
\end{equation}
where $d\theta_k$ is a solid angle in momentum space, viz. $d^pk = k^{p-1} d\theta_k dk$. With this definition,
\begin{equation}
\label{E:EnergyDensity}
        E_I  = \int_0^{\infty} E(k) dk\,,
\end{equation}
and also
\begin{equation}
        \Omega_I = \int_0^{\infty} k^2 E(k) dk\,.
\end{equation}
If $E_I$ is to decrease while $\Omega_I$ is to remain constant then $E(k)$ must distribute itself in such a way that energy will flow from lower momentum modes to higher ones. This process is referred to as the Kolmogorov cascade. If we constantly supply energy into the system and assume that $E(k)$ depends on $e_0$ and $k$ we find the celebrated $-5/3$ law,
\begin{equation}
\label{E:Kmgvscaling}
        E(k) \sim e_0 k^{-5/3}\,.
\end{equation}
Various numerical and experimental verifications of \eqref{E:Kmgvscaling} can be found in \cite{chen1999direct,kang2003decaying}.

An exception to \eqref{E:Kmgvscaling} arises when $p=2$ \cite{kraichnan1967inertial, leith1968diffusion, batchelor1969computation}. In two spatial dimensions we may treat $\omega_{ij}$ as a volume form. It is then straightforward to show that the vortex stretching term vanishes for incompressible flow. In this case the enstrophy is bound from above by its value at $t=0$ so that energy is conserved when the Reynolds number becomes large. In this case the palinstrophy plays the same role that enstrophy played in three dimensional flow. One then expects that the energy distribute itself towards lower wavenumber. If energy is continuously pumped into large scales via a driving force then in order for the system to remain in  a steady state some type of large scale friction needs to be introduced into the system. This friction introduces an energy scale and yields a power law behavior as in \eqref{E:Kmgvscaling} referred to as the inverse cascade. In addition to the inverse cacade there is a direct cascade associated with the enstrophy production term. An analysis similar to the one that led to \eqref{E:Kmgvscaling} implies
\begin{equation}
\label{E:Krchnscaling}
        E(k) \sim w_0 k^{-3}\,.
\end{equation}
See \cite{rutgers1998forced} for various discussions.

In the current work we will study decaying turbulence where the typical velocity scale $U$ and length scale $L_0$ vary with time. In the context of the Navier-Stokes equation, a power law behavior associated with a direct cascade, c.f., \eqref{E:Kmgvscaling}, was observed in numerical simulations of decaying turbulence in $p=3$ dimensions \cite{kida1987kolmogorov,hughes2001multiscale, vreman1992finite}. In $p=2$ dimensions numerical simulations of the Navier-Stokes equations usually lead to either an inverse or direct cascade. Numerical simulations of decaying turbulence with no slip boundary conditions may exhibit both a direct cascade and an inverse cascade, as in \eqref{E:Kmgvscaling} and \eqref{E:Krchnscaling} \cite{clercx2000energy,clercx2001two}. In periodic domains a direct cascade as in \eqref{E:Krchnscaling} is often observed \cite{santangelo1989generation,bracco2000revisiting}. Recent simulations exhibit both a direct and an inverse cascade in such scenarios \cite{PhysRevE.87.033002}.

Let us return our attention to \eqref{eq:EOMdimless}. Using intuition gained from the analysis at small Mach number we focus on the evolution of the energy and enstrophy.  We find that our equations of motion \eqref{eq:EOMdimless} imply
\begin{equation}
        \frac{\partial}{\partial t} E_C  = \frac{1}{M^2} \int p  \, (\vec{\nabla} \cdot \vec{u}) \, d^px  -\frac{1}{4 Re}  \int p  \left( \omega_{ij} \omega^{ij} + \sigma_{ij} \sigma^{ij} \right)d^px 
\end{equation}
where we now define
\begin{equation}
        E_C = \int p \, |\vec{u}|^2 d^px
\end{equation}
and 
\begin{equation}
        \omega_{ij} = \partial_i u_j - \partial_j u_i
        \qquad
        \sigma_{ij} = \partial_i u_j + \partial_j u_i \,.
\end{equation}
Note that we have used the letter $p$ for both the number of spatial dimensions and the dimensionless energy density. In the equations above and in the remainder of this section the number of dimensions $p$ will appear only in the measure $d^px$.

Similar to the case of small Reynolds number, we find that $E_C$ is approximately conserved for large $M$ and large $Re$. %\footnote{Since $E_C$ is approximately conserved for both large and small $M$, one may posit that $E_C$ is approximately conserved independantly on $M$. We will see in the next section some numerical evidence that this is indeed the case. \AY{I'm feeling a bit hesitant about this footnote. I'd rather remove it.} }
Furthermore, similar to \eqref{E:icenstrophy}, we find the following equation for the dynamics of the enstrophy
\begin{multline}
\label{E:OmegaC}
        \frac{\partial}{\partial t} \Omega_C = \int \frac{1}{p}  \omega_{ij}\omega^{jk} \left( \sigma_k^{i} - \delta_k^{i} \vec{\nabla} \cdot \vec{u} \right) d^px 
        + \frac{1}{Re} \int 
        \frac{|\vec{\nabla} p |^2}{p^{3}} \omega^{2} 
        - p \partial_k \frac{\omega_{ij}}{p} \partial_k \frac{\omega_{ij}}{p} 
        \\
        - \frac{2}{p^2} \sigma_{kl} \omega^{ lj} \left( \partial_j \partial_k p - \frac{\partial_j p \partial_k p}{p} \right)
        - \frac{2}{p^2} \omega_{kl} \omega^{ lj} \left( \partial_j \partial_k p - \frac{\partial_j p \partial_k p}{p} - \delta_{jk} \left( \nabla^{ 2}p - |\vec{\nabla} p |^2 \right)\right) d^px\,,
\end{multline}
with
\begin{equation}
        \Omega_C = \int \frac{\omega_{ij} \omega^{ij}}{p}\,d^px\,.
\end{equation}
We will refer to the first term on the right hand side of \eqref{E:OmegaC} as a vortex stretching term. This term vanishes for $p=2$ (as does the last term on the right hand side of \eqref{E:OmegaC}). Thus $\Omega_C$ is conserved at high Reynolds number and we may expect an inverse cascade in such a configuration. When $Re$ is finite, it is difficult to determine the sign of the rate of change of $\Omega_C$ and therefore difficult to assess a priori whether a scaling regime exists or not. If it does exist, we may expect the same power law behavior as in \eqref{E:Krchnscaling}. Since \eqref{E:OmegaC} is less manageable than \eqref{E:icenstrophy}, to make headway we must resort to numerical methods. We carry out such an analysis in the next section.

\vspace{10pt}
\section{Analysis of Turbulent Flows}
\label{S:numerical}

We now turn to a numerical analysis of turbulent flows. While the analysis of section \ref{S:fluidflow} was valid for unbound domains, our simulations focus on flows in a bound, toroidal domains of length $L \xi$ in each direction. We comment on the influence of these boundary conditions on our results when relevant. 

%To check for turbulent behavior of the fluid equations \eqref{eq:EOM} which follows from the ensuing analysis of section \ref{S:fluidflow}, we resort to numerical techniques. Spatial discretization was achieved by placing the theory in a periodic domain of length $L$ and using a Fourier transform to evaluate derivatives in the spatial direction. Time evolution was carried out using a third order Adams Bashforth time step evolution. 

We solved the equations of motion \eqref{collEOM} using a variety of methods including Fourier spectral methods and finite differencing in the spatial directions.
We evolved the variables $a$ and $f_a= \epsilon u_a$ forward in time using third order Adams-Bashforth or explicit Runge-Kutta.
The initial conditions we used for our simulations were ``perturbed shear flows", i.e. constant density flows where the
velocity field is perpendicular to its gradient.  Such flows solve \eqref{eq:EOMdimless} in the limit of infinite Reynolds number.

More specifically, for $p=2$ we used
\begin{subequations}
\label{E:inic}
\begin{equation}
\label{eq:init2D}
	f_{x} =E c_s \delta f_{x}\left(\vec{x}\right),\qquad 
	f_{y} =E c_s \cos \left(\frac{2\pi n}{L} x\right)+\delta f_{y}\left(\vec{x}\right),\qquad 
	a = a_0 E
\end{equation}
whereas for $p=3$ we used
\begin{eqnarray}\label{eq:init3D}
	f_{x}  & = & E c_s \cos \left(\frac{2\pi n}{L} y\right) + \delta f_{x}\left(\vec{x}\right),\qquad 
	f_{y}  =E c_s \cos \left(\frac{2\pi n}{L} z\right) + \delta f_{y}\left(\vec{x}\right)\nonumber \\
	f_{z} & = & E c_s \cos \left(\frac{2\pi n}{L} x\right) + \delta f_{z}\left(\vec{x}\right),\qquad 
	a  = a_0 E\,.
\end{eqnarray}
\end{subequations}
Here $E$ is an energy scale which drops out of the equations of motion. The parameter $a_0$ in both \eqref{eq:init2D} and \eqref{eq:init3D} is constant in space and the $\delta f_i$ denote perturbations of the form
\begin{equation}
\label{E:random}
	\delta f_{i}\left(\vec{x}\right)=\sum\limits_{\vec{m}} A_{i,\vec{m}}Ec_s\cos\left(\Delta\phi_{i,\vec{m}}+\frac{2\pi(\vec{m}\cdot\vec{x})}{L}\right)
\end{equation}
where $\vec{m}$, $A_{i,\vec{m}}$ and $\Delta\phi_{i,\vec{m}}$ are chosen from a random sample. The sum in \eqref{E:random} included 10 to 100 random modes. Typical simulations of two dimensional flow included 20 modes and typical three dimensional flow included 40 or 80 modes. The components of the wavenumber $\vec{m}$ was chosen from a uniform distribution of integers running from $1$ to $64$ for a typical two dimensional flow and $1$ to $16$ or $1$ to $32$ for a typical three dimensional flow. The amplitude $A_{i,\vec{m}}$ was chosen from a uniform distribution ranging from $0$ to $A$ with $A$ a pre-determined parameter. The phase $\Delta\phi_{i,\vec{m}}$ was chosen from a uniform distribution ranging from 0 to $2\pi$.

Since our setup does not involve a driving force, the flow we generate does not reach steady state turbulence. Nevertheless, we may associate a Reynolds number and a Mach number to the initial conditions of the flow. Choosing $U = \hbox{max}(u) - \langle u \rangle$ and $L_0 = L \xi/n$, we find that
\begin{equation}
	Re = \frac{L}{a_0 n}\,,
	\qquad
	M=\frac{1}{a_0}\,.
\end{equation}
If the initial Reynolds number is sufficiently large we expect a turbulent instability to emerge. From a numerical standpoint these instabilities are trigered by the modes characterized by $A_{i,\vec{m}}$ in \eqref{E:inic}. For sufficiently large $Re$ we found that numerical roundoff error was sufficient to trigger the instability even for $A_{i,\vec{m}}=0$.

\begin{figure}[hbtp]
        \centering
        \includegraphics[scale=0.325]{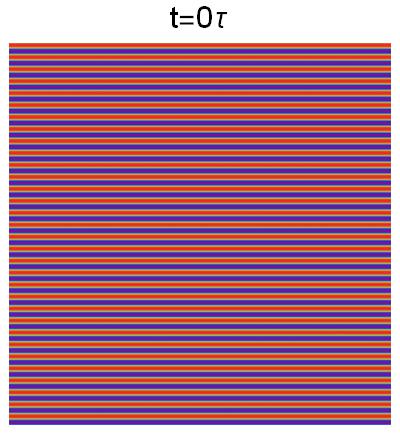}\includegraphics[scale=0.325]{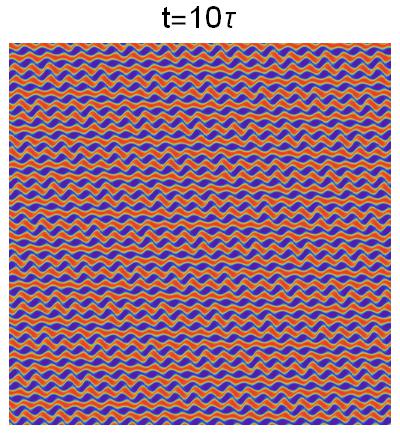}\includegraphics[scale=0.325]{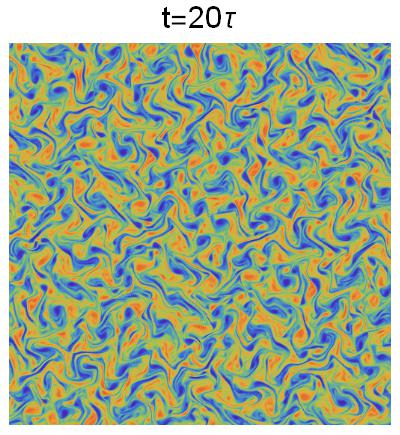}
        \includegraphics[scale=0.335]{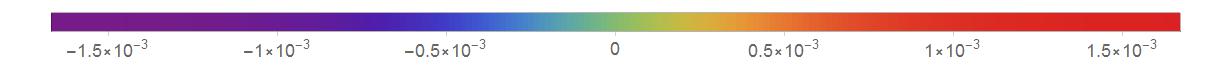}
        \includegraphics[scale=0.325]{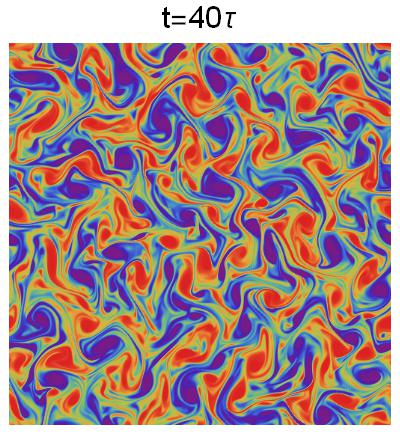}\includegraphics[scale=0.325]{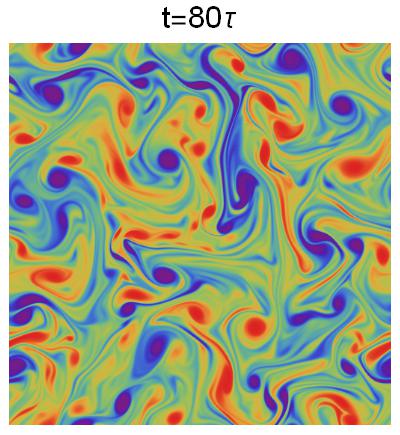}\includegraphics[scale=0.325]{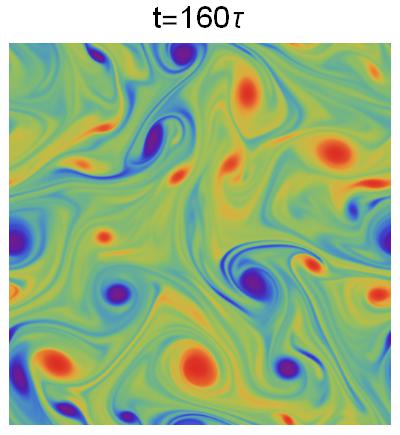}
        \includegraphics[scale=0.335]{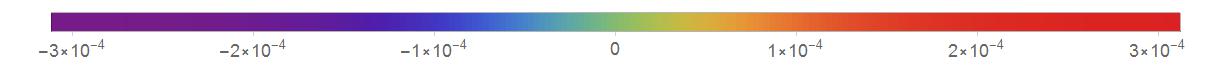}
        \includegraphics[scale=0.325]{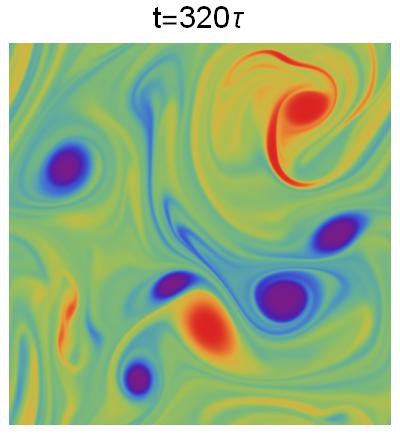}\includegraphics[scale=0.325]{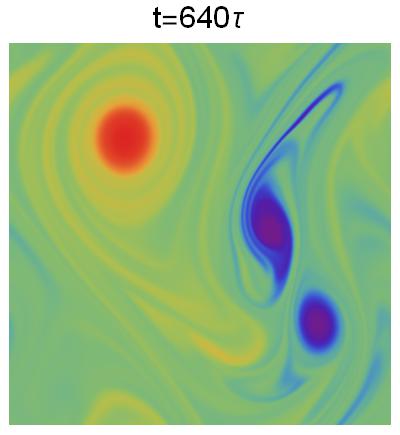}\includegraphics[scale=0.325]{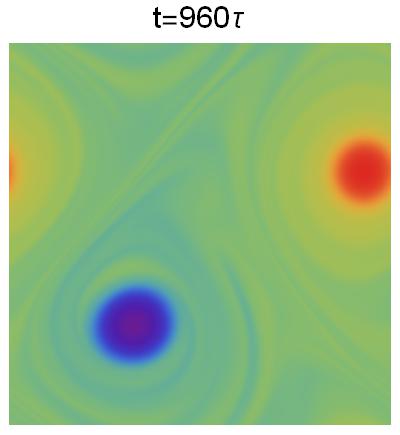}
        \includegraphics[scale=0.335]{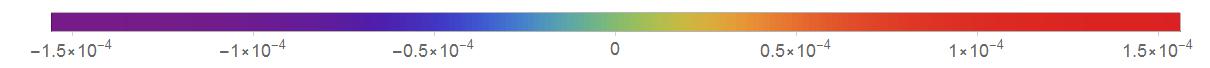}
        \caption{The vorticity field $\omega = \partial_x u_y  - \partial_y u_x$ of a two dimensional turbulent regime with initial Reynolds number $Re = 1562.5$, Mach number $M=0.5$, initial mode $n=32$, and noise scale $A=10^{-5}$ at various times. The instability  becomes apparent around $t \approx 10\tau$, followed by a chaotic behavior generating small eddies shown at $t=20\tau$. The inverse cascade is apparent in the "merging" of co-rotating eddies to form larger ones starting around $t \approx 20\tau$ and ending at $t \approx 960\tau$. The termination of the inverse cascade is caused by the finite length of the box. The final stage shown at $t = 960\tau$ is of slow decay of two counter rotating eddies.}
        \label{fig:vorticity2D}
\end{figure}

\begin{figure}[hbtp]
        \centering
        \begin{tabular}{cccc}
        \textbf{\large \ } & \textbf{\large $\omega_x$} & \textbf{\large $\omega_y$} & \textbf{\large $\omega_z$}\par\medskip \\
        \textbf{\rotatebox{90}{\ \ \ \ \ \ \ \ \ \ \ \ {\large $t=0\tau$}}} &\includegraphics[scale=0.205]{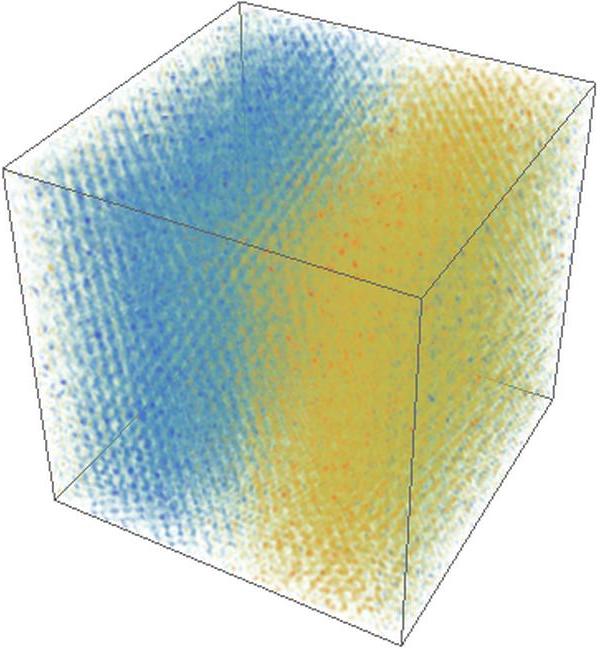} & \includegraphics[scale=0.205]{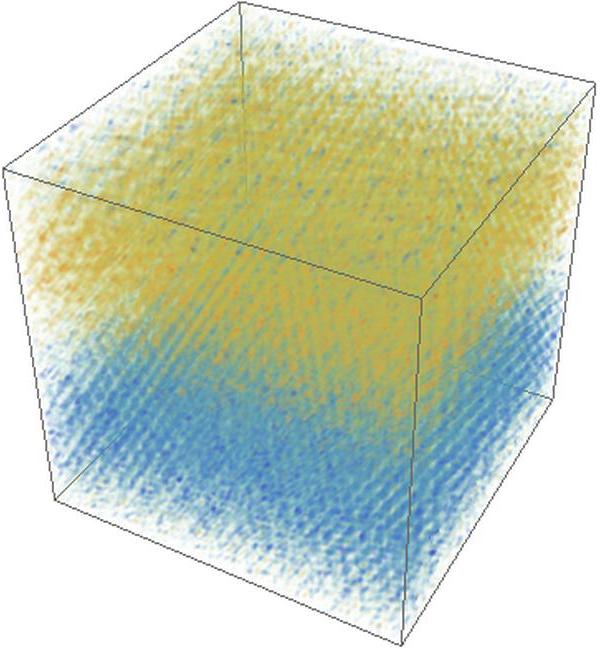} & \includegraphics[scale=0.205]{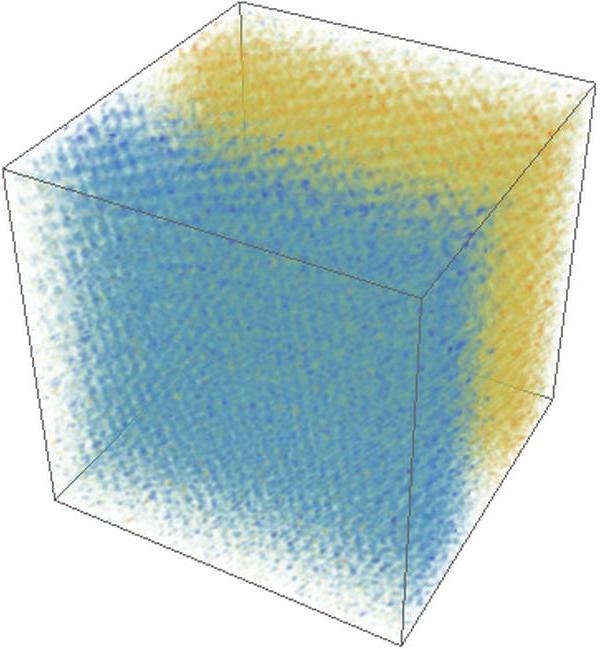} \\
        & \multicolumn{3}{c}{\includegraphics[scale=0.205]{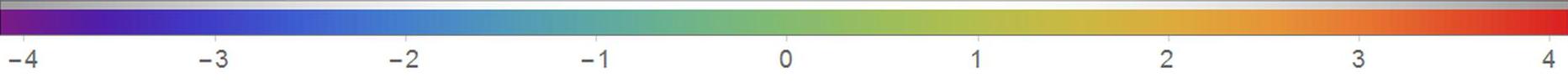}} \\
        \textbf{\rotatebox{90}{\ \ \ \ \ \ \ \ \ \ \ \ {\large $t=0.5\tau$}}} &\includegraphics[scale=0.205]{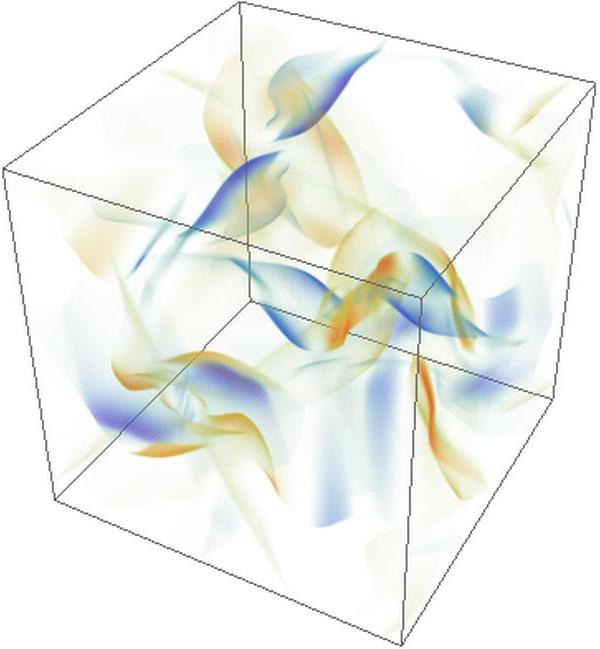} & \includegraphics[scale=0.205]{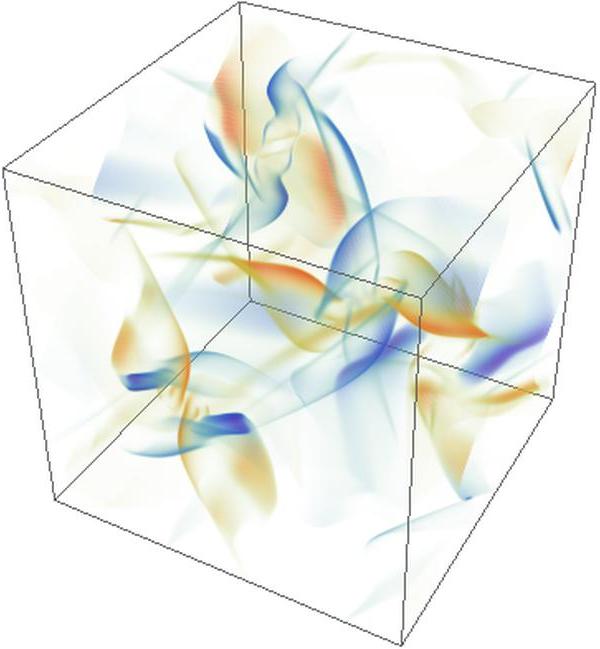} & \includegraphics[scale=0.205]{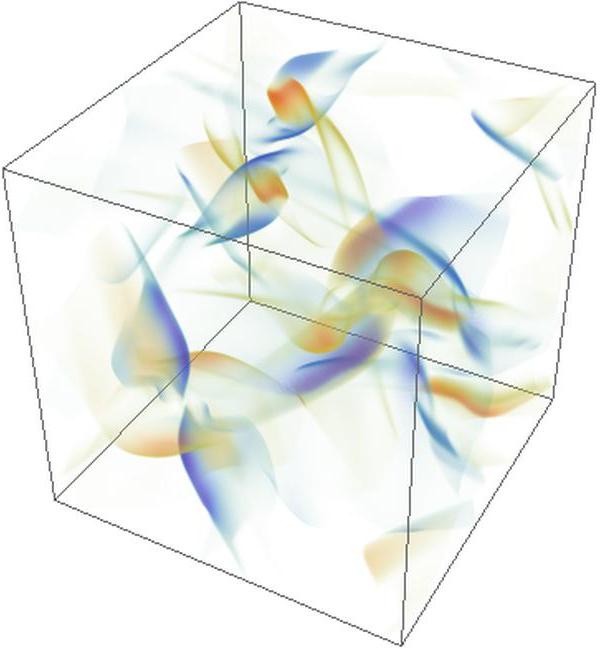} \\
        & \multicolumn{3}{c}{\includegraphics[scale=0.205]{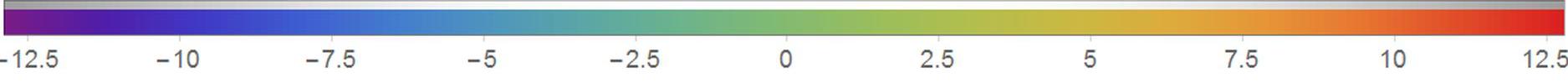}} \\
        %\textbf{\rotatebox{90}{\ \ \ \ \ \ \ \ \ \ \ \ {\large $t=2\tau$}}} &\includegraphics[scale=0.155]{3DvorticityX200n1.jpg} & \includegraphics[scale=0.155]{3DvorticityY200n1.jpg} & \includegraphics[scale=0.155]{3DvorticityZ200n1.jpg} \\
       % & \multicolumn{3}{c}{\includegraphics[scale=0.155]{3DvorticityBar200n1.jpg}} \\
        \textbf{\rotatebox{90}{\ \ \ \ \ \ \ \ \ \ \ \ {\large $t=10\tau$}}} &\includegraphics[scale=0.205]{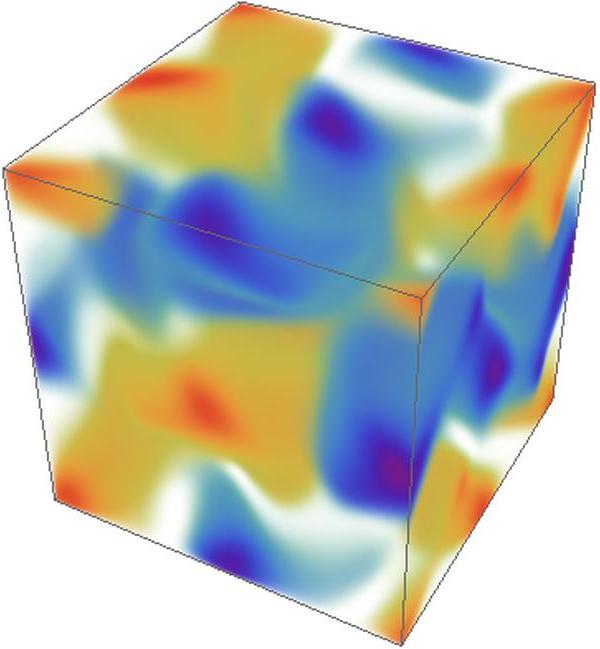} & \includegraphics[scale=0.205]{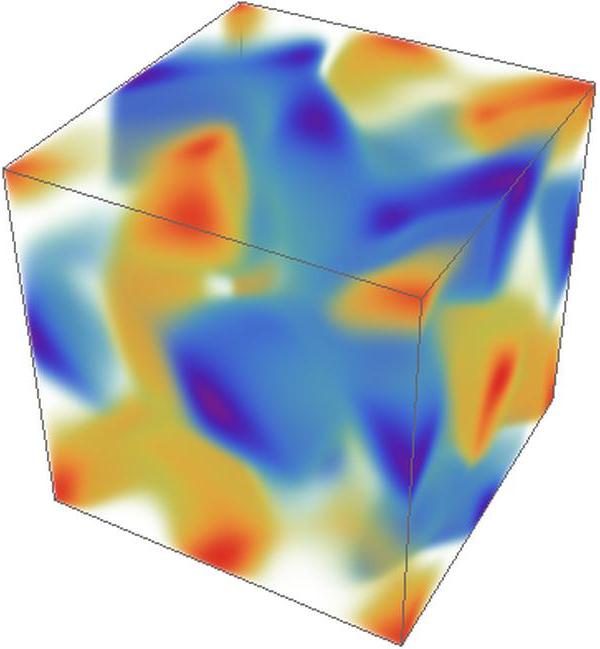} & \includegraphics[scale=0.205]{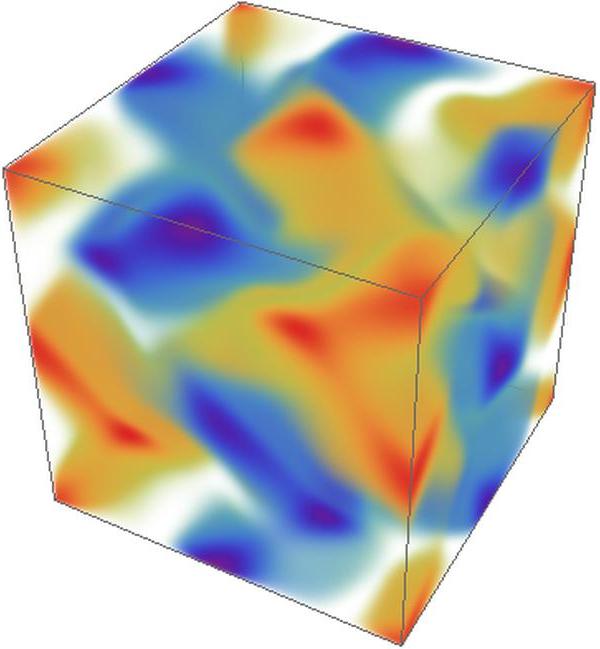} \\
        & \multicolumn{3}{c}{\includegraphics[scale=0.205]{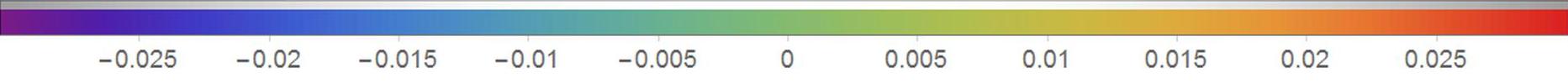}}
        \end{tabular}
        \caption{The vorticity field $\vec{\omega} = \nabla\times\vec{u}$ of  three dimensional flow with initial Reynolds number $Re=750$, initial Mach number $M=1$, initial mode $n=1$ and noise with amplitude $A=10^{-2}$ at different times. The instability driven modes become visible around $t \approx 0.15\tau$. At later times one observes emergent small scale structure ($t = 0.5\tau$). As time progresses, the Reynolds number decreases and we eventually reach the final stage of the flow ($t = 10\tau$) where $Re<1$ and the initial ($n=1$) modes are observed.}
        \label{fig:vorticity3D}
\end{figure}

In presenting our results we find it convenient to use scaled coordinates
\begin{equation}
	\tau = \frac{L_0}{U} = \frac{Re}{M^2} \frac{\xi}{c_s}
	\quad
	e=\frac{a U^2}{L_0^{p-1}}= \frac{M^p}{Re^{p-1}}\frac{E}{\xi^{p-1}} \,,
\end{equation}
for time and energy power spectrum respectively. 
Typical results of two dimensional and three dimensional fluid flow with relatively high Reynolds number can be found in figures \ref{fig:vorticity2D} and \ref{fig:vorticity3D}. 

In both two and three dimensions we found that the dynamical behavior of the fluid can be characterized by three stages. An initial stage where an instability drives the shear flow into a chaotic configuration. Typical flow in this stage can be found in the first row of figures \ref{fig:vorticity2D} and \ref{fig:vorticity3D}. Once the instability sets in, we observe a turbulent regime where the energy power spectrum exhibits power law behavior. Typical turbulent behavior can be seen in the second row of figures \ref{fig:vorticity2D} and \ref{fig:vorticity3D}. The distinction between two dimensional flow and three dimensional flow is most apparent (visually) in the turbulent phase: in two dimensional flow large scale structure is formed while in three dimensional flow, vortices break up into smaller ones. In the final stage the flow decays into an equilibrated configuration. This behavior can be observed in the last row of figures \ref{fig:vorticity2D} and \ref{fig:vorticity3D}. In the remainder of this section we will discuss each of these regimes in detail.

\vspace{10pt}
\subsection{Initial Phase: Onset of Instability}
\label{S:InitPhase}

The initial phase is characterized by the development of instabilities. To quantify these instabilities, and later also turbulence, we analyze the energy spectrum $E_C$, defined similar to $E_I$ in 
%\AY{Replaced $u_{\epsilon}$ with $u_p$ throughout.}
\eqref{E:EnergyDensity},
\begin{equation}\label{E:ModifiedES}
	E_C(t,k) = \frac{\partial}{\partial k}\int_{\left|\vec{k}'\right|\leq k}\frac{d^{p}k'}{\left(2\pi\right)^{p}}\left|\vec{u}_p\left(t,\vec{k}'\right)\right|^{2}
\end{equation}
with $\vec{u}_p\left(t,\vec{k}\right) = \int d^{p}x\sqrt{p(t,\vec{x})}\vec{u}\left(t,\vec{x}\right)e^{-i\vec{k}\cdot\vec{x}}$. When the Reynolds number is small we find that the shear flow quickly decays to an equilibrated configuration. When the Reynolds number is sufficiently large we find that instabilities dominate the spectrum and lead to turbulent berhavior. %See figure \ref{F:instabilitycoarse}. 
\begin{figure}[hbt]
        \centering
        \includegraphics[scale=0.175]{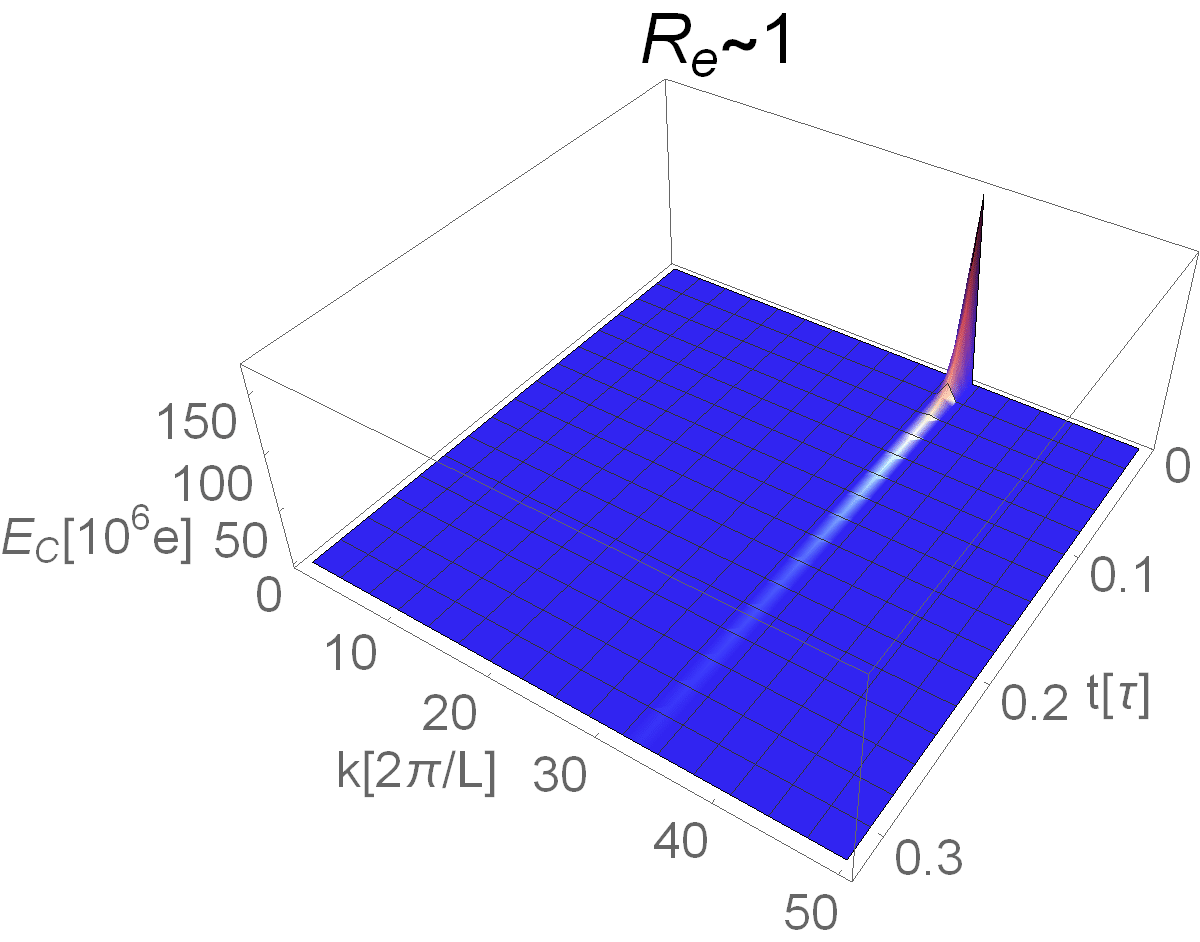}      \includegraphics[scale=0.28]{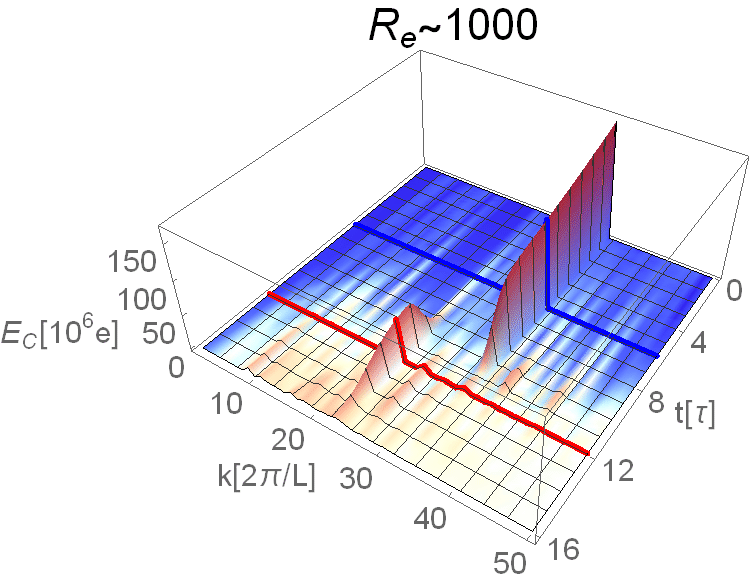}
        \includegraphics[scale=0.28]{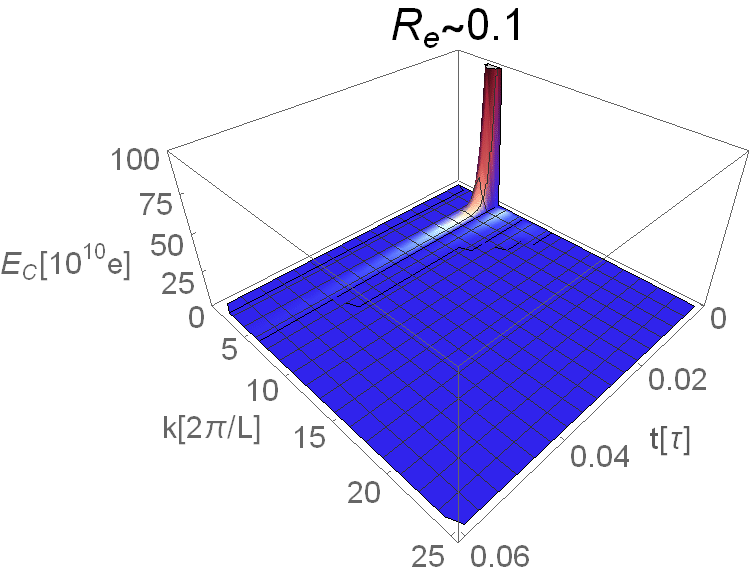}      \includegraphics[scale=0.28]{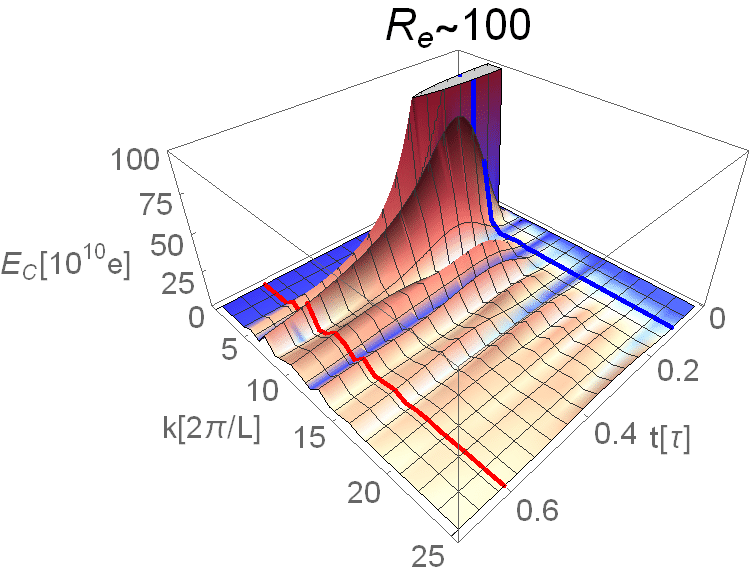}
        \caption{The energy spectrum $E_C$ for two dimensional and three dimensional fluid flow (top row and bottom row, respectively). At low Reynolds number (left column) we find that $E_C$ decays quickly into an equilibrated phase. For two dimensional flow and high Reynolds number (top right) we find that instabilities with lower wavenumber grow and dominate the flow. In three dimensional flow with large Reynolds number (bottom right) turbulent instabilities develop at large wave numbers. A detailed energy spectrum at times marked by the red and blue lines can be found in figures \ref{F:instabilitydetailed2D} and \ref{F:Instabilitydetailed3D}. }
        \label{F:instabilitycoarse}
\end{figure}

Typical behavior of turbulent instabilities in two dimensional fluid flow can be observed in the top right panel of figure \ref{F:instabilitycoarse}. This particular run was carried out with $Re=1562.5$, $M =0.5$, $n=32$ and an amplitude $A=10^{-5}$ for the noise. We have carried out similar runs with Mach numbers ranging from $M=0.005$ to $M=2$ and Reynolds number of up to $Re\sim 1562$. As the Mach number increased the amplitude of the random noise required to generate turbulence became higher; for low Mach number numerical noise was sufficient to generate the instability while for $M=2$ we had to set $A=0.01$ to generate a turbulent instability. It is possible that at very high Mach numbers and the initial conditions \eqref{E:inic} a turbulent instability will not form.

\begin{figure}[hbt]
        \centering
        \includegraphics[scale=0.3]{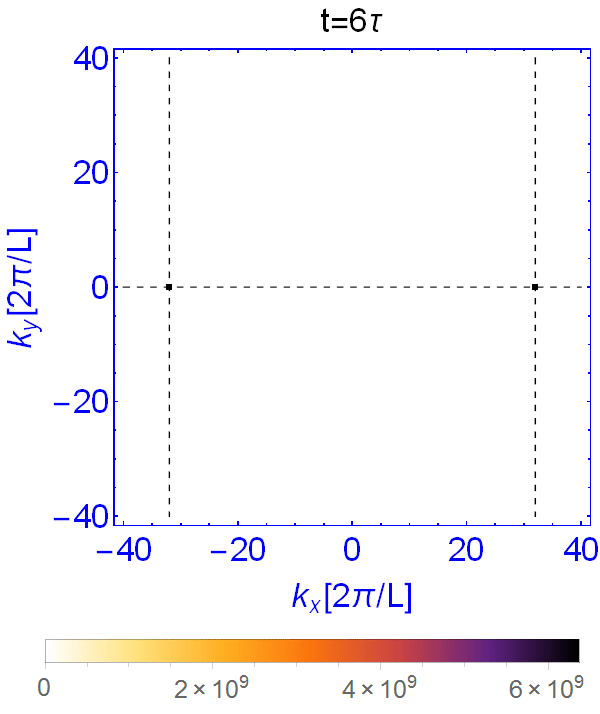}\hspace{1cm} \includegraphics[scale=0.3]{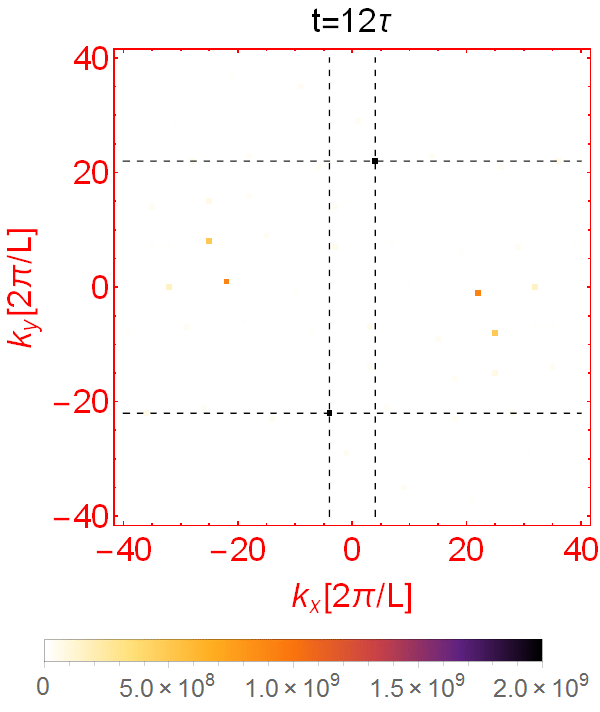}
        \caption{The spectral decomposition of the squared absolute value of $\vec{u}_p(t,\vec{k}) = \int d^{p}x\sqrt{p(t,\vec{x})}\vec{u}\left(t,\vec{x}\right)e^{-i\vec{k}\cdot\vec{x}}$  for the two dimensional flow in the top right panel of figure \ref{F:instabilitycoarse}. At $t=6\tau$ (left) the initial disturbance at $k_x=32\times 2\pi/L$, $k_y=0$ dominates the spectrum. At $t=12\tau$ an instability with $k_x \sim 0$ and $k_y = 22 \times 2 \pi/L$, orthogonal to the initial disturbance but with approximately half the initial wavenumber has set in and dominates the spectrum. Additional weaker instabilities with $k_x \sim 20 \times 2 \pi/L$ and $k_y\sim 0$ can also be observed.}
        \label{F:instabilitydetailed2D}
\end{figure}

Going back to figure \ref{F:instabilitycoarse}, around $t=10\tau$ the initial disturbance with $k=32\times 2\pi/L$ has decayed and an unstable mode with $k=22\times 2\pi/L$ can be observed. A more detailed analysis of the spectrum shows that the unstable mode is a shear mode orthogonal to the first. Indeed, in figure \ref{F:instabilitydetailed2D} we plot the Fourier decomposition of $|u_{p}|^2$ at different times. At $t=6\tau$, the instability has not set in and $|u_{p}|$ receives support from the initial shear mode which we have set as input into the system. At $t=12\tau$ some remnants of the initial shear mode are observed (in red), but most of the support for $|u_{p}|$ comes from a transverse shear mode.
The fact that instabilities with lower wavenumber become excited is suggestive of the inverse cascade present in two dimensional flow.

In three dimensions, the instabilities are associated with higher wave numbers. Typical behavior for instabilities in three dimensions can be found in the bottom right corner of figure \ref{F:instabilitycoarse}.  The data for the bottom right plot in figure \ref{F:instabilitycoarse} was obtained for a flow with $Re=162.5$, $M=1$ and an initial perturbation associated with the fourth harmonic, $n=4$. We have carried out similar runs with Mach numbers of up to $10$. Here, as opposed to the two dimensional case, a turbulent instability seems to emerge even for high Mach numbers. For the initial conditions associated with figure \ref{F:instabilitycoarse}, around $t=0.1\tau$ we observe a growing unstable mode which is roughly double the wavelength of the original.

\begin{figure}[hbt]
        \centering
        \includegraphics[scale=0.24]{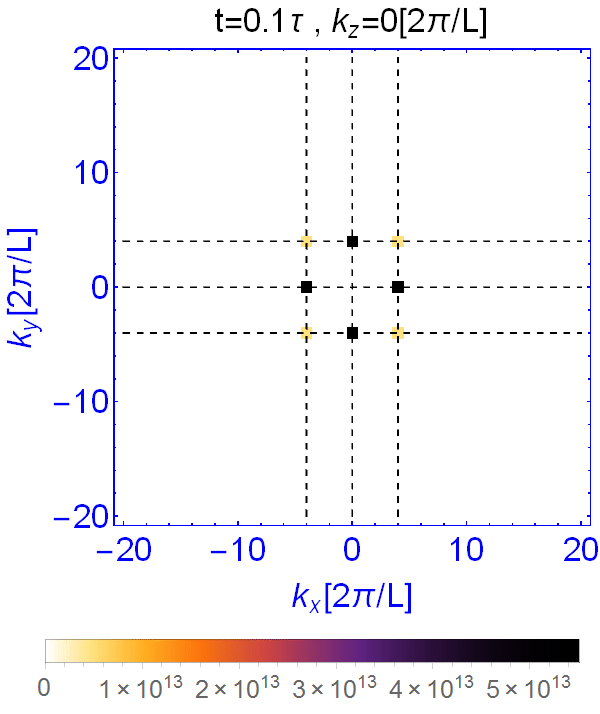}\hspace{1cm} \includegraphics[scale=0.24]{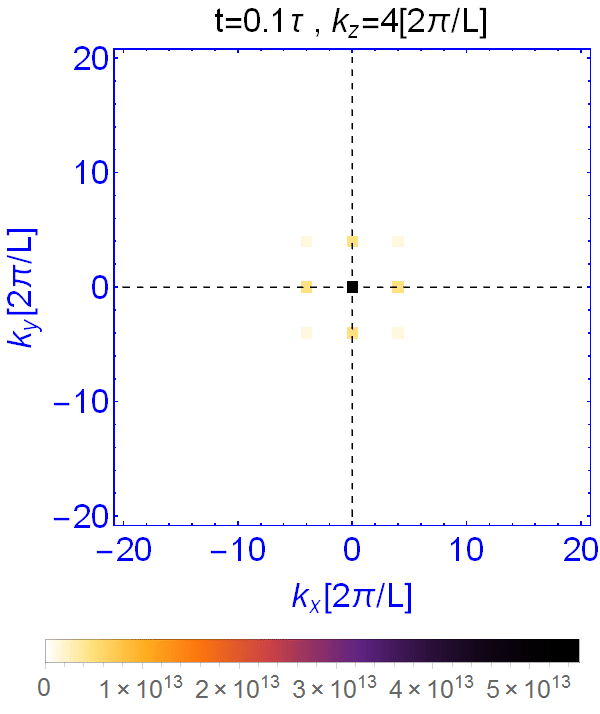}%\hspace{1cm} \includegraphics[scale=0.2]{3DCascadeM1_t=01_z=8.png}
        \\
        \includegraphics[scale=0.24]{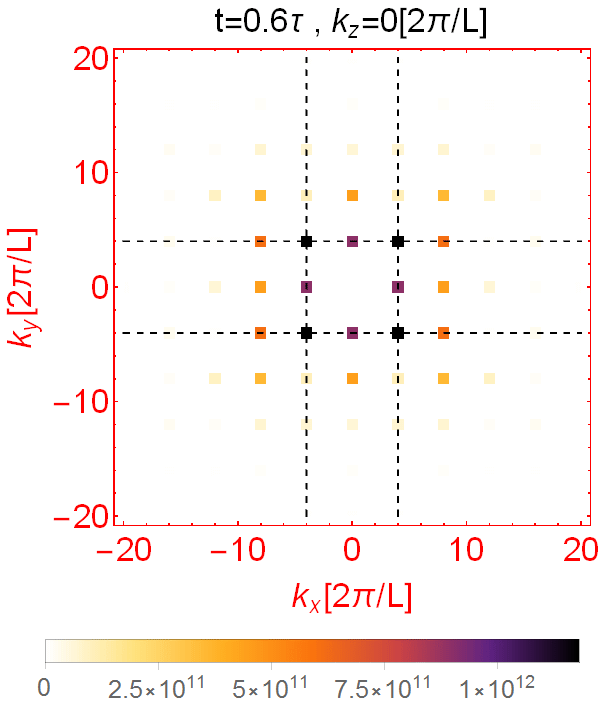}\hspace{1cm} \includegraphics[scale=0.24]{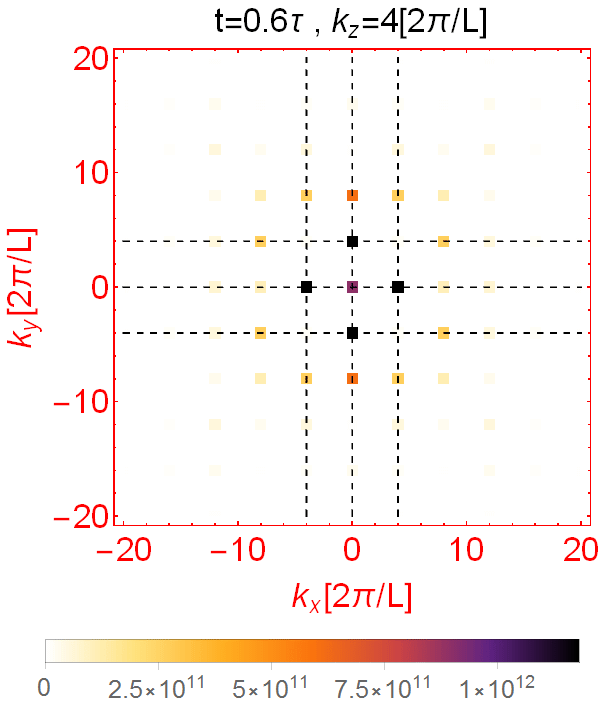}%\hspace{1cm} \includegraphics[scale=0.2]{3DCascadeM1_t=06_z=8.png}
        \caption{Constant $k_z$ slices of the squared absolute value of $\vec{u}_p(t,\vec{k}) = \int d^{p}x\sqrt{p(t,\vec{x})}\vec{u}\left(t,\vec{x}\right)e^{-i\vec{k}\cdot\vec{x}}$ for the fluid flow depicted in the  bottom right panel of figure \ref{F:instabilitycoarse} at $t=0.1\tau$ (blue) and $t=0.6\tau$ (red). 
        At $t=0.1\tau$ the modes $\vec{k} = \pm(4,0,0),\,\pm(0,4,0),\,\pm(0,0,4)$ dominate the spectrum (note the change of scale for the plots on the third column. At $t=0.6 \tau$ the initial modes have decayed and the modes $\vec{k} = \pm(4,4,0),\,\pm(-4,4,0),\,\pm(4,0,4),\,\pm(-4,0,4),\,\pm(0,4, 4),\,\pm(0,-4, 4)$ dominate the spectrum. 
       }
        \label{F:Instabilitydetailed3D}
\end{figure}

A typical Fourier decomposition of the velocity field for three dimensional flow can be found in figure \ref{F:Instabilitydetailed3D} where we show constant $k_z$ slices of the spectrum of $|\vec{u}|$ at different times. The spectrum of the instabilities suggests that the dominant unstable modes take the form $u_i  \sim \cos\left( \frac{2\pi n}{L}(i \pm j)\right)$
where $i,j\in{x,y,z}$, $j$ is the initial mode direction of $f_i$ in \eqref{E:inic}. (For example: if the initial shear flow was of the form $f_x=E c_s \cos \left(\frac{2\pi n}{L} y\right)$, then the associated unstable mode is $u_x  \sim \cos\left( \frac{2\pi n}{L}(x \pm y)\right) $. )

The authors of \cite{green2014holographic} define a critical Reynolds number as the minimal Reynolds number required for a turbulent instability to exist. Put differently, having the Reynolds number above the critical one is a necessary condition for the appearance of turbulence. An analysis similar to that of \cite{green2014holographic} implies that the critical Reynolds number associated with our initial conditions is of the order of 15, roughly two orders of magnitude smaller than the initial Reynolds number needed for turbulence to develop. We provide more details regarding the computation of the critical Reynolds number in appendix \ref{A:critical}.

\vspace{10pt}
\subsection{Turbulent Phase}
\label{S:TurbPhase}
\begin{figure}[hbt]
        \centering
        \includegraphics[scale=0.235]{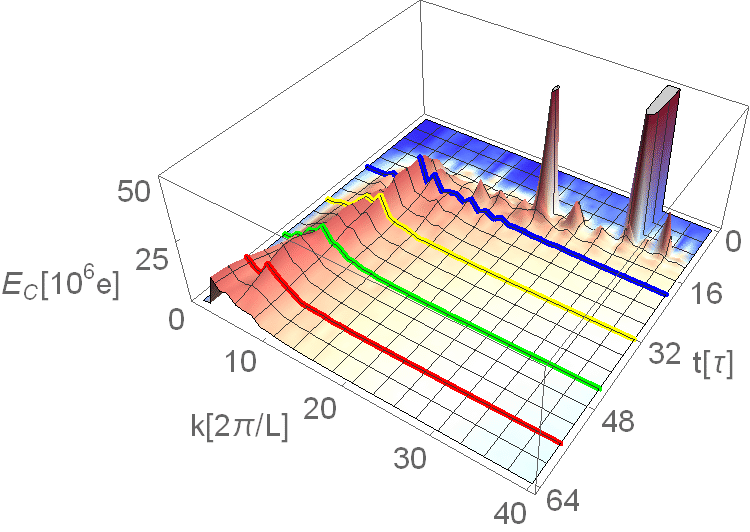}\hspace{1cm}\includegraphics[scale=0.235]{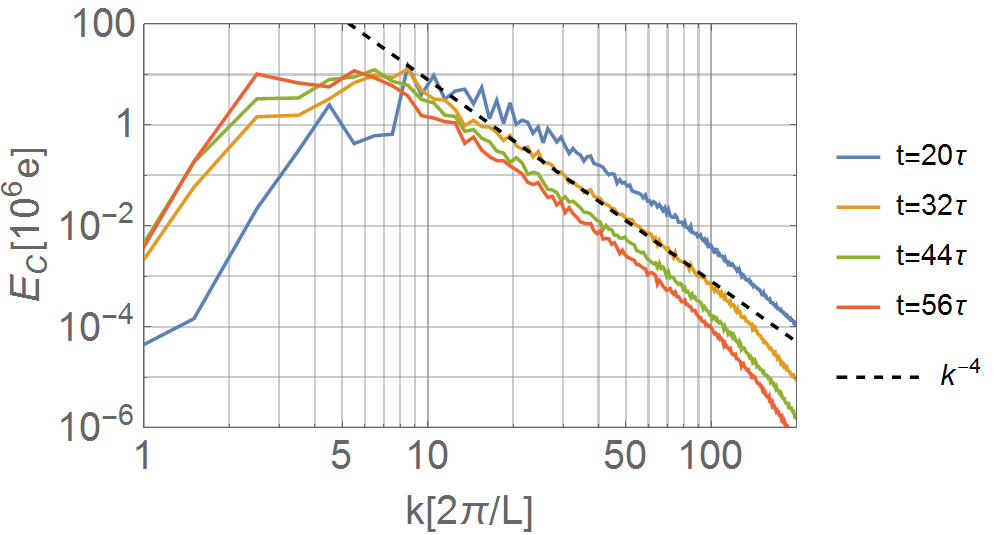}
        \caption{The energy spectrum for two dimensional flow with an initial Reynolds number of $Re=1562.5$ initial Mach number given by $M=0.5$ and initial mode $n=32$. The right plot provides an overlay of the energy spectrum at various times. An emergent power law $E_C(t,k) \sim k^{-4}$ is observed from $k\sim 10\times 2\pi/L$ to $k \sim 100 \times 2\pi/L$. By restricting oneself to lower wavenumber it is possible to fit the data to other power law behavior though the merit in doing so is unclear.}
        \label{F:2dtypical}
\end{figure}

\begin{figure}[hbt]
        \centering
        \includegraphics[scale=0.235]{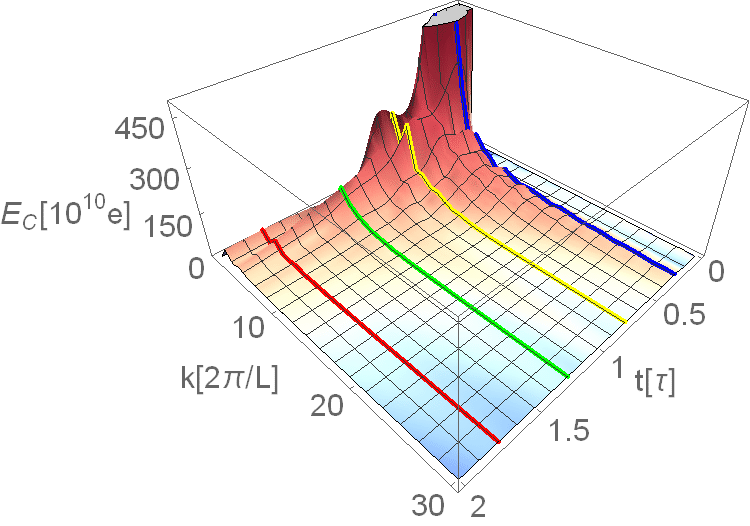}\hspace{1cm}\includegraphics[scale=0.235]{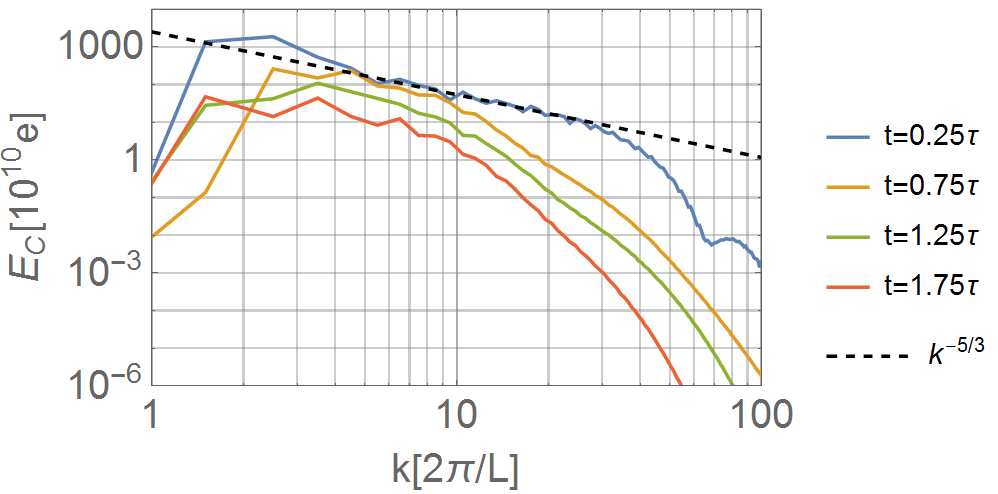}
        \caption{The energy spectrum for three dimensional flow with an initial Reynolds number of $Re\sim 750$ initial Mach number of $M = 1$ and initial mode $n=1$. The right plot is an overlay of the energy spectrum at various times. An emergent power law $E_C(t,k) \sim k^{-5/3}$ is observed at early times and later decays for intermediate values of $k$.}
        \label{F:3dtypical}
\end{figure}

Once the instability of the initial phase is strong enough, it drives the fluid to a turbulent regime which we characterize using the energy density power spectrum, $E_C$, defined in \eqref{E:ModifiedES}. We present a detailed analysis of the power spectrum during the turbulent phase in appendix \ref{A:extraplots}. Here we confine ourselves to summary of the salient features of that analysis. Representative plots of $E_C$ for two and three dimensional fluid flow can be found in figures \ref{F:2dtypical} and \ref{F:3dtypical} respectively.

For two dimensional fluid flow we found a consistent $k^{-4}$ power law for Mach numbers between $0.005$ and $2$ and Reynolds number between 781 and 1562. As the Reynolds number decreases the range for which power law behavior is observed becomes smaller and vanishes completely around $Re \sim 390$. See appendix \ref{A:extraplots}. While it is clear that lower modes get populated as time progresses, indicative of an inverse cascade, the expected $k^{-5/3}$ and (or) $k^{-3}$ law is absent from the simulations we have studied. Early simulations of  non- relativistic decaying turbulence displayed similar scaling \cite{santangelo1989generation,bracco2000revisiting}. Perhaps increasing the initial Reynolds number of our flow will ameliorate this problem, as it does in the non-relativistic case.

% and is associated with can also be understood as follows. The inverse cascades pushes energy from the initial disturbance into lower wavenumbers. Since there is no constant injection of energy into the system, the original disturbance becomes depleted of energy and the system can not sustain a $k^{-3}$ power law. 

In contrast to two dimensional fluid flow, in three dimensions we found remarkable agreement with an $E_C \sim k^{-5/3}$ power law. Our runs include Mach numbers between $M=0.1$ and $M=10$ and Reynolds numbers between $Re=81.25$ and $Re=750$, and initial data involving initial modes $n=1$ and $n=4$. Simulations with initial modes with $n>4$ and reasonably high (initial) Reynolds number are expensive. Indeed, for the $n=4$ run the initial Reynolds number was rather low, $Re = 162.5$, which apparently manifested itself as a visible $\lambda = L/4$ periodic behavior of the flow even in the ``turbulent'' regime where Kolmogorov scaling was observed. %The appearance of the Kolmogorov power law behavior for decaying turbulence may be understood as follows %using the same reasoning for the deviation from the Kraichnan scaling $E_C \sim k^{-3}$ in two dimensions. 
%The cascade drives energy from the initial disturbance to higher modes. While there is no constant injection of energy into the system, if the initial disturbance is sufficiently large the higher modes may be populated so that a $k^{-5/3}$ power law may be sustained.

\vspace{10pt}
\subsection{Final Phase}
\begin{figure}[hbt]
        \centering
        \includegraphics[scale=0.23]{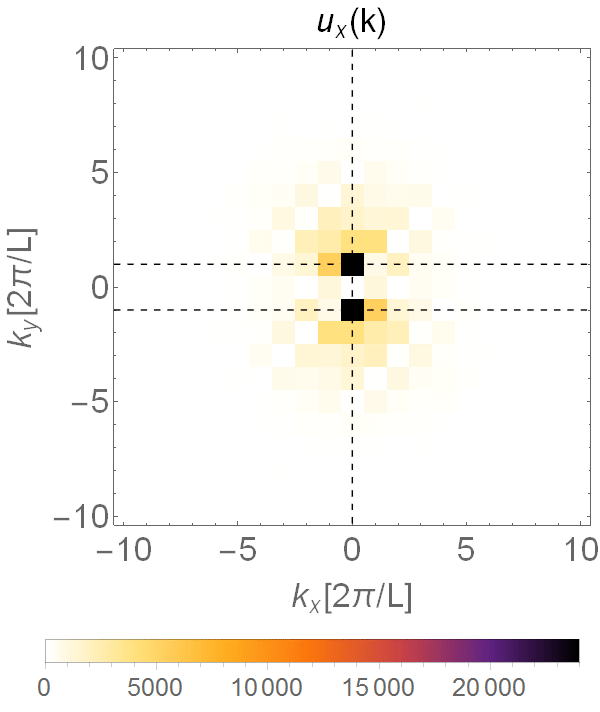}\hspace{0.4cm}\includegraphics[scale=0.23]{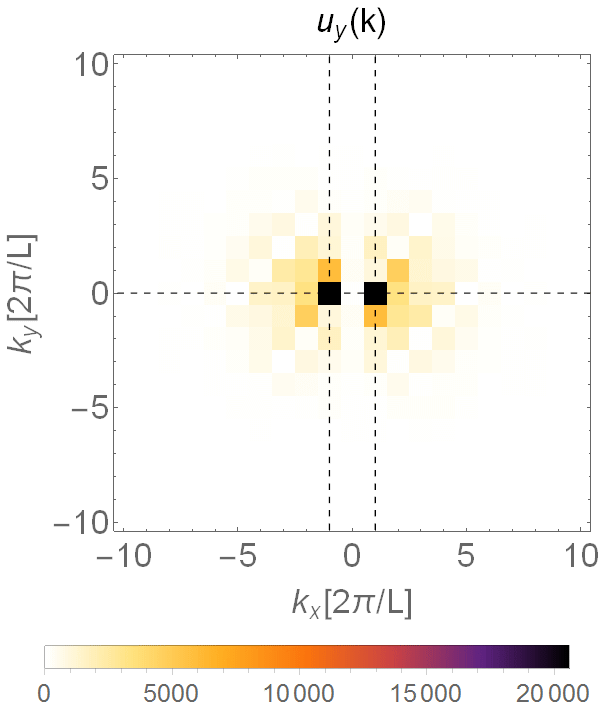}\hspace{0.4cm}\includegraphics[scale=0.23]{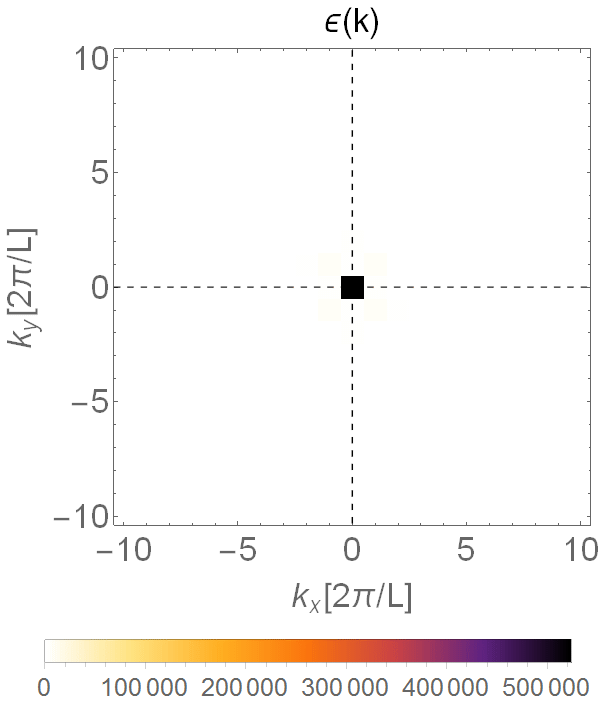}
        \caption{The k-space plot of the velocity field (left) and the energy density (right) at the final phase of the two dimensional turbulent flow ($t=960\tau$) from figure \ref{fig:vorticity2D}. One can notice that the velocity field is dominated by the lowest transverse modes as shown in \eqref{E:LateTime}, while the energy density is dominated completely by the zero mode ($\epsilon \sim \mathcal{E}_0$).}
        \label{F:LateTimeK}
\end{figure}

\begin{figure}[hbt]
        \centering
        \includegraphics[scale=0.36]{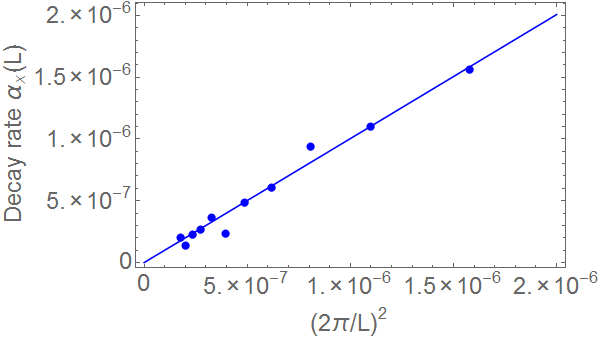}\hspace{0.2cm}\includegraphics[scale=0.36]{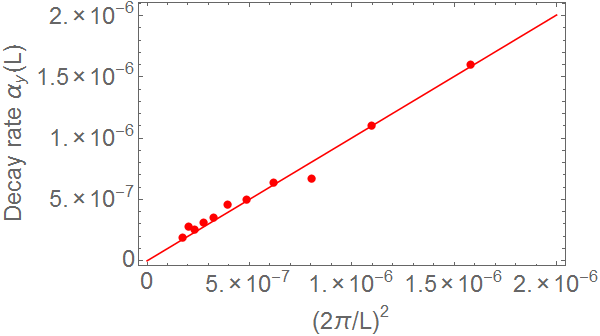}
        \caption{The decay rate of the velocity fields $u_x$ and $u_y$, denoted by $\alpha_x$ (left) and $\alpha_y$ (right) at late times as a function of $(2\pi/L)^2$ with $L$ the length of the torus. Box lengths ranged from $L=5000$ to $L=15000$. The solid lines denote a linear fit}
        \label{fig:decayRate}
\end{figure}

Since there is no driving force the fluid is expected to reach equilibrium at late times. In two dimensional non-relativistic and incompressible fluid flow on $\mathbb{R}^2$ the late time behavior of the velocity field is given by the Oseen Vortex solution which is an attractor of the Navier-Stokes equation \cite{wayne2011vortices}. Since we are placing our fluid on a torus the late time behavior of the fluid is somewhat different. In particular, we find that due to the inverse cascade the lowest lying mode dominates the flow, such that at late times
\begin{eqnarray}\label{E:LateTime}
\epsilon(x,y) & = & \mathcal{E}_0 \nonumber \\
	u_{x}(x,y) & = & U_{x\,0}e^{- \left(\frac{2\pi}{L}\right)^{2}t}\cos\left(\left(\frac{2\pi}{L}\right)y+\phi_{1}\right)\nonumber \\ 
	u_{y}(x,y) & = & U_{y\,0}e^{- \left(\frac{2\pi}{L}\right)^{2}t}\cos\left(\left(\frac{2\pi}{L}\right)x+\phi_{2}\right)\,,
\end{eqnarray}
where $\mathcal{E}_0$, $U_{x\,0}$ and $U_{y\,0}$ are constants.
One can check that \eqref{E:LateTime} solves \eqref{eq:EOMdimless} for low Mach number up to exponentially suppressed corrections. 
Typical flow of the form \eqref{E:LateTime} is depicted in the lower right corner of figure \ref{fig:vorticity2D}. A spectral analysis of the flow can be found in figure \ref{F:LateTimeK}.
To get a quantitative handle on \eqref{E:LateTime} we have fitted the late time behavior of $u_x$ and $u_y$ to an exponential decay law $u_i \sim e^{-\alpha_i(L) t} = e^{-\nu_i \left(\frac{2\pi}{L}\right)^{2}t}$. Let us denote the $L_2$ norm of a quantity $X$ by $L_2(X)$. We evaluate $\alpha_i(L)$ by fitting the dependence of $L_2(u_x)$ and $L_2(u_y)$ on time to a power law fall off, obtaining
\begin{equation}
	\nu_x = 1.004 \pm 0.066,\qquad \nu_y = 1.002 \pm 0.053.
\end{equation}
See figure \ref{fig:decayRate}.

\begin{figure}[hbt]
        \centering
        \begin{tabular}{cccc}
        \includegraphics[scale=0.164]{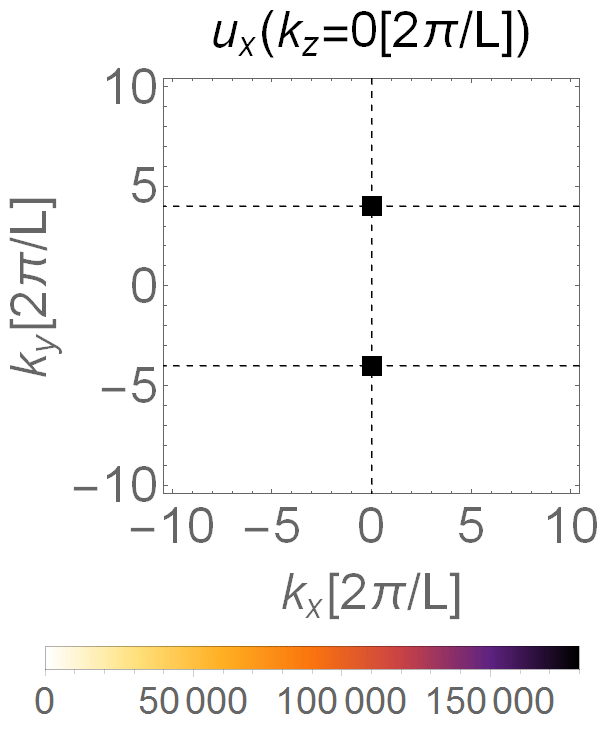} & \includegraphics[scale=0.164]{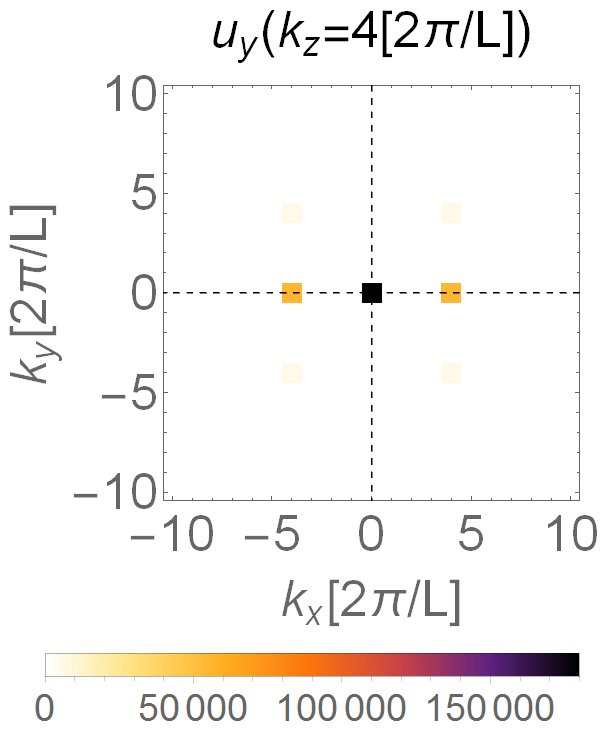} & \includegraphics[scale=0.164]{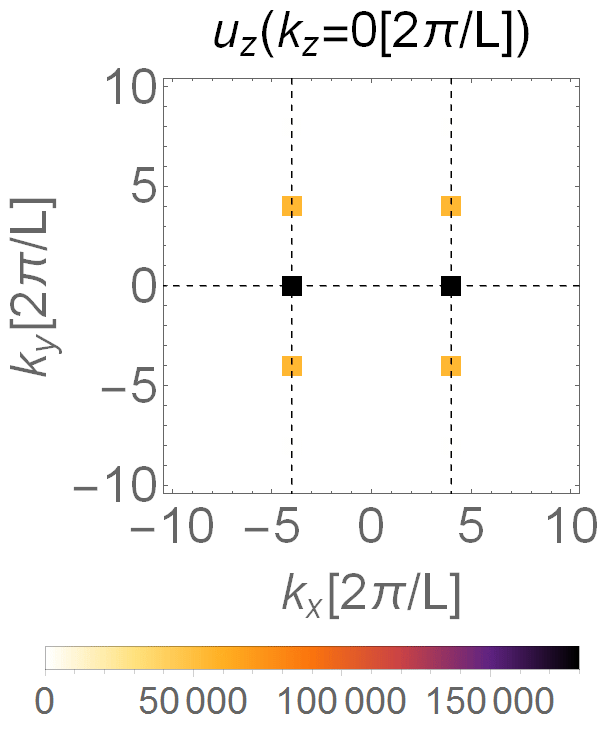} & \includegraphics[scale=0.164]{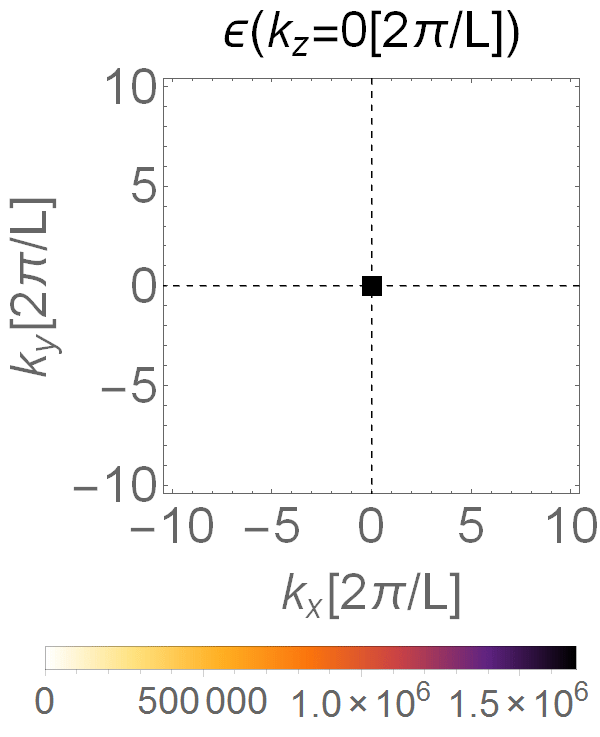}
        \end{tabular}
        \caption{Constant $k$ slices of the velocity field and the energy density at the final ($t=10\tau$) phase of a three dimensional flow with initial mode $n=4$. Other modes, not exhibited in these slices are sub-dominant. The dominant modes at late times are the same as the ones injected at $t=0$. See equation \eqref{E:LateTime3D} .}
        \label{F:LateTimeK3D}
\end{figure}

Since simulating three dimensional fluid flow at late times is expensive, we have not carried out a full analysis of its late time behavior as we have done for two dimensional flow. There is some indication that the late time solution will be dominated by the modes associated with the initial condition \eqref{E:inic}. For the runs we have carried out in three dimensions with $n=1$ and $n=4$ (see \eqref{E:inic}) we find that the late time behavior takes the form
\begin{eqnarray}\label{E:LateTime3D}
	u_{x}  & = & U_x e^{- \left(\frac{2\pi n}{L}\right)^{2}t} \cos \left(\frac{2\pi n}{L} y\right),\qquad 
	u_{y}  = U_y e^{- \left(\frac{2\pi n}{L}\right)^{2}t} \cos \left(\frac{2\pi n}{L} z\right)\nonumber \\
	u_{z} & = & U_z e^{- \left(\frac{2\pi n}{L}\right)^{2}t} \cos \left(\frac{2\pi n}{L} x\right),\qquad 
	\epsilon  = \mathcal{E}_0\,.
\end{eqnarray}
where $n$ is the 
mode number injected into the system in \eqref{eq:init3D}.
Typical flow of the form \eqref{E:LateTime3D} is exhibited on the bottom row of figure \ref{fig:vorticity3D}. A k-space view of the velocity and energy density fields matching \eqref{E:LateTime3D} can be found in figure \ref{F:LateTimeK3D}. A possible explanation for \eqref{E:LateTime3D} may be that the initial perturbation still holds most of the energy at late times and due to the direct cascade is slowest to decay. It is also possible that our three dimensional simulations do not exhibit full turbulent behavior which would be in line with the periodic behavior we observe for three dimensional flow at $n=4$, as mentioned earlier.

\vspace{10pt}
\section{Geometrizing Turbulence}
\label{S:geometry}

As should be clear from our discussion so far, a full understanding of turbulence, even in the incompressible limit, is far from complete. It would be favorable if it were possible to utilize the geometric tools available from numerous studies of black hole dynamics to address turbulence. The AdS/CFT correspondence opens the possibility for such a procedure \cite{Bhattacharyya:2008jc} and there are numerous suggestions for identifying an appropriate geometric quantity which captures the turbulent behavior of the dual fluid \cite{Adams:2013vsa,Eling:2015mxa,Westernacher-Schneider:2015gfa}. In what follows we will focus on the work of \cite{Adams:2013vsa} where the authors argue that the horizon power spectrum, $\mathcal{A}$, to be defined shortly, is proportional to the energy power spectrum defined in \eqref{E:EnergyEnstrophyDef} and \eqref{E:EnergyDensity}. By appealing to the large $d$ limit we obtain an analytic handle over such a relation and can study its regime of validity.

Recall that the leading order contribution to the black brane metric is given by 
\begin{equation}\label{eq:metric}
	\mathrm{d}s^{2}=\mathrm{d}t \left(-\left(1-\frac{a\left(\zeta^{\mu}\right)}{R}\right)\mathrm{d}t+\frac{2}{nR}\mathrm{d}R - \frac{2f_{a}\left(\zeta^{\mu}\right)}{nR}\mathrm{d}\zeta^{a} \right)+\frac{\delta_{ab}}{n} \mathrm{d}\zeta^{a}\mathrm{d}\zeta^{b}+\frac{1}{n}\mathrm{d}\vec{\chi}_{\perp}^{2}\,.
\end{equation}
The only null surface associated with the metric \eqref{eq:metric} which agrees with a black brane topology is given by
\begin{equation}\label{eq:horizon}
	R=a\left(t,\zeta^a \right). 
\end{equation}
which must therefore be identified with the event horizon.

The extrinsic curvature on the event horizon is given by
\begin{equation}
\Theta_{MN}\equiv\Pi_{\ M}^{P}\Pi_{\ N}^{Q}\nabla_{P}n_{Q}\ , \qquad \Pi_{\ N}^{M}\equiv\delta_{\ N}^{M}+\ell^{M}n_{N}
\end{equation}
where Latin indices run over all $d=n+p+2$ dimensions, $n_M$ is the null normal to the horizon, and $\ell_M$ is an auxiliary null vector, which satisfies $\ell_M n^M=-1$. 
Following \cite{Adams:2013vsa} we consider the rescaled traceless horizon curvature, defined by
\begin{equation}
\label{E:tracelesstheta}
	\theta_{\:j}^{i}\equiv\sqrt{\frac{\gamma}{\kappa^{2}}}\Sigma_{\:j}^{i}\ , \qquad \Sigma_{\:j}^{i}\equiv\Theta_{\;j}^{i}-\frac{1}{d-2}\Theta_{\;n}^{n}\delta_{\;j}^{i},
\end{equation}
where $i,j$ run over the $n+p$ spatial dimensions of the horizon, $\sqrt{\gamma} \equiv \sqrt{\det(g_{ij})}$ is the area element on a spatial slice of the event horizon, and $\kappa$ is defined by the geodesic equation $n^{M}\nabla_{M}n_{Q}=\kappa n_{Q}$. 
The horizon curvature power spectrum is defined by
\begin{equation}
\label{E:horizonpower}
	\mathcal{A}\left( t,k \right) \equiv \frac{\partial}{\partial k} \int\limits_{|\mathbf{k'}|\le k} \frac{\mathrm{d}^pk'}{(2\pi)^p}  \tilde{\theta}_{\ \ j}^{*i}\left( t,\mathbf{k'} \right) \tilde{\theta}_{\ i}^{j}\left( t,\mathbf{k'} \right) ,
\end{equation}
where
\begin{equation}
\label{E:Fourierdef}
	\tilde{\theta}_{\ j}^{i}\left( t,\mathbf{k} \right) \equiv \int \mathrm{d}^px\ \theta_{\ j}^{i}\left( t,\mathbf{x} \right) e^{-i \mathbf{k\cdot x}}
\end{equation}
and $p$ is the number of spatial dimensions where the dynamics take place. 

An explicit computation gives us
\begin{align}
	n_M \mathrm{d}x^M = \mathrm{d}R - \partial_t a \mathrm{d}t - \partial_b a \mathrm{d}\zeta^b\,,
	\qquad
	\ell_M \mathrm{d} x^M = - \frac{1}{nR}\mathrm{d}t \,.
\end{align}
and 
\begin{equation}
	\sqrt{\gamma}=a\left(\zeta^{\mu}\right)n^{-(n+p)/2}\,,
	\qquad
	\kappa=a\left(\zeta^{\mu}\right)n^2/2
\end{equation}
from which
\begin{equation}
\label{E:thetaExact}
	\theta_{\:j}^{i} = \frac{a\delta^{il}}{n^{(p+n)/2}}\left(\partial_{j}\left(\frac{f_{l}}{a}-\frac{\partial_{l}a}{a}\right)+\partial_{l}\left(\frac{f_{j}}{a}-\frac{\partial_{j}a}{a}\right)\right)
\end{equation}
follows. Substituting $a=\epsilon$, $f_{a}=\epsilon\beta_{a}$ and taking the incompressible limit, we find
\begin{equation}
\label{E:thetaAproximate}
	\theta_{\:j}^{i}=\frac{\epsilon}{n^{(p+n)/2}}\left(\partial_{j}\beta^{i}+\partial^{i}\beta_{j}\right)\,.
\end{equation}

The Fourier transform $\tilde{\theta}$ will involve a convolution of $\tilde{\epsilon}$ and $\tilde{\beta}$ which are the Fourier transform of $\epsilon$ and $\beta$ respectively, a 'la \eqref{E:Fourierdef}. However, if we are working in the limit of small Mach number then $\epsilon$ is approximately constant and we find
\begin{equation}
\label{E:thetalowmach}
	\tilde{\theta}_{\:j}^i \eqsim \frac{\epsilon}{n^{(p+n)/2}}\left(i k_{j}\tilde{\beta}^{i}+i k^{i}\tilde{\beta}_{j}\right)\,.
\end{equation}
Inserting \eqref{E:thetalowmach} into \eqref{E:horizonpower} and comparing to \eqref{E:ModifiedES} we find that at low Mach number,
\begin{equation}
\label{E:AElowMach}
	\mathcal{A}(k)/E(k) = \frac{2 \epsilon}{n^{(p+n)}} k^2
\end{equation}
as predicted in \cite{Adams:2013vsa}. If, on the other hand, the Mach number is not small a relation of the form $\mathcal{A} \sim k^2 E$ will not hold. 

\begin{figure}[hbtp]
        \centering
        \textbf{$Re=1562.5\ \ M=0.005\ \ (2d)$}\hspace{2.9 cm}
        \textbf{$\ \ Re=1562.5\ \ M^{}=2\ \ (2d)\ \ $}
        \includegraphics[scale=0.217]{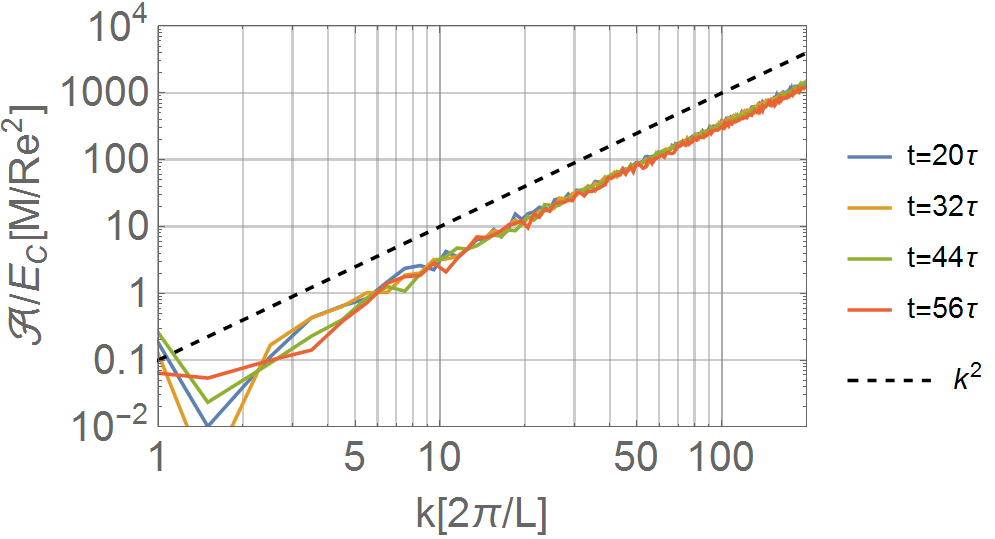}\hspace{0.4 cm}
        \includegraphics[scale=0.21]{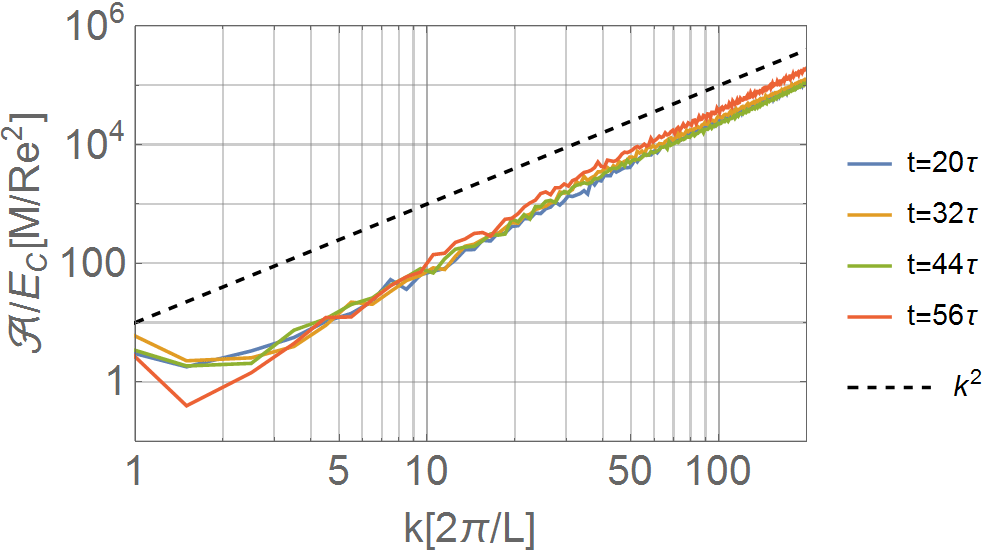} \\
	\vspace{0.5 cm}\textbf{$Re =750\ \ M =10\ \ (3d)$} \\
        \includegraphics[scale=0.22]{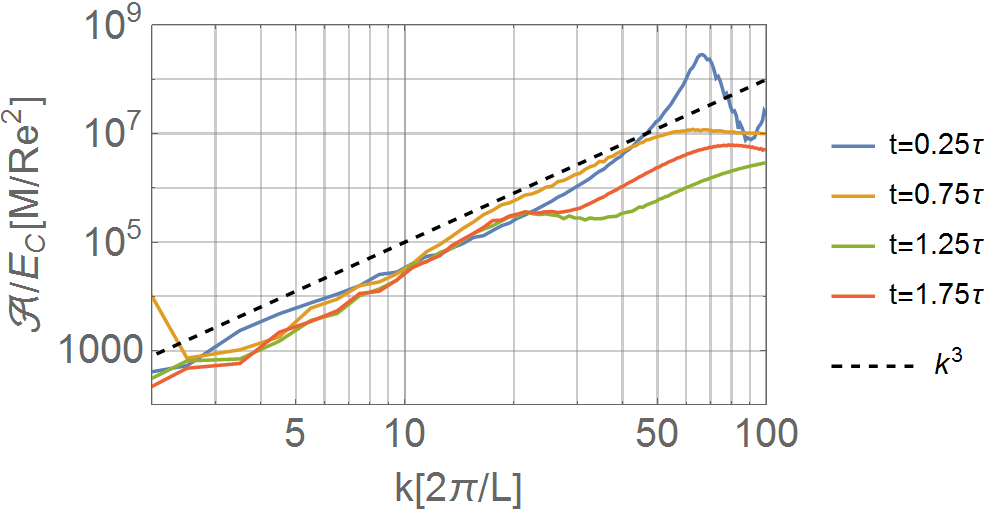}
        \caption{Plots of the ratio of the horizon area power spectrum to the energy power spectrum, $\mathcal{A}(t,k)/E_C(t,k)$, for two and three dimensional turbulent flows at intermediate times. The dashed line specifies a $k^{2}$ or $k^3$ behavior which is in line with the expectations of section \ref{S:geometry}.}
        \label{F:HorizSpec}
\end{figure}

Indeed, in figure \ref{F:HorizSpec} we show typical results for the horizon power spectrum for high and low initial Mach number in two and three dimensions. As expected, $\mathcal{A} \sim k^2 E$ holds only for very low Mach number. A more refined analysis can be found in appendix \ref{A:Horizonpower}.

\vspace{10pt}
\section{Summary and Outlook}
\label{S:Outlook}

In this paper we have discussed relativistic hydrodynamics in the limit where the number of dimensions is large,  revealing the simplifications that occur in this limit. We have focused our attention particularly on holographic theories whose dynamics surprisingly follow these simplified equations. We have analyzed turbulent flows of these dynamical systems, and their relation to the geometry of black hole horizons.

We have seen that three dimensional flows exhibit a Kolmogorov cascade with the expected power law behavior, $k^{-5/3}$, for a wide range of Mach numbers. While promising, we remind the reader that the initial conditions for our three dimensional simulations where carried out with a low wavenumber ($n=1$ and $n=4$). Higher wave numbers would require more computational resources. Moreover, in the $n=4$ simulations we observed a periodicity with wavelength $\lambda = L/4$ throughout the flow. We expect that such behavior will disappear for sufficiently high initial Reynolds number which we have not yet reached.

In contrast to three dimensional flows, two dimensional flows did not exhibit either a $k^{-5/3}$ power law or a $k^{-3}$ power law, albeit displaying a tendency to form large scale structures, associated with the expected inverse cascade. The behavior we observe is probably due to an insufficiently high Reynolds number. It would be worthwhile to improve on this point; simulations of decaying turbulence which exhibited both a direct and inverse cascade usually require a grid much larger than the one we have used \cite{PhysRevE.87.033002}. Preliminary runs do indicate that increasing the Reynolds number may result in a canonical power law.

Recall that the power law behavior of the energy power spectrum has been evaluated for sustained turbulence in which case a driving force continuously supplies energy into the system. It would be interesting to study sustained turbulence in our setup, which would allow for more robust scaling relations to be observed in the steady state. In order to reach that steady state, one would need to add stochastic force to supply energy at the appropriate range of wave numbers, and for the scenario of inverse cascade to include friction as a sink of energy at large scales (keeping in mind that in the direct cascade energy is dissipated at small scales.)

In addition to the Kolmogorov scaling discussed in this work, steady state turbulence would allow us to investigate real space scaling relations of the sort recently described in \cite{Westernacher-Schneider:2017snn}. In the present context of decaying turbulence, those relations are not stable enough to be clearly visible, but we expect that in sustained turbulence we would be able to investigate them more reliably. Similar comments apply to the holographic study of superfluid turbulence. While holographic superfluidity was studied in the large dimension limit \cite{Emparan:2013oza}, the reduction to horizon equations does not hold in the presence of a scalar field. We hope to return to this problem in the future.

While in principle these additional elements are possible in the large dimension limit, it seems we are no longer afforded the simplifications of that limit, with those additional elements. The dimensional reduction of the equations, allowing us to discuss the horizon fluid first and deduce the resulting spacetime subsequently, is no longer in effect when adding external dials, since those are imposed at infinity. Nevertheless, perhaps there is another simplifying limit that would allow for investigation of sustained turbulence in the present context.

Finally, using the simplicity of the large d black brane metric we were able to compare our analytic expression for the area power spectrum to the energy power spectrum. We have found that the power law behavior of these two quantities is related only at very low Mach number at which point there is very good agreement with \cite{Adams:2013vsa}. It would be interesting to identify a geometric entity which captures the Kolmogorov power law behavior. With such a quantity at hand one may be able to use the powerful tools of general relativity to compute this quantity in a more precise manner.

\vspace{10pt}
\section*{Acknowledgements}
The work of AY and ES is supported by an ISF grant and an ISF-UGC grant. The work of MR is supported by a Discovery grant from NSERC.

\vspace{10pt}
\begin{appendix}
\vspace{10pt}
\section{Critical Reynolds Number}
\label{A:critical}

Following \cite{green2014holographic} one may quantify the critical Reynolds number, above which the instability grows. 
Given the initial conditions \eqref{E:inic}, we define a critical Reynolds number, $Re^c$, for these conditions, as the instantaeous Reynolds number at the moment where the amplitude of the unstable mode reaches a maximal value which is lower than the amplitude of the initial shear mode. That is, starting with \eqref{E:inic} we look for a maximum of the amplitude of the growing perturbed mode. We have carried out the analysis for several values of $n$ (mode number), $a_0$ (Mach number) and $L$ (box size). 

\begin{figure}[hbt]
        \centering
        \includegraphics[scale=0.25]{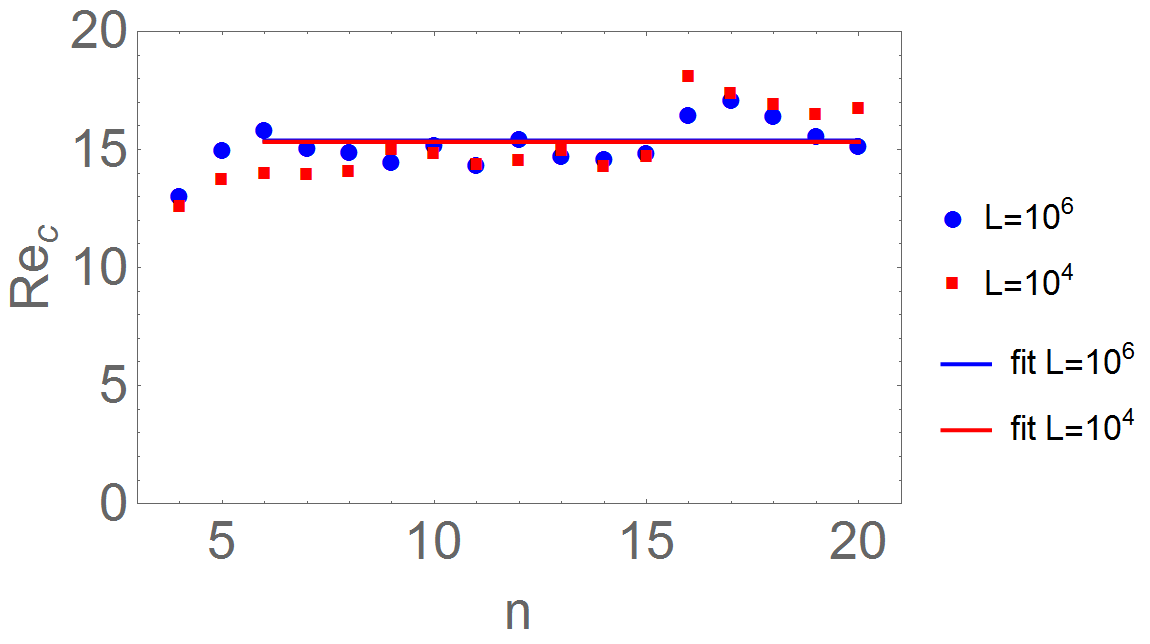}
        \caption{The critical Reynolds number of a two dimensional flow vs. the initial wave number $n$. In black the numerical results for $M \sim 0.01 \ll 1$ ($L=10^6$) resulting in $Re^c=15.38 \pm 0.81$ while ignoring the two points on the left. In blue the numerical results for $M \sim 1$ ($L=10^4$) resulting in $Re^c=15.33 \pm 1.38$ where we again ignored the two points on the left.}
        \label{fig:criticalRe2D}
\end{figure}

Our results for two dimensional flow can be found in figure \ref{fig:criticalRe2D}. We find
\begin{equation}
Re^c = 15.38 \pm 0.81.
\end{equation}
independent of $n$ for $L=10^6$ and $M=0.01$ and
\begin{equation}
Re^c = 15.33 \pm 1.38.
\end{equation}
for $L=10^4$ and $M=1$.
As described in the main text, once the Mach number is much larger than 1, it is difficult to generate a turbulent instability using the initial condition \eqref{E:inic}.

\begin{figure}[hbt]
        \centering
        \includegraphics[scale=0.25]{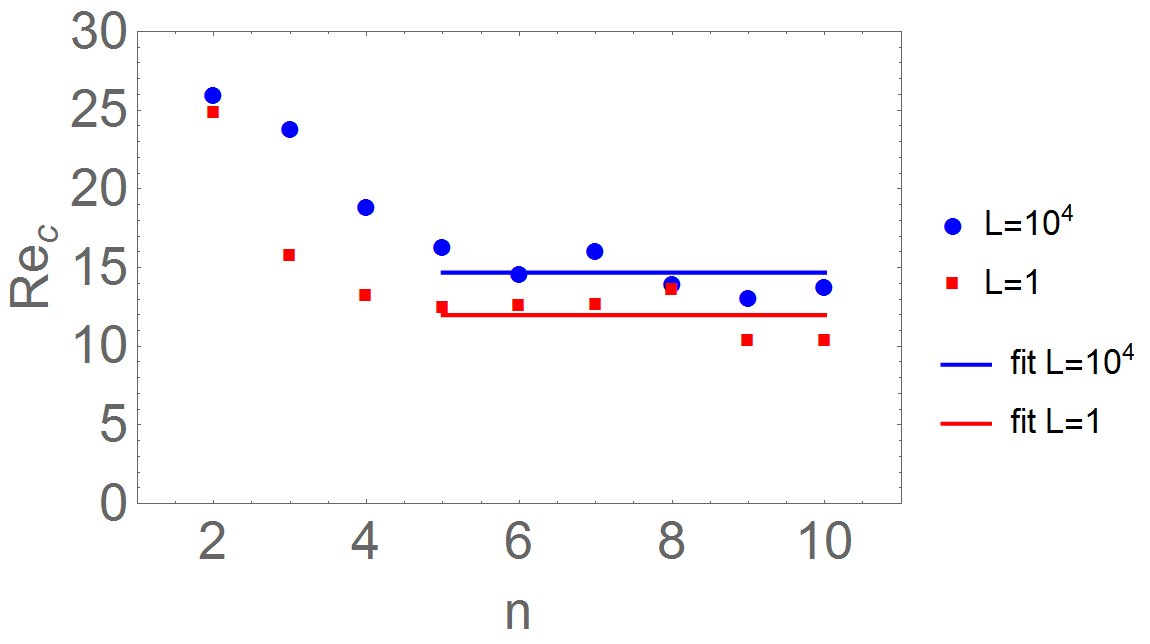}
        \caption{The critical Reynolds number of a three dimensional flow vs. the initial wave number $n$. In black the numerical results for $M \sim 0.01 \ll 1$ ($L=10^4$) resulting in $Re^c=15.2 \pm 2.6$ ignoring the first two points on the left. In blue the numerical results for $M \sim 100 \gg 1$ ($L=1$) resulting in $Re^c=12.2 \pm 1.1$ ignoring the two leftmost points.}
        \label{fig:criticalRe3D}
\end{figure}

Results for three dimensional flow can be found in figure \ref{fig:criticalRe3D}. 
Here, we find that 
\begin{equation}
	Re^c = 14.7 \pm 1.3.
\end{equation}
for $M=0.01$ and $L=10^4$ and
\begin{equation}
Re^c = 12 \pm 1.3
\end{equation}
for $M=100$ and $L=1$. While it is relatively simple to generate three dimensional flow with large Mach number, it is difficult to numerically simulate a flow with initial conditions \eqref{E:inic} and large $n$. The main complication in simulating large $n$ is related to the direct cascade where the initial disturbance is pushed to large wave numbers. The fluids tendency to populate modes with large wavenumber require a large number of grid points to simulate.

\vspace{10pt}
\section{Detailed Analysis of the Power Spectrum}
\label{A:extraplots}

\begin{figure}[hbt]
        \centering
        \begin{tabular}{ccc}
        \textbf{\rotatebox{90}{\ \ \ \ \ \ \ \ \ \ $M=2$}} & \includegraphics[scale=0.23]{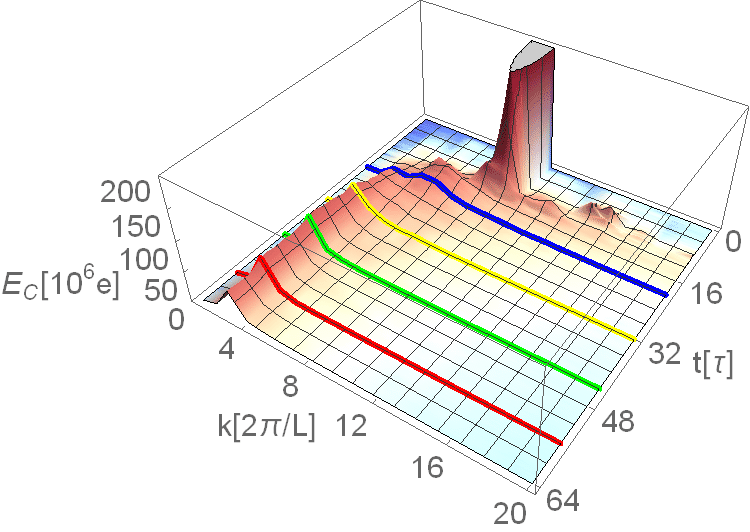} & \includegraphics[scale=0.23]{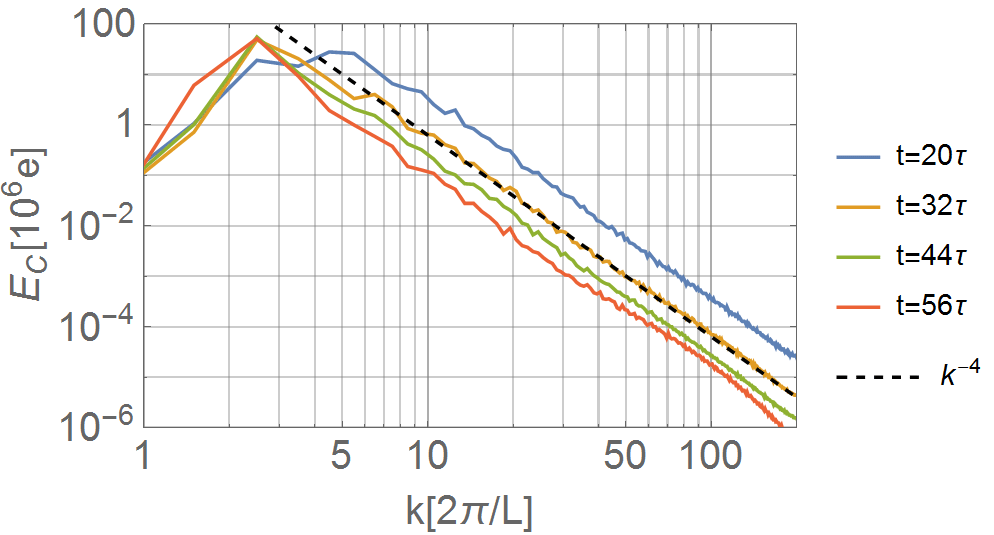} \\
        \textbf{\rotatebox{90}{\ \ \ \ \ \ \ \ \ \ $M=0.5$}} &\includegraphics[scale=0.23]{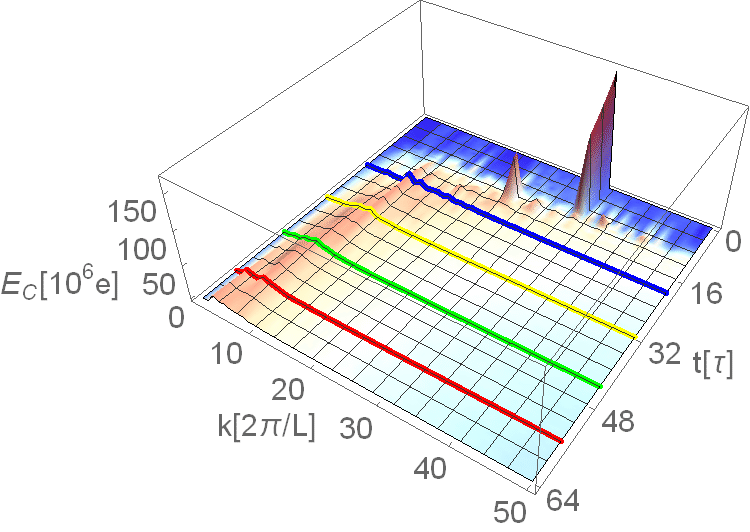} & \includegraphics[scale=0.23]{2DSpectrumM05.png} \\
        \textbf{\rotatebox{90}{\ \ \ \ \ \ \ \ \ \ $M=0.05$}} &\includegraphics[scale=0.23]{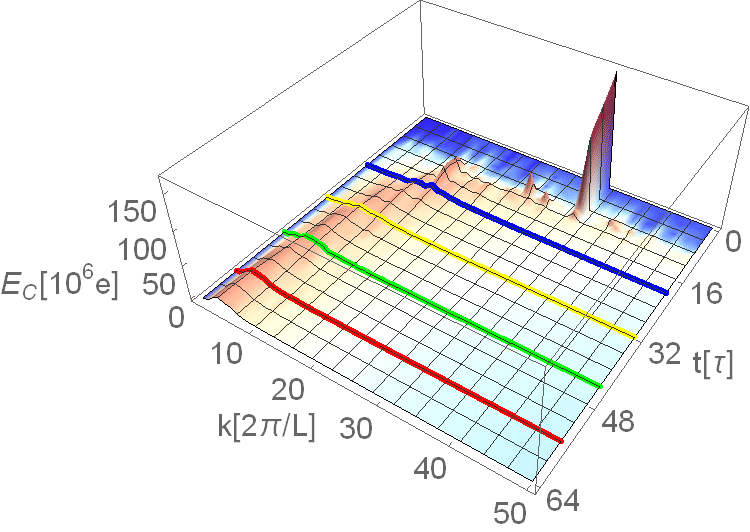} & \includegraphics[scale=0.23]{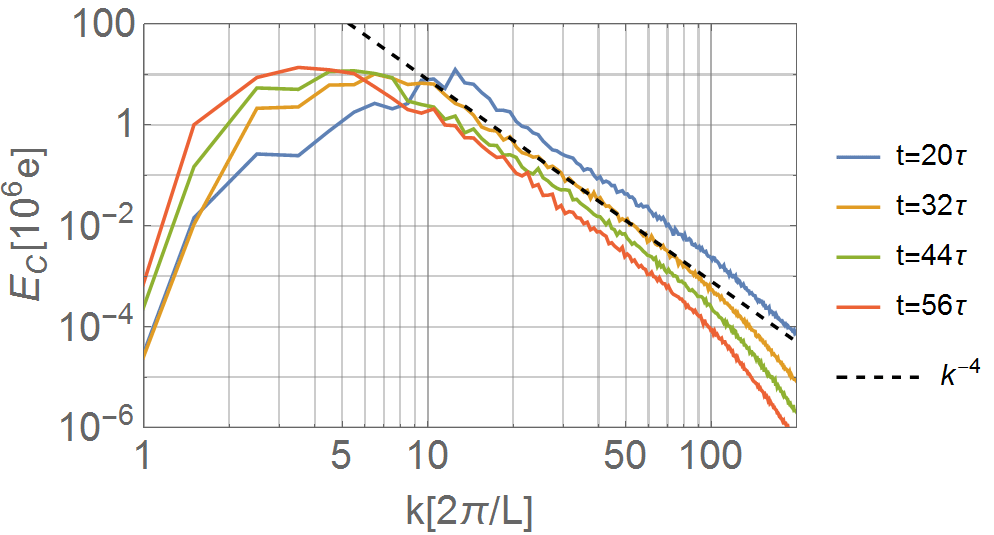} \\
        \textbf{\rotatebox{90}{\ \ \ \ \ \ \ \ \ \ $M=0.005$}} &\includegraphics[scale=0.23]{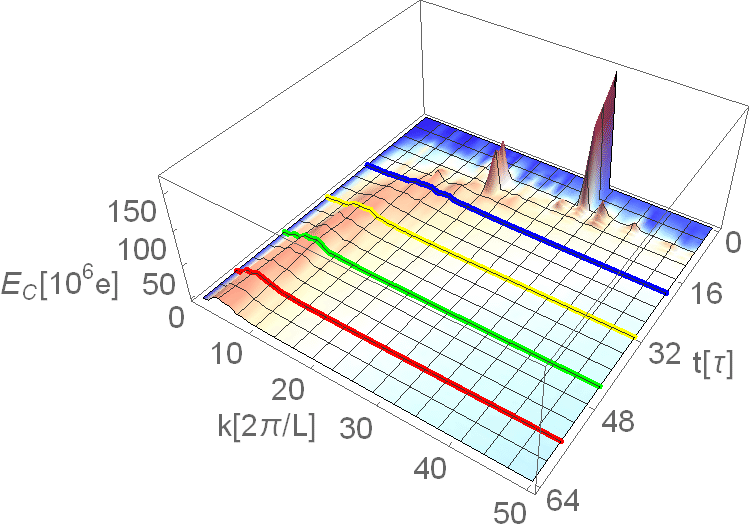} & \includegraphics[scale=0.23]{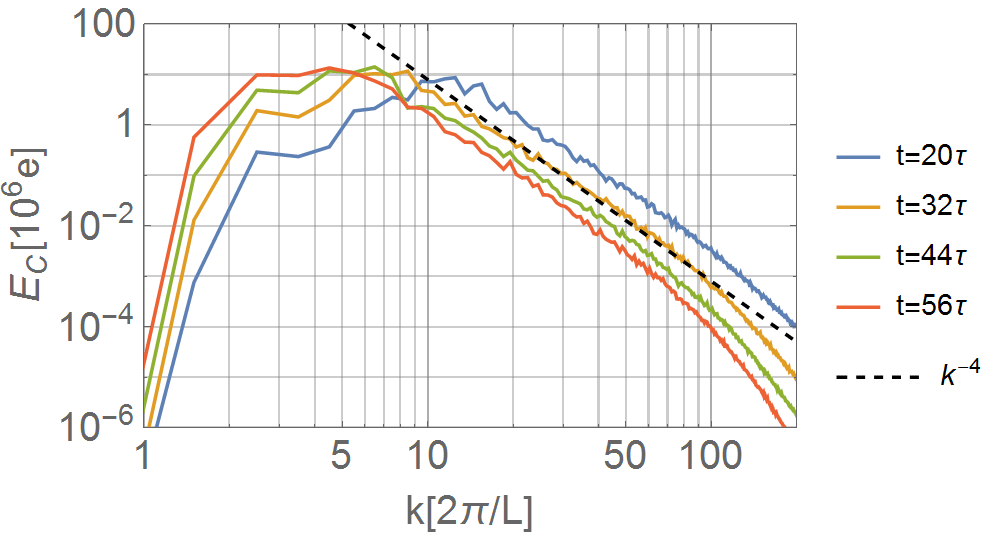}
        \end{tabular}
        \caption{The energy spectrum of two dimensional flow with an initial Reynolds number of $Re=1562.5$ for different initial Mach numbers. The right column provides detailed information on the power spectrum for four selected times.}
        \label{F:2DSpectrumPL}
\end{figure}

\begin{figure}[hbt]
        \centering
        \begin{tabular}{ccc}
        \textbf{\rotatebox{90}{\ \ \ \ \ \ \ \ \ \ $Re=1562.5$}} & \includegraphics[scale=0.23]{2DCascadeM05t0-64.png} & \includegraphics[scale=0.23]{2DSpectrumM05.png} \\
        \textbf{\rotatebox{90}{\ \ \ \ \ \ \ \ \ \ $Re=781.25$}} &\includegraphics[scale=0.23]{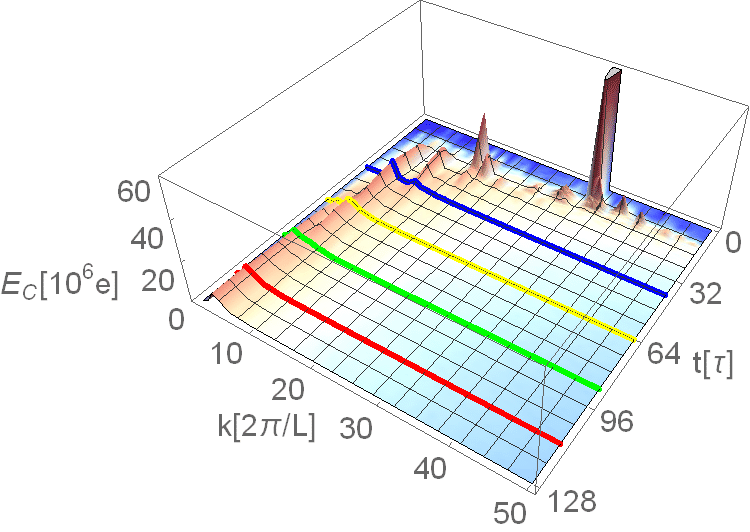} & \includegraphics[scale=0.23]{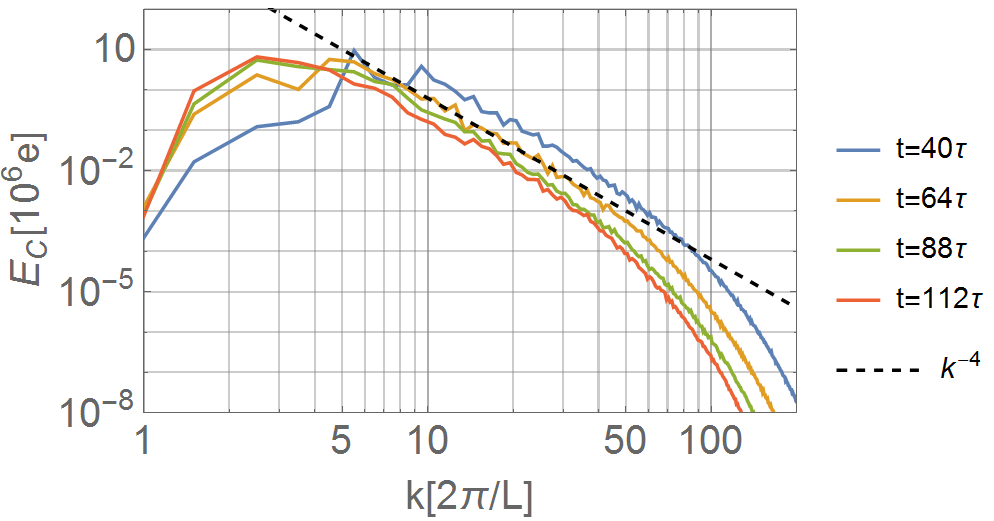} \\
        \textbf{\rotatebox{90}{\ \ \ \ \ \ \ \ \ \ $Re=390.625$}} &\includegraphics[scale=0.23]{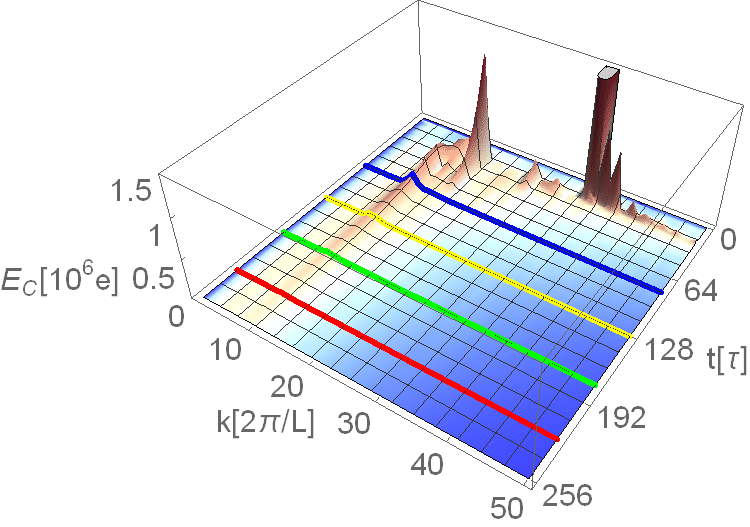} & \includegraphics[scale=0.23]{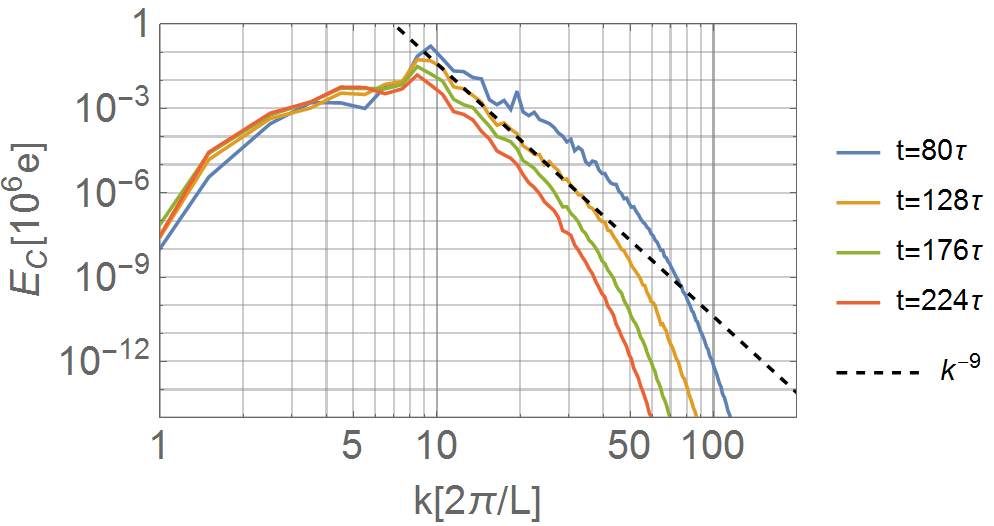}
        \end{tabular}
        \caption{The energy power spectrum for two dimensional fluid flow with an initial Mach number $M=0.5$ and various initial Reynold number. We note that the size of the inertial range, the power law behavior and the times at which power law behavior is observed are dependent on the Reynolds number. This behavior should be contrasted with that exhibited in figure \ref{F:2DSpectrumPL} .}
        \label{F:2DSpectrumPL2}
\end{figure}

As discussed in section \ref{S:TurbPhase}, given a sufficiently high Reynolds number the flow eventually reaches a turbulent regime in which the energy power spectrum $E_C$ has power law behavior.  In two dimensional decaying turbulence we have observed a $k^{-4}$ power law behavior for a variety of initial Mach numbers and initial Reynolds numbers. In figure \ref{F:2DSpectrumPL} we have plotted the energy power spectrum for fixed initial Reynolds number and varying Mach number. In figure \ref{F:2DSpectrumPL2} we have plotted the energy power spectrum for fixed Mach number and varying Reynolds number. As the initial Reynolds number decreases, the time $t$ at which the power law behavior may be observed increases, and the size of the inertial range decreases.

\begin{figure}[hbt]
        \centering
        \begin{tabular}{cccc}
        \textbf{\rotatebox{90}{\ \ \ \ \ \ \ $Re=162.5$}} & \textbf{\rotatebox{90}{\ \ \ \ \ \ \ \ $M=10$}} & \includegraphics[scale=0.22]{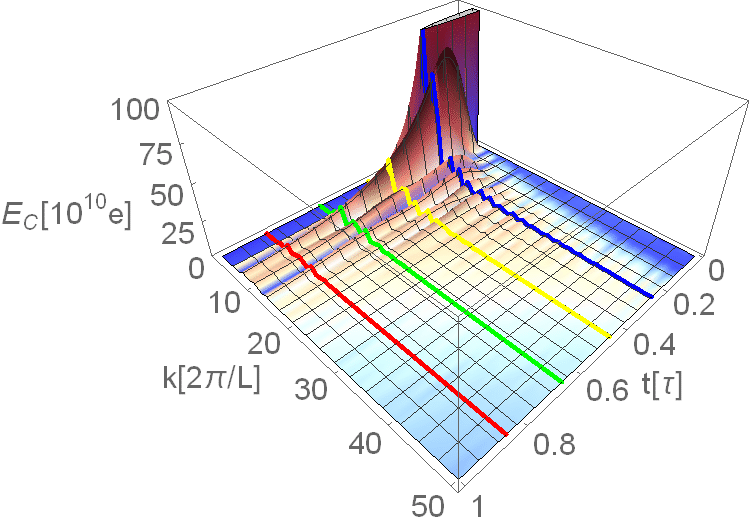} & \includegraphics[scale=0.22]{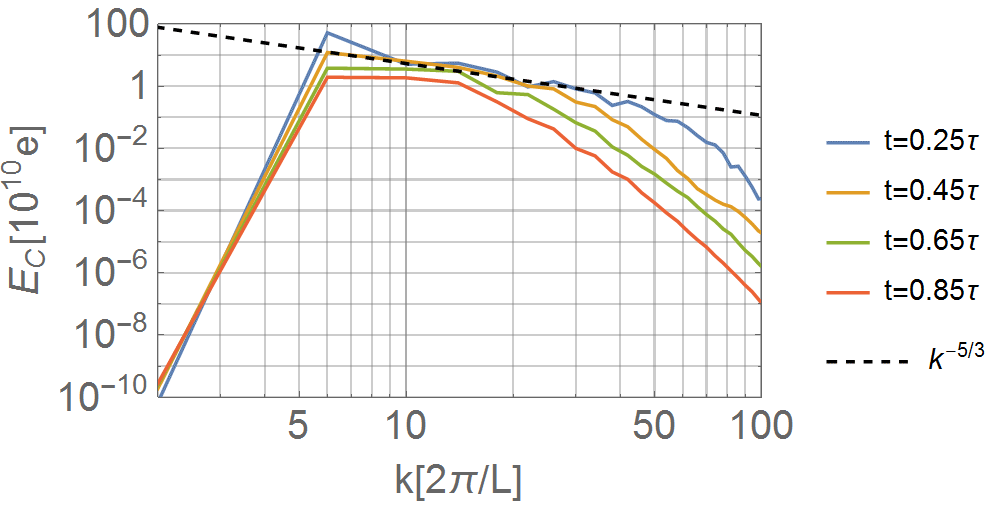} \\
        \textbf{\rotatebox{90}{\ \ \ \ \ \ \ $Re=162.5$}} &\textbf{\rotatebox{90}{\ \ \ \ \ \ \ \ $M=1$}} &\includegraphics[scale=0.22]{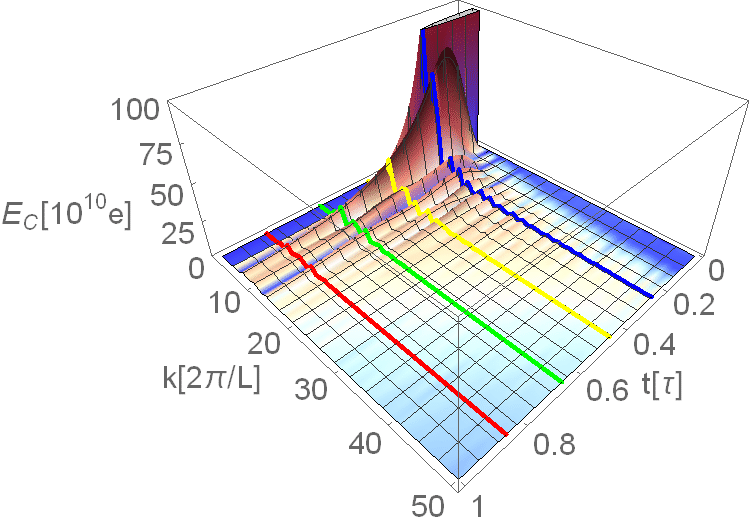} & \includegraphics[scale=0.22]{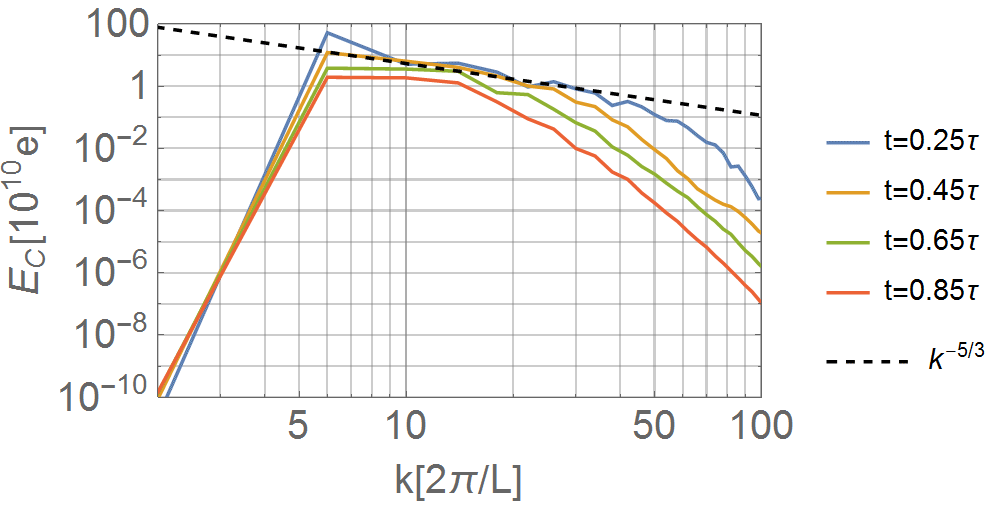} \\
        \textbf{\rotatebox{90}{\ \ \ \ \ \ \ $Re=162.5$}} &\textbf{\rotatebox{90}{\ \ \ \ \ \ \ \ $M=0.1$}} &\includegraphics[scale=0.22]{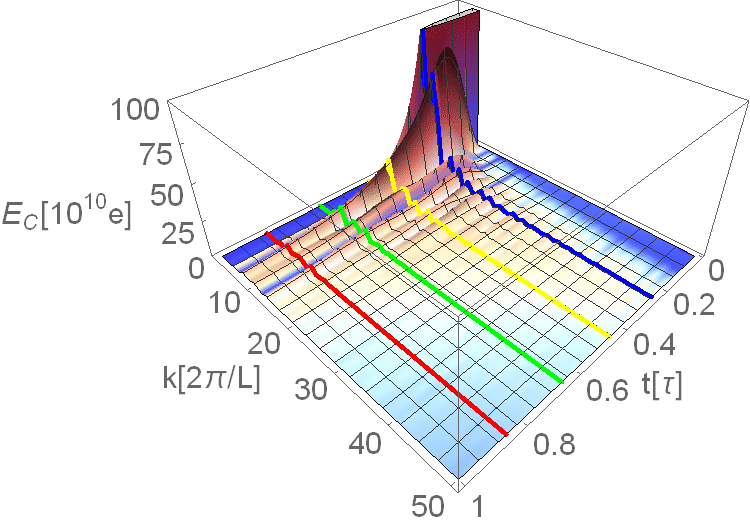} & \includegraphics[scale=0.22]{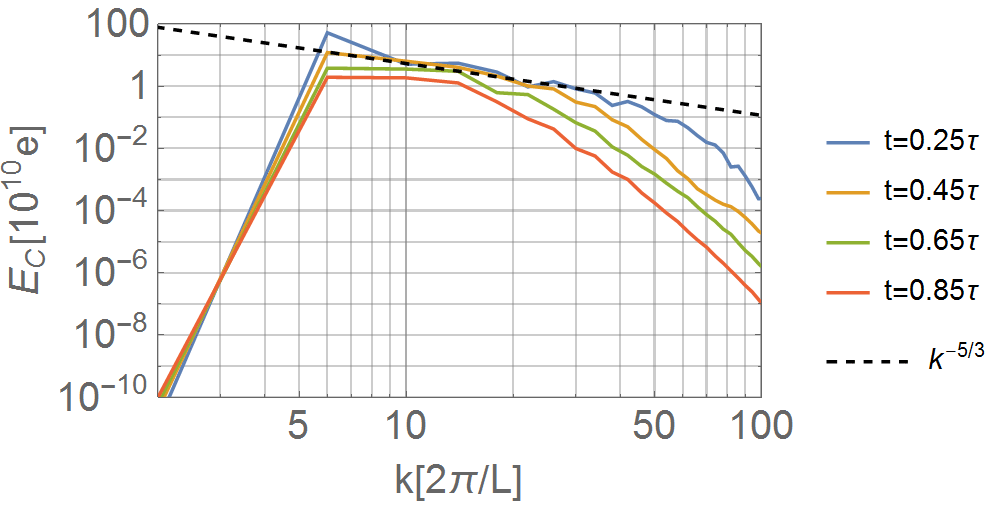} \\
        \textbf{\rotatebox{90}{\ \ \ \ \ \ \ $Re=81.25$}} &\textbf{\rotatebox{90}{\ \ \ \ \ \ \ \ $M=0.1$}} &\includegraphics[scale=0.22]{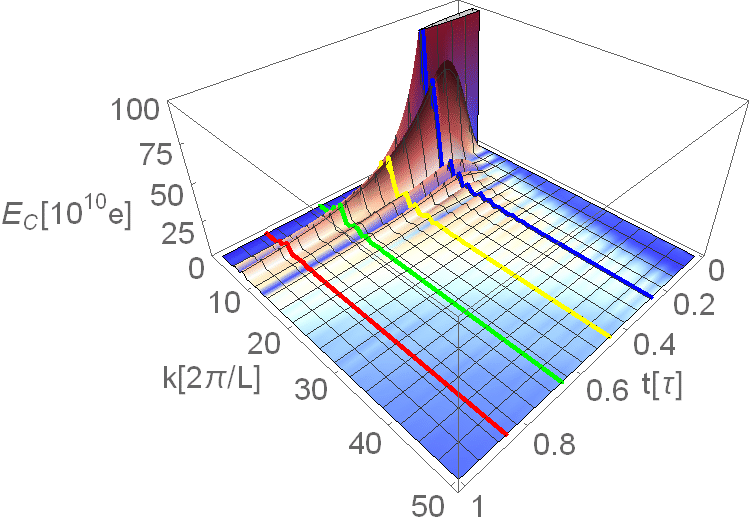} & \includegraphics[scale=0.22]{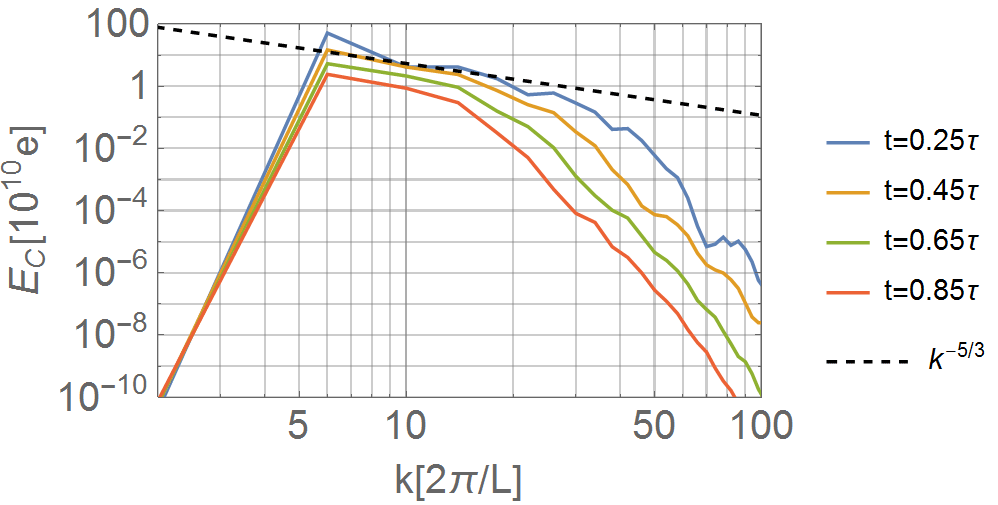}
        \end{tabular}
        \caption{On the left, the energy spectrum of three dimensional flows with initial mode $n=4$ at intermediate times. On the right, log log plots of selected times of same flows shown in color on the full temporal spectrum (on the left), showing the fitting scaling law for each set of parameters. The results on the right are averaged in jumps of four k points, since the initial mode $n=4$ sets the jump scale. One can see the Mach number has little to no effect on the spectrum. All the plots display a weak signature of the expected power law $E_C(t,k) \sim k^{-5/3}$. The lowering of the Reynolds number has a significant effect on the power law scaling range, shortening it from $k \in (10,30)$ when $Re=162.5$ to $k \in (10,18)$ when $Re=81.25$}
        \label{F:3DSpectrumPLn4}
\end{figure}

\begin{figure}[hbt]
        \centering
        \begin{tabular}{cccc}
        \textbf{\rotatebox{90}{\ \ \ \ \ \ \ $Re=750$}} & \textbf{\rotatebox{90}{\ \ \ \ \ \ \ \ $M=10$}} & \includegraphics[scale=0.22]{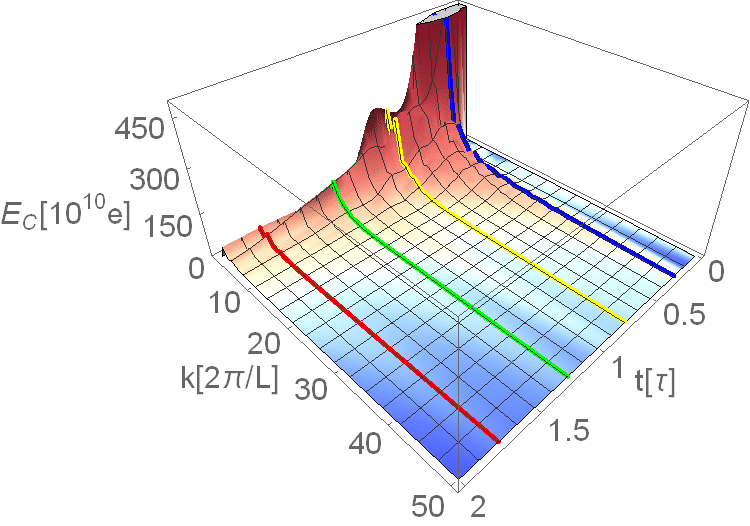} & \includegraphics[scale=0.22]{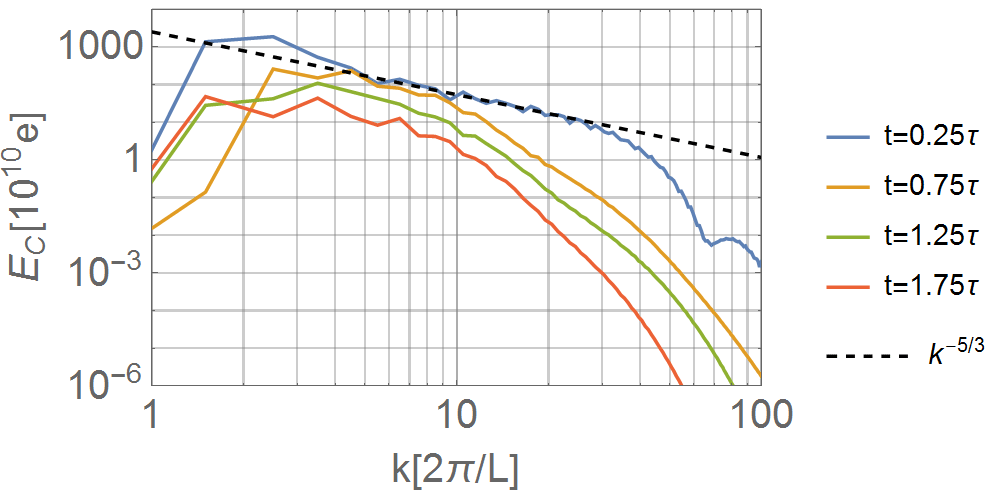} \vspace{5pt}\\
        \textbf{\rotatebox{90}{\ \ \ \ \ \ \ $Re=750$}} &\textbf{\rotatebox{90}{\ \ \ \ \ \ \ \ $M=1$}} &\includegraphics[scale=0.22]{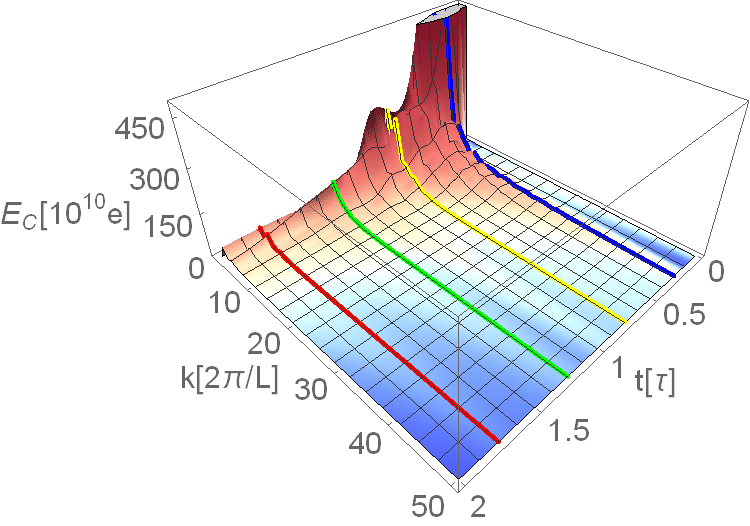} & \includegraphics[scale=0.22]{3DSpectrumM1n1.png} \vspace{5pt}\\
        \textbf{\rotatebox{90}{\ \ \ \ \ \ \ $Re=750$}} &\textbf{\rotatebox{90}{\ \ \ \ \ \ \ \ $M=0.1$}} &\includegraphics[scale=0.22]{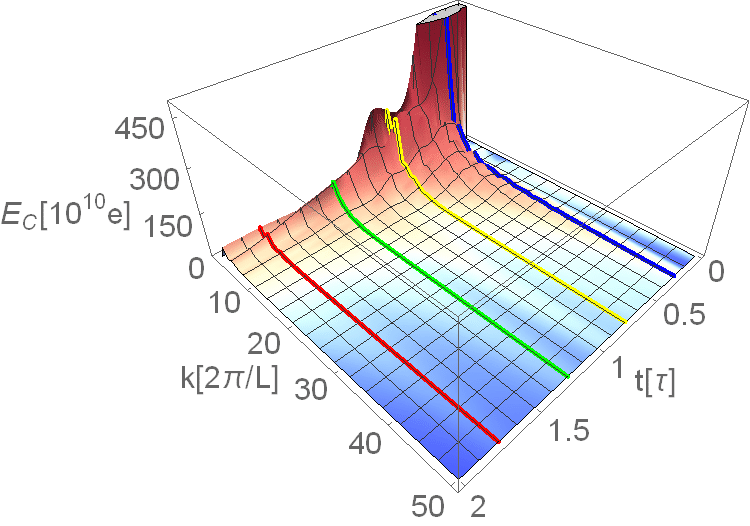} & \includegraphics[scale=0.22]{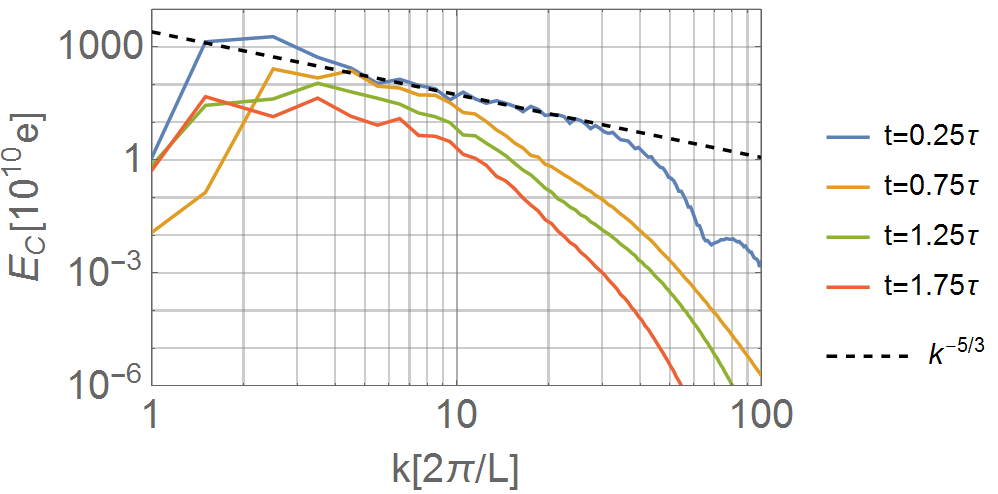} \vspace{5pt}\\
        \textbf{\rotatebox{90}{\ \ \ \ \ \ \ $Re=375$}} &\textbf{\rotatebox{90}{\ \ \ \ \ \ \ \ $M=0.1$}} &\includegraphics[scale=0.22]{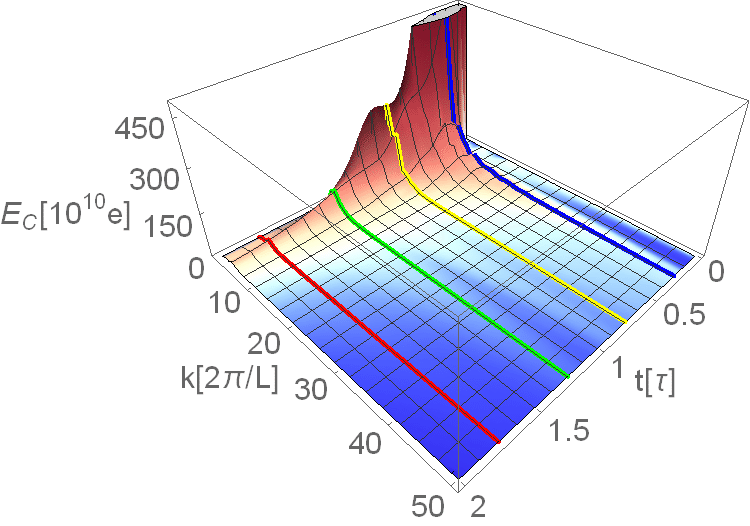} & \includegraphics[scale=0.22]{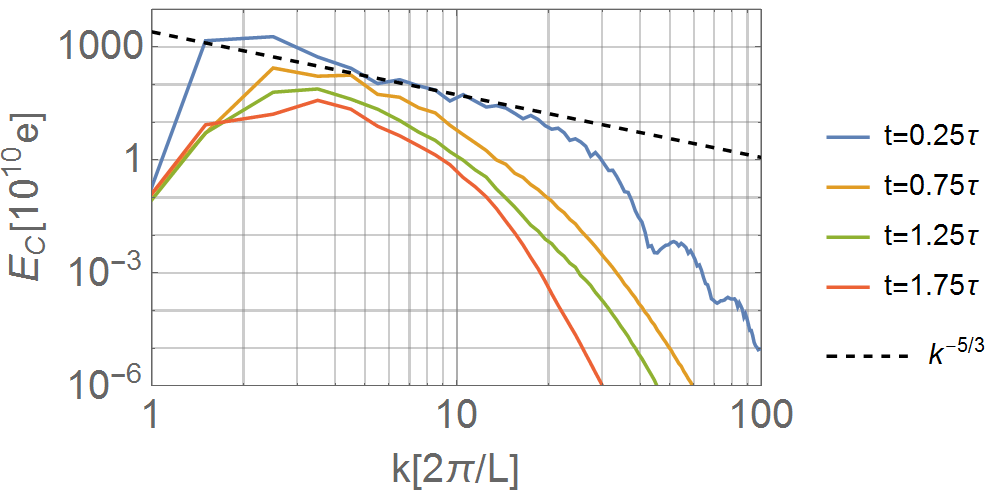}
        \end{tabular}
        \caption{On the left, the energy spectrum of three dimensional flows with initial mode $n=1$ at intermediate times. On the right, log log plots of selected times of same flows shown in color on the full temporal spectrum (on the left). The Mach number has no apparent effect on the spectrum. All the plots display a clear signature of the expected power law $E_C(t,k) \sim k^{-5/3}$. The lowering of the Reynolds number shortens the $k^{-5/3}$ scaling  range from $k \in (4,30)$ when $Re=750$ to $k \in (4,18)$ when $Re=375$}
        \label{F:3DSpectrumPLn1}
\end{figure}

A spectral analysis of $E_C$ at intermediate times, in three spatial dimensions, can be found in figures \ref{F:3DSpectrumPLn4} and \ref{F:3DSpectrumPLn1}. In figure \ref{F:3DSpectrumPLn4} we have plotted $E_C$ for a variety of initial Reynolds and Mach numbers and an initial disturbance \eqref{E:inic} with $n=4$. Figure \ref{F:3DSpectrumPLn1} describes fluid flow for the same initial Reynolds and Mach numbers but an initial disturbance with $n=1$. Somewhat surprisingly, the Kolmogorov power law $E_C \sim k^{-5/3}$, seems to be robust and holds also for large Mach number when the incompressible approximation is no longer valid \cite{Falkovich:2009mb}.

\vspace{10pt}
\section{Analysis of the Horizon Area Power Spectrum}
\label{A:Horizonpower}

In \eqref{E:AElowMach} we have argued that the ratio of the horizon area power law spectrum to the energy spectrum grows like $k^2$ only for flows with low Mach number. Recall that the traceless part of the expansion, $\theta^i{}_j$  is given by equation \eqref{E:thetaExact}. Once the Mach number is small enough the flow becomes incompressible and we may approximate \eqref{E:thetaExact} by \eqref{E:thetaAproximate}  which yields $\mathcal{A}/E_C \sim k^2$. A numerical analysis of $\mathcal{A}/E_C$ for various initial Mach and Reynolds number is displayed in figure \ref{F:HorizSpecCheck2D}. As expected, the dependence of $\mathcal{A}/E_C$ on $k$ seems to deviate from $k^2$ once the Mach number becomes large. 
\begin{figure}[hbtp]
        \centering
        \textbf{$Re=1562.5\ \ M=0.005$}\hspace{4.2cm}\textbf{$Re=1562.5\ \ M=0.05\ $}
        \includegraphics[scale=0.21]{2DHorizSpectrumM0005.png}\hspace{0.5cm}\includegraphics[scale=0.21]{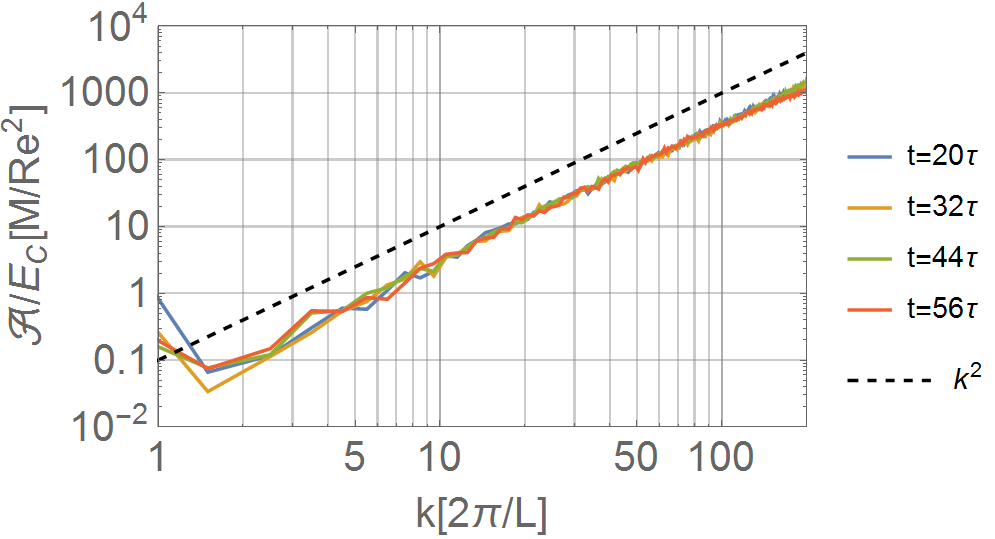}\vspace{10pt}
        \textbf{$Re=1562.5\ \ M=0.5$}\hspace{4.2cm}\textbf{$Re=1562.5\ \ M=2\ \ $}
        \includegraphics[scale=0.21]{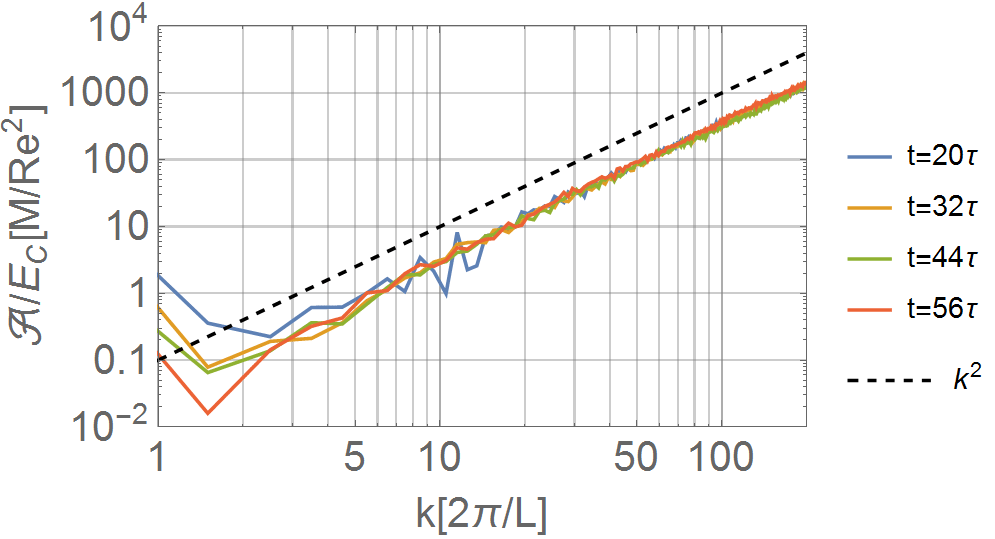}\hspace{0.5cm}\includegraphics[scale=0.21]{2DHorizSpectrumM2.png}\vspace{10pt}
        \textbf{$Re=781.25\ \ \ M=0.5$}\hspace{4.2cm} \textbf{$Re=390.625\ \ M=0.5$}
        \includegraphics[scale=0.21]{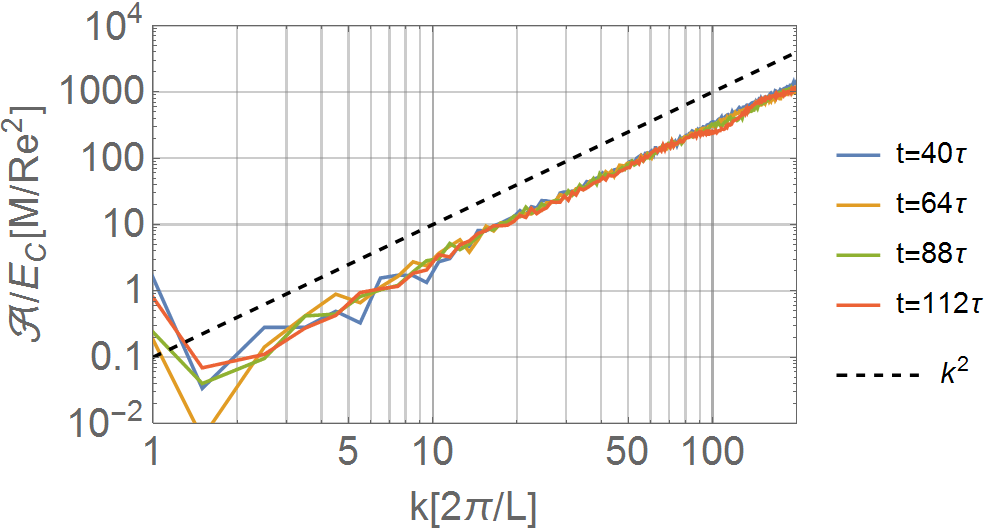}\hspace{0.5cm}\includegraphics[scale=0.21]{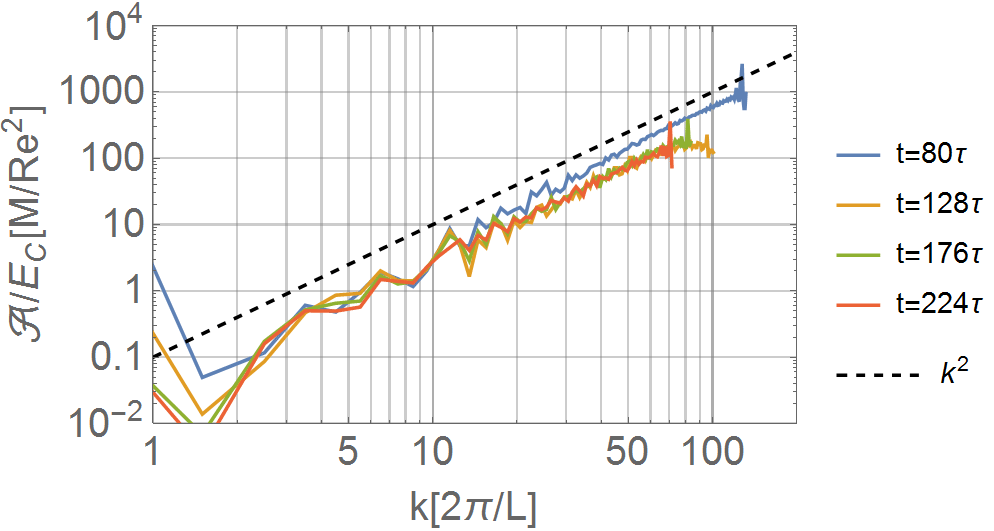}
        \caption{The ratio $\mathcal{A}(t,k)/E_C(t,k)$ for two dimensional turbulent flows at intermediate times. The dashed line scales like $k^{2}$. The ratio $\mathcal{A}/E_C$ deviates from the expected $k^2$ scaling for $M=2$.}
        \label{F:HorizSpec2D}
\end{figure}

To quantify the deviation of $\mathcal{A}/E_c$ from a $k^2$ behavior we go back to the compressible contributions to the traceless expansion \eqref{E:thetaExact}.
Let us denote the Fourier transform of a quantity $X$ by $\mathcal{F}(X)$. Then, in order for the incompressible terms in \eqref{E:thetaExact} to dominate over the compressible ones, we need $\frac{\partial}{\partial k}\int dk^p\left|\mathcal{F}\left[\partial_{j}(f_{\ell}/a)\right]\right|^2$ to be much larger than $\frac{\partial}{\partial k}\int dk^p\left|\mathcal{F}\left[\partial_{j}(\partial_{\ell}a/a)\right]\right|^2$. In order quantify this relation we define the following matrix,
\begin{equation}
	M_{j\ell} = \frac{\frac{\partial}{\partial k}\int dk^p\left|\mathcal{F}\left[\partial_{j}(\partial_{\ell}a/a)\right]\right|^2}{\frac{\partial}{\partial k}\int dk^p\left|\mathcal{F}\left[\partial_{j}(f_{\ell}/a)\right]\right|^2},
\end{equation}
which should become negligible whenever $\mathcal{A}/E_C \propto k^2$ is valid. In figure \ref{F:HorizSpecCheck2D} we have plotted $M_{ij}$ for a two dimensional flow with Mach numbers ranging from $5\times 10^{-3}$ to $2$. While $M_{j=\ell}$ remains small, $M_{j\neq l}$ becomes non negligible at large Mach number.
\begin{figure}[hbtp]
        \centering
        \includegraphics[scale=0.32]{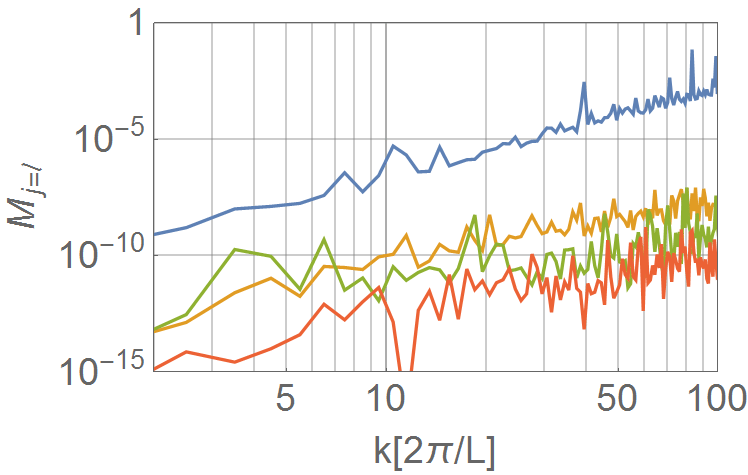}\hspace{0.5cm}\includegraphics[scale=0.24]{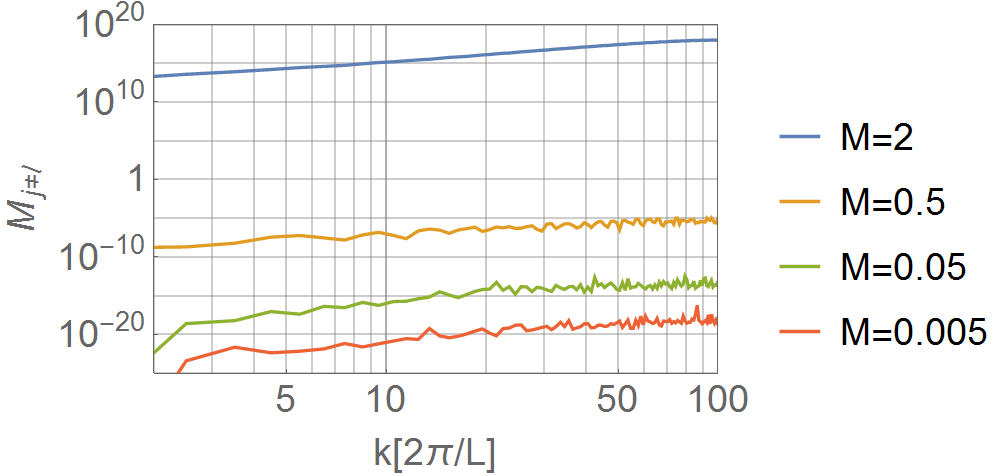}
        \caption{The relation between $\frac{\partial}{\partial k}\int dk^p\left|\mathcal{F}\left[\partial_{j}(\partial_{\ell}a/a)\right]\right|^2$ and $\frac{\partial}{\partial k}\int dk^p\left|\mathcal{F}\left[\partial_{j}(f_{\ell}/a)\right]\right|^2$ denoted by $M_{j\ell}$ for two dimensional turbulent flows with $Re=1562.5$ appearing in figure \ref{F:HorizSpec2D} at time $t=32\tau$. When the Mach number is very large $M_{j\ne\ell}$ becomes very large signifying the breakdown of $\mathcal{A}/E_c \propto k^2$.}
        \label{F:HorizSpecCheck2D}
\end{figure}

\begin{figure}[hbtp]
        \centering
        \textbf{$Re=162.5\ \ M=10$}\hspace{4.7cm}\textbf{$Re=162.5\ \ M=1\ $}
        \includegraphics[scale=0.21]{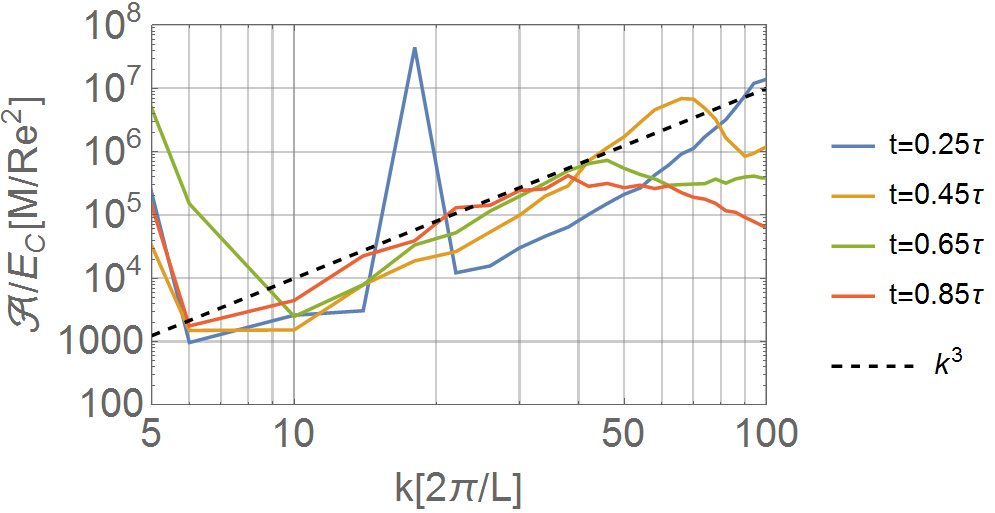}\hspace{0.5cm}\includegraphics[scale=0.21]{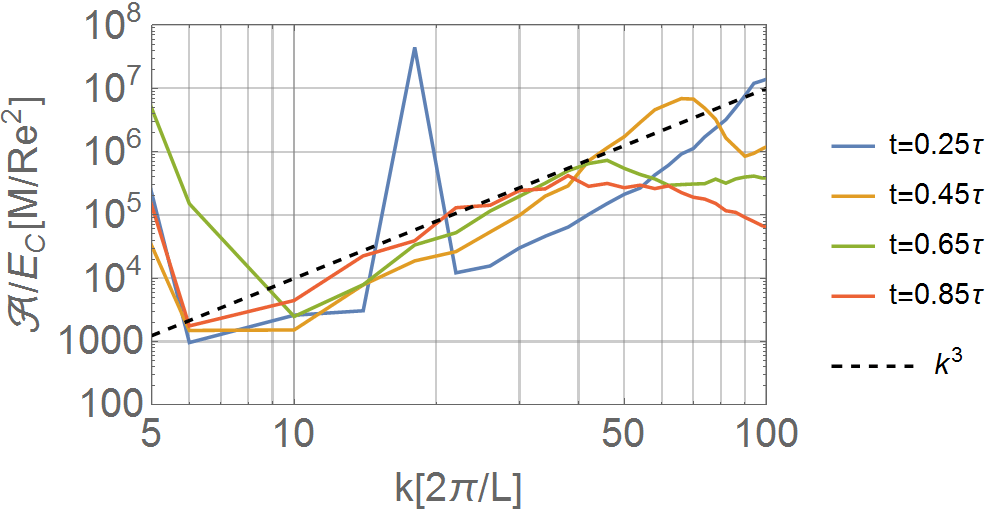}\vspace{10pt}
        \textbf{$Re=162.5\ \ M=0.1$}\hspace{4.7cm}\textbf{$Re=81.25\ \ M=0.1\ $}
        \includegraphics[scale=0.21]{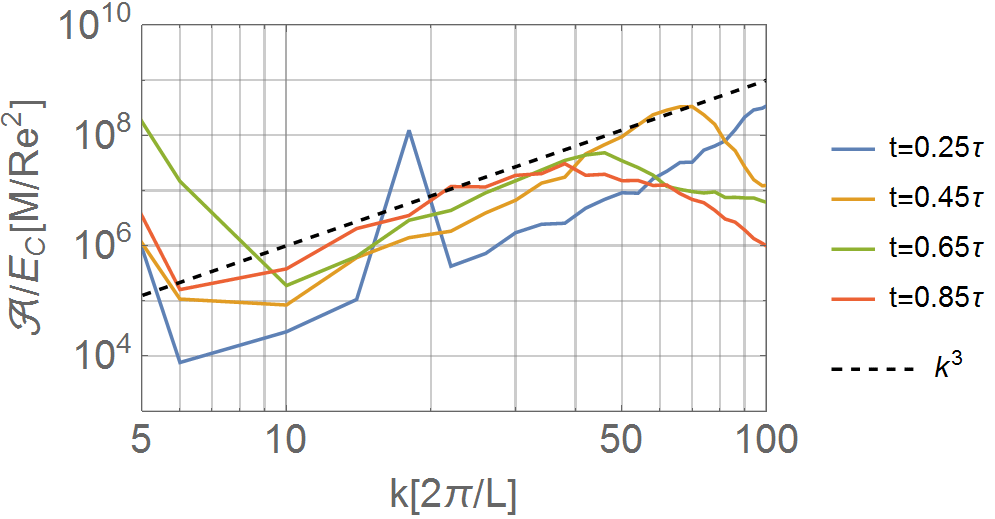}\hspace{0.5cm}\includegraphics[scale=0.21]{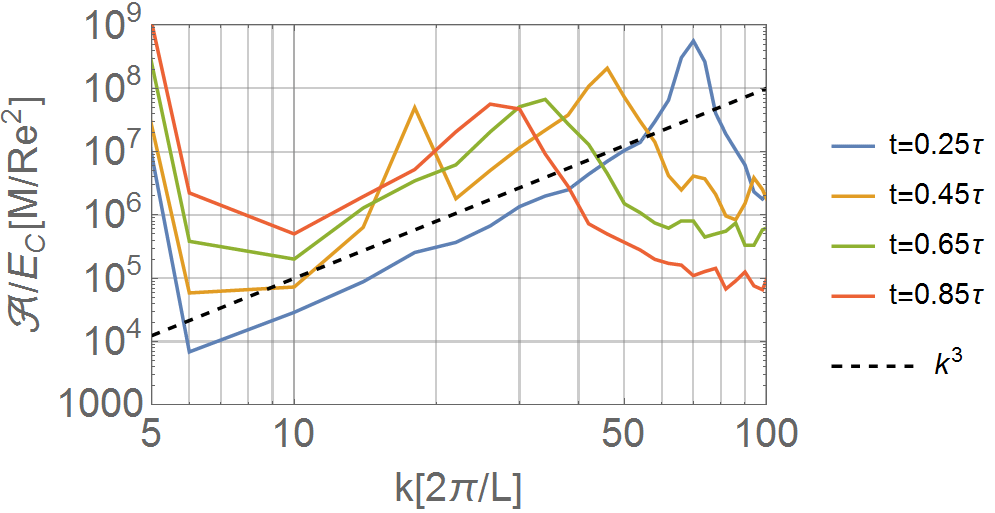}
        \caption{$\mathcal{A}(t,k)/E_C(t,k)$ for three dimensional turbulent flows with initial mode $n=4$ at intermediate times. The dashed line is proportional to $k^{3}$.  }
        \label{F:HorizSpec3Dn4}
\end{figure}

\begin{figure}[hbtp]
        \centering
        \textbf{$Re=750\ \ M=10$}\hspace{5cm}\textbf{$Re=750\ \ M=1\ $}
        \includegraphics[scale=0.21]{3DHorizSpectrumM10n1.png}\hspace{0.5cm}\includegraphics[scale=0.21]{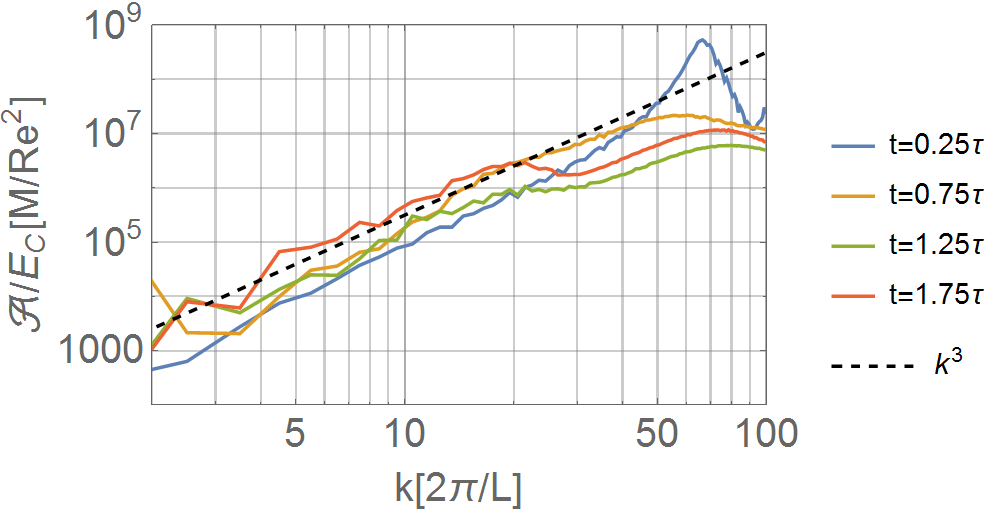}\vspace{10pt}
        \textbf{$Re=750\ \ M=0.1$}\hspace{4.8cm}\textbf{$Re=375\ \ M=0.1\ $}
        \includegraphics[scale=0.21]{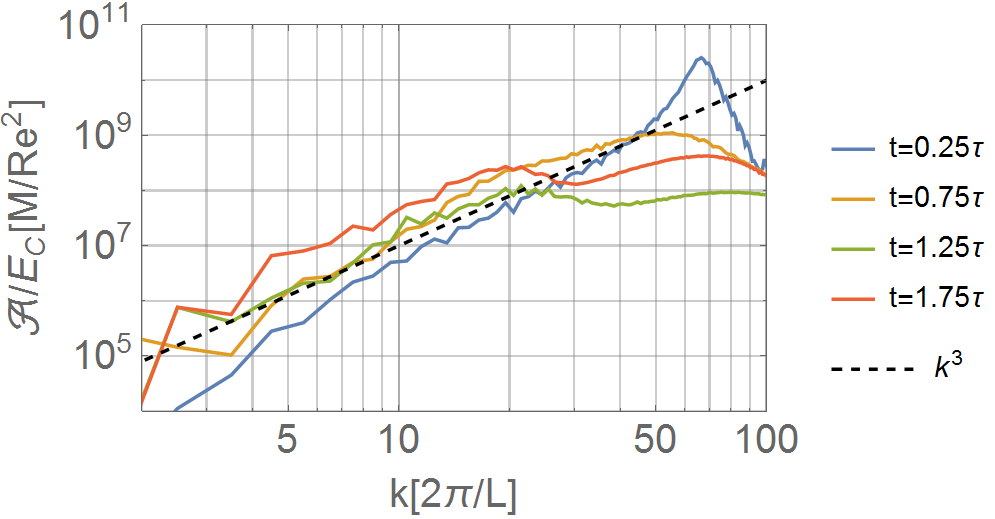}\hspace{0.5cm}\includegraphics[scale=0.21]{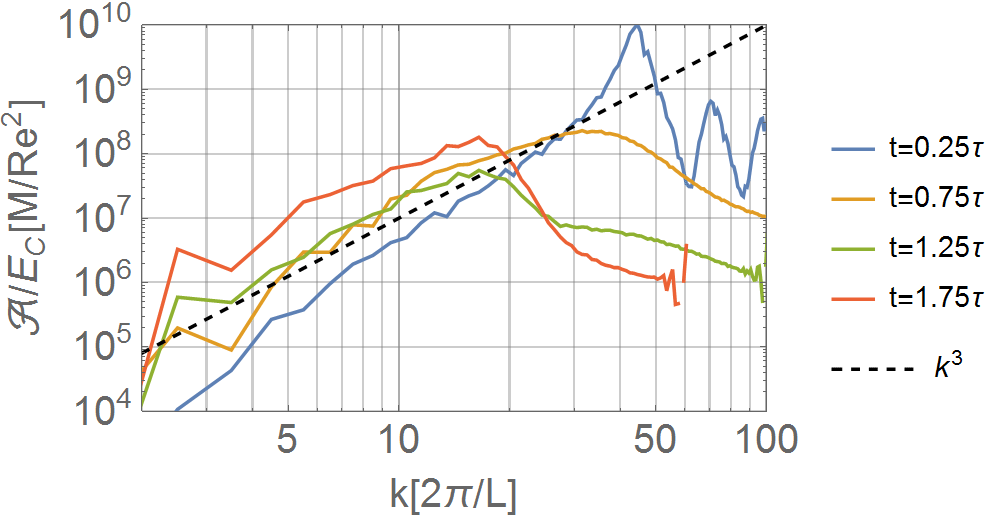}
        \caption{$\mathcal{A}(t,k)/E_C(t,k)$ for three dimensional turbulent flows with initial mode $n=1$ at intermediate times. The dashed line corresponds to a $k^{3}$ behavior.}
        \label{F:HorizSpec3Dn1}
\end{figure}

\begin{figure}[hbtp]
        \centering
        \includegraphics[scale=0.34]{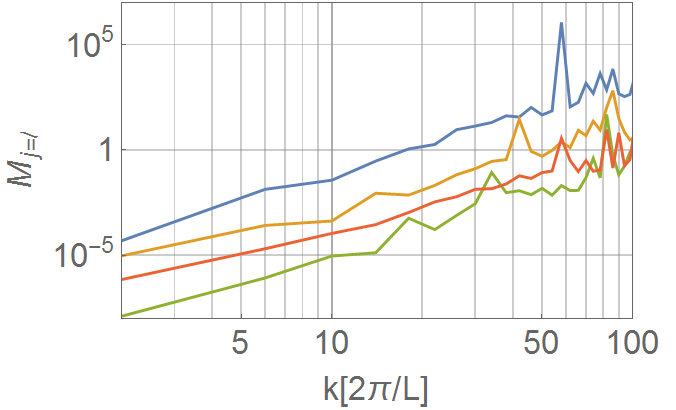}\hspace{0.5cm}\includegraphics[scale=0.25]{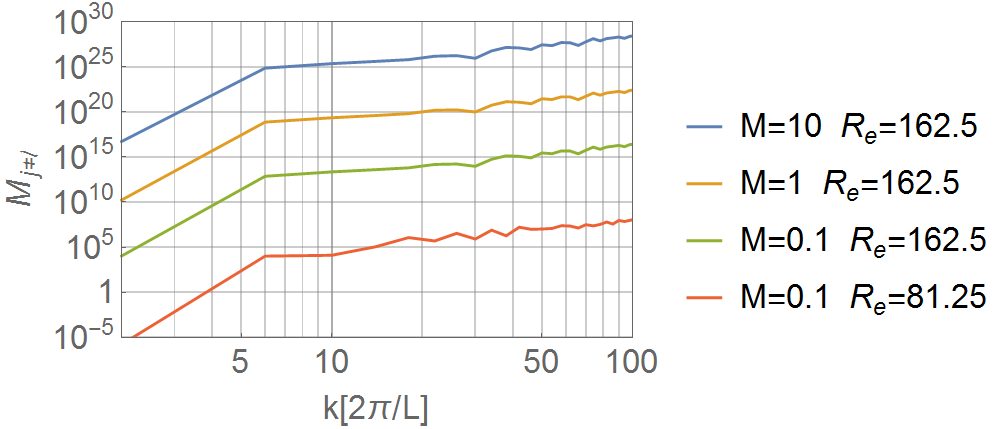}\vspace{10pt}
        \includegraphics[scale=0.34]{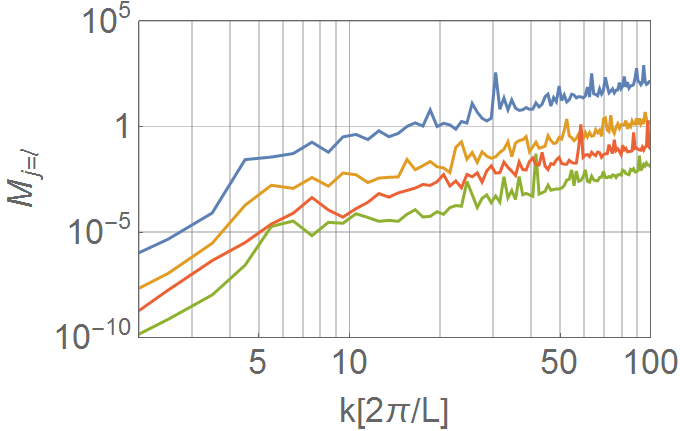}\hspace{0.5cm}\includegraphics[scale=0.25]{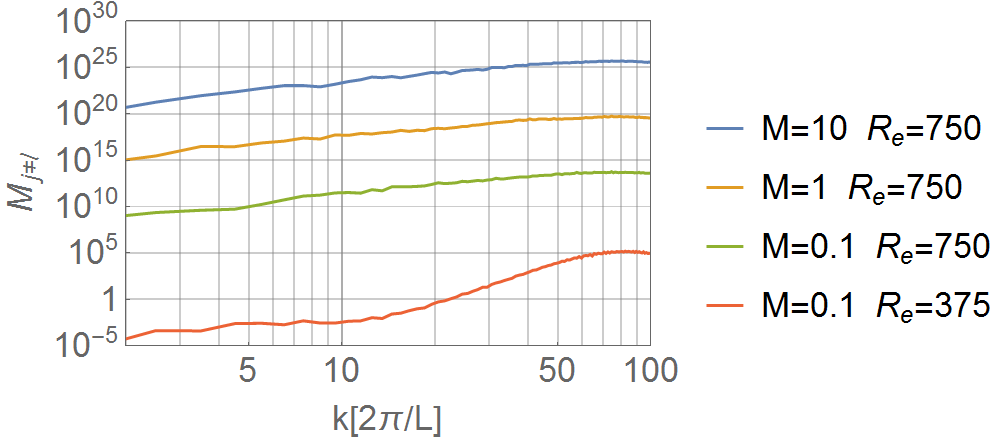}
        \caption{The relation between $\frac{\partial}{\partial k}\int dk^p\left|\mathcal{F}\left[\partial_{j}(\partial_{\ell}a/a)\right]\right|^2$ and $\frac{\partial}{\partial k}\int dk^p\left|\mathcal{F}\left[\partial_{j}(f_{\ell}/a)\right]\right|^2$ denoted by $M_{j\ell}$ for three dimensional turbulent flows. Above, $M_{j\ell}$ for the flows with initial mode $n=4$ appearing in figure \ref{F:HorizSpec3Dn4} at time $t=0.45\tau$. Below, $M_{j\ell}$ for the flows with initial mode $n=1$ appearing in figure \ref{F:HorizSpec3Dn1}. The right plots shows the typical relation for the diagonal terms $M_{j=\ell}$. The left plots depicts the typical relation for the off diagonal terms $M_{j\ne\ell}$. The variance between Mach numbers, rather high or low appears almost linear on the log scale for $M_{j\ne\ell}$, and is drastically different from the two dimensional flows on figure \ref{F:HorizSpec2D}. This is responsible for the lack of accuracy of the relation \eqref{E:AElowMach}. The aforementioned variance between Mach numbers leads to the conclusion that a much lower Mach number than $M \sim 0.1$ is required in order to drop high order derivative terms in \eqref{E:thetaExact}.}
        \label{F:HorizSpecCheck3D}
\end{figure}

A similar analysis for three dimensional systems can be found in figures \ref{F:HorizSpec3Dn4} and \ref{F:HorizSpec3Dn1} which display $\mathcal{A}/E_C$ for flows with initial conditions \eqref{E:inic} and $n=4$ or $n=1$ respectively. For both types of flows a $k^3$ scaling law is observed, even for the lowest Mach number which we could numerically access, $M=0.1$.
Extrapolating the data regarding the ratios of $\frac{\partial}{\partial k}\int dk^p\left|\mathcal{F}\left[\partial_{j}(f_{l}/a)\right]\right|^2$ to $\frac{\partial}{\partial k}\int dk^p\left|\mathcal{F}\left[\partial_{j}(\partial_{l}a/a)\right]\right|^2$ from figure \ref{F:HorizSpecCheck3D} we estimate that in order for the flow to become incompressible we need $M \sim 10^{-4}$ with $Re=750$ and $n=1$ or $Re=162$ and $n=4$. Three dimensional flows with such a low Mach number are expensive and will not be covered in this work.

\end{appendix}

\vspace{20pt}
\bibliographystyle{JHEP}
\bibliography{largedbib}

\end{document}